\documentstyle[12pt,psfig]{livrev}

\newcommand\eg{{\it e.g.\ }}
\newcommand\etc{{\it etc.\ }}
\newcommand\ie{{\it i.e.\ }}
\newcommand\etal{{\it et~al.\ }}
\newcommand\PRD{{\it Phys. Rev. D}}
\newcommand\PRL{{\it Phys. Rev. Lett.}}

\begin{document}

\title{The Confrontation between General Relativity and Experiment}

\author{Clifford M. Will\\
McDonnell Center for the Space Sciences \\
 Department of Physics \\
 Washington University, St. Louis MO 63130}

\date{}
\maketitle
\begin{abstract}
The status of experimental tests of general relativity and of
theoretical frameworks for analysing them are reviewed.
Einstein's equivalence principle (EEP) is well supported by experiments
such as the E\"otv\"os experiment, tests of special relativity, and
the gravitational redshift experiment.  Future tests of EEP and of the
inverse square law will
search for new interactions arising from unification or quantum
gravity.  Tests of general
relativity at the post-Newtonian level
have reached high precision, including the light
deflection, the Shapiro time delay, the perihelion advance of
Mercury, and the Nordtvedt effect in lunar motion.  Gravitational
wave damping has been detected to half a percent using the binary
pulsar, and new binary pulsar systems may yield further improvements.
When direct observation of gravitational radiation from astrophysical
sources begins, new tests of general relativity will be possible.
\end{abstract}

%========================================================
%  Section 1
%========================================================

\section{Introduction}\label{S1}

At the time of the birth of general relativity (GR), experimental
confirmation was almost a side issue.  Einstein did
calculate observable effects of general relativity, such as the
perihelion advance of Mercury, which he knew to be an unsolved
problem, and the
deflection of light, which was subsequently verified, but compared to the inner
consistency and elegance of the theory, he regarded such empirical
questions as almost peripheral.  But today, experimental
gravitation is a major component of the field, characterized by
continuing efforts to test the theory's predictions, to search
for gravitational imprints of high-energy particle interactions,
and to detect gravitational
waves from astronomical sources.  

The modern history of experimental relativity can be divided
roughly into four periods, Genesis, Hibernation, a Golden Era,
and the Quest for Strong Gravity.  The Genesis (1887--1919) comprises
the period of the two great experiments which were the foundation
of relativistic physics---the Michelson-Morley experiment and the
E\"otv\"os experiment---and the two immediate confirmations of
GR---the deflection of light and the
perihelion advance of Mercury.  Following this was a period of
Hibernation (1920--1960) during which theoretical work temporarily
outstripped technology and experimental possibilities, and, as a
consequence 
the field stagnated and was relegated to the
backwaters of physics and astronomy.

But beginning around 1960, astronomical discoveries (quasars,
pulsars, cosmic background radiation) and new experiments pushed
GR to the forefront.  Experimental gravitation
experienced a Golden Era (1960--1980) during which a systematic,
world-wide effort took place to understand the observable
predictions of GR, to compare and contrast them
with the predictions of alternative theories of gravity, and to
perform new experiments to test them.  The period began with an
experiment to confirm the gravitational frequency shift of light
(1960) and ended with the reported decrease in the orbital period
of the binary pulsar at a rate consistent with the general
relativity prediction of gravity-wave energy loss (1979).  The
results all supported GR, and most alternative
theories of gravity fell by the wayside (for a popular review, see
\cite{WER}).

Since 1980, the field has entered what might be termed a Quest for
Strong Gravity.  Many of the remaining interesting weak-field predictions of
the theory are extremely small and difficult to check, in some
cases requiring further technological development to bring them
into detectable range.  The sense of a systematic assault on the
weak-field
predictions of GR has been supplanted to some
extent by an opportunistic approach in which novel and unexpected
(and sometimes inexpensive) tests of gravity have arisen from new
theoretical ideas or experimental techniques, often from unlikely
sources.  Examples include the use of laser-cooled atom and ion
traps to perform ultra-precise tests of special relativity; 
the proposal of a ``fifth'' force, which led to a host
of new tests of the weak equivalence principle; and recent ideas of
large extra dimensions, which have motived new tests of the inverse
square law of gravity at sub-millimeter scales.  Several major ongoing
efforts also continue, principally the Stanford Gyroscope
experiment, known as Gravity Probe-B.  

Instead, much of the focus has shifted to experiments
which can probe the effects of strong gravitational fields.  The
principal figure of merit that distinguishes
strong from weak gravity is the quantity $\epsilon \sim GM/Rc^2$,
where $G$ is the
Newtonian gravitational constant, $M$ is the characteristic mass scale
of the phenomenon, $R$  is the characteristic distance scale, and $c$
is the speed of light.  Near the
event horizon of a non-rotating black hole, or for the expanding
observable
universe,
$\epsilon \sim 0.5$; for neutron stars, $\epsilon \sim 0.2$.  These
are
the regimes of strong gravity.  For the solar system $\epsilon <
10^{-5}$;
this is the regime of weak gravity.  At one
extreme are the strong gravitational fields associated with
Planck-scale physics.  Will unification of the forces, or
quantization of gravity at this scale leave observable effects
accessible by experiment?  Dramatically improved tests of the
equivalence principle or of the inverse square law are being
designed, to search for or bound the imprinted effects of Planck-scale
phenomena.  At the other extreme are the strong fields associated with
compact objects such as black holes or neutron stars.  Astrophysical
observations and gravitational-wave detectors are being planned to
explore and test GR in the strong-field,
highly-dynamical regime associated with the formation and dynamics of
these objects.

In this living review, we shall survey the theoretical frameworks
for studying experimental gravitation, summarize the current
status of experiments, and attempt to chart the future of the
subject.
We shall not provide complete references to early work done in 
this field but instead will refer the reader to the appropriate
review articles and monographs, specifically to {\it Theory and
Experiment in Gravitational Physics} \cite{tegp}, hereafter referred
to as TEGP.  Additional recent reviews in this subject are
\cite{Will300,ijmp,sussp,slac,damourreview,shapiro}.
References to TEGP will be by chapter or section, \eg ``TEGP 8.9''.

%================================================================
% Section 2
%================================================================

\section{Tests of the Foundations of Gravitation \break Theory}\label{S2}

%---------------------------------------
% Subsection 2.1
%---------------------------------------

\subsection{The Einstein Equivalence Principle}\label{eep}

The principle of equivalence has historically played an important
role in the development of gravitation theory.  Newton regarded
this principle as such a cornerstone of mechanics that he devoted
the opening paragraph of the {\it Principia} to it.  In 1907,
Einstein used the principle as a basic element of general
relativity.  We now regard the principle of equivalence as the
foundation, not of Newtonian gravity or of GR,
but of the broader idea that spacetime is curved.

One elementary equivalence principle is the kind Newton had in
mind when he stated that the property of a body called ``mass'' is
proportional to the ``weight'', and is known as the weak
equivalence principle 
(WEP)\index{WEP,weak equivalence principle}.
  An alternative statement of WEP is
that the trajectory of a freely falling body (one not acted upon
by such forces as electromagnetism and too small to be affected
by tidal gravitational forces) is independent of its internal
structure and composition.  In the simplest case of dropping two
different bodies in a gravitational field, WEP states that the
bodies fall with the same acceleration (this is often termed the
Universality of Free Fall, or UFF).

A more powerful and far-reaching equivalence principle is
known as the Einstein equivalence principle 
(EEP)\index{EEP, Einstein equivalence principle}.
It states that
\begin{enumerate}
\item 
WEP is valid.
\item
The outcome of any local
non-gravitational experiment is independent of the velocity of
the freely-falling reference frame in which it is performed.
\item
The outcome of any local non-gravitational experiment is
independent of where and when in the universe it is performed.
\end{enumerate}

The second piece of EEP is called local Lorentz invariance 
(LLI)\index{LLI, local Lorentz invariance},
and the third piece is called local position invariance 
(LPI)\index{LPI, local position invariance}.

For example, a measurement of the electric force between two
charged bodies is a local non-gravitational experiment; a
measurement of the gravitational force between two bodies
(Cavendish experiment) is not.

The Einstein equivalence
principle is the heart and soul of gravitational theory, for it
is possible to argue convincingly that if EEP is valid, then
gravitation must be a ``curved spacetime'' phenomenon, in other
words, the effects of gravity must be equivalent to the effects of
living in a curved spacetime.  As a consequence of this argument,
the only theories of gravity that can embody EEP are those that
satisfy the postulates of ``metric theories of gravity'', which
are 
\begin{enumerate}
\item
Spacetime is endowed with a symmetric metric.
\item
The
trajectories of freely falling bodies are geodesics of that
metric.
\item
In local freely falling reference frames, the
non-gravitational laws of physics are those written in the
language of special relativity.  
\end{enumerate}

The argument that leads to this
conclusion simply notes that, if EEP is valid, then in local
freely falling frames, the laws governing experiments must be
independent of the velocity of the frame (local Lorentz
invariance), with constant values for the various atomic
constants (in order to be independent of location).  The only
laws we know of that fulfill this are those that are compatible
with special relativity, such as Maxwell's equations of
electromagnetism.  Furthermore, in local freely falling frames,
test bodies appear to be unaccelerated, in other words they move
on straight lines; but such ``locally straight'' lines simply
correspond to ``geodesics'' in a curved spacetime (TEGP 2.3).

General relativity is a metric theory of gravity, but then so are
many others, including the Brans-Dicke theory.  The nonsymmetric
gravitation theory (NGT) of Moffat is not a metric theory.  Neither, in
this narrow sense, is
superstring theory (see Sec. \ref{newinteractions}), which, while
based fundamentally on a spacetime metric, introduces additional
fields (dilatons, moduli)
that can couple to material stress-energy in a 
way that can lead to violations, say, of WEP.  So
the notion of curved spacetime is a very general and fundamental
one, and therefore it is important to test the various aspects of
the Einstein Equivalence Principle thoroughly.  

A direct test of WEP is the comparison of the acceleration of two
laborat\-ory-sized bodies of different composition in an external
gravitational field\index{E\"otv\"os experiment}.  
If the principle were violated, then the
accelerations of different bodies would differ.  The simplest
way to quantify such possible violations of WEP in a form
suitable for comparison with experiment is to suppose that for
a body with inertial mass $m_I$, the passive gravitational
mass $m_P$ is no longer equal to $m_I$, so that in a
gravitational field $g$, the acceleration is given 
by $m_I a= m_P g$.  Now the inertial mass of a typical
laboratory body is made up of several types of mass-energy:  rest
energy, electromagnetic energy, weak-interaction energy, and so
on.  If one of these forms of energy contributes to $m_P$
differently than it does to $m_I$, a violation of WEP would
result.  One could then write
\begin{equation}\label{E1}
    m_P = m_I + \sum_A \eta^A E^A /c^2 \,,
\end{equation}
where $E^A$ is the internal energy of the body generated by
interaction $A$, and $\eta^A$ is a dimensionless parameter that
measures the strength of the violation of WEP induced by that
interaction, and $c$ is the speed of light.  A measurement or limit
on the fractional difference in acceleration between two bodies
then yields a quantity called the ``E\"otv\"os ratio'' given by
\begin{equation}\label{E2}
 \eta \equiv {{2 | a_1  -  a_2 |} \over {| a_1 +  a_2 |}} 
     = \sum_A \eta^A
 \left (   {{E_1^A} \over {m_1 c^2}}
      -  {{E_2^A} \over {m_2 c^2} }
\right )  \,,
\end{equation}
where we drop the subscript I from the inertial masses.
Thus, experimental limits on $\eta$ place limits on the
WEP-violation parameters $\eta^A$.

\begin{figure}
\begin{center}
\leavevmode
\psfig{figure=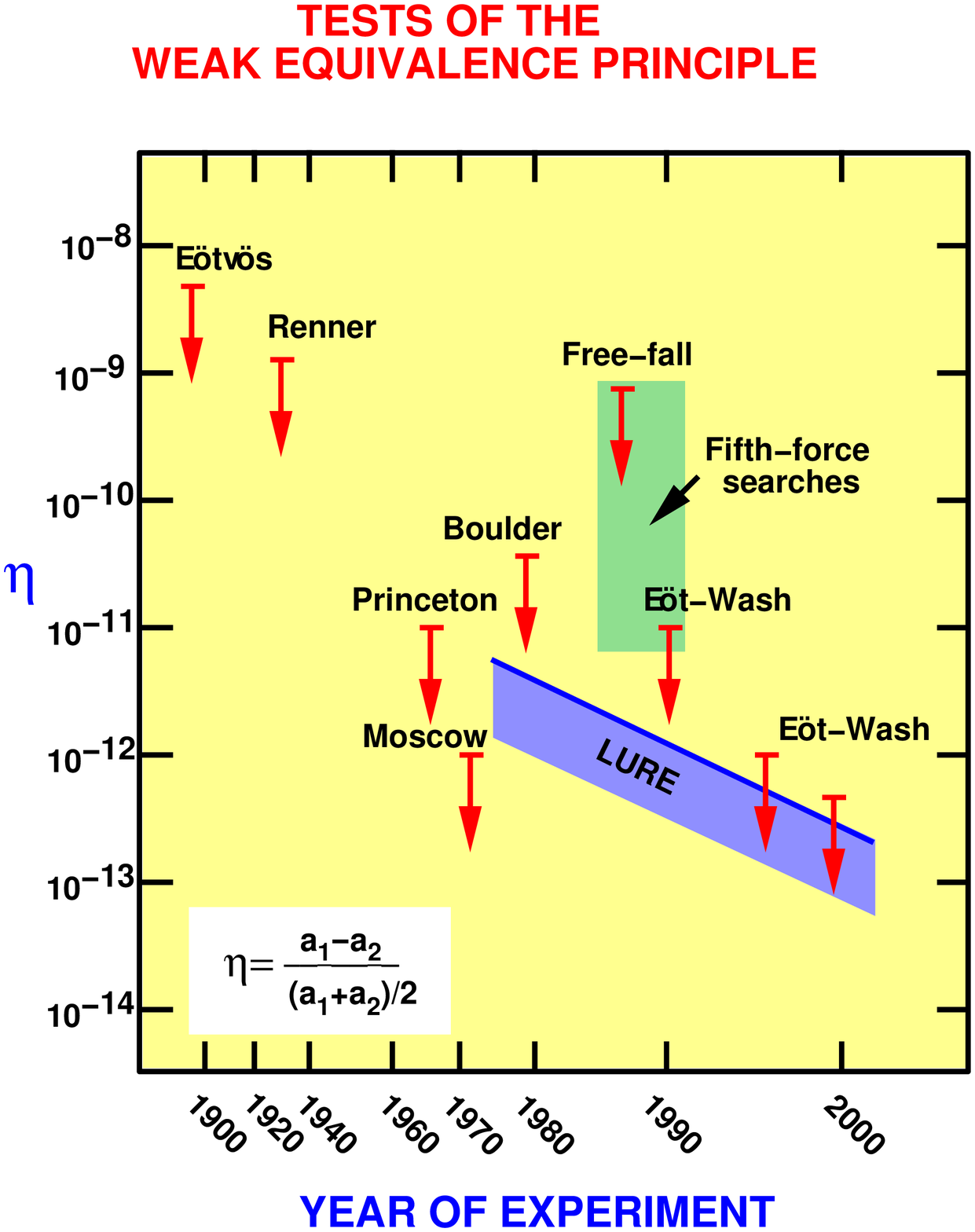,height=6.0in}
\end{center}
\caption{Selected tests of the Weak Equivalence Principle,
showing bounds on $\eta$, which measures fractional difference in
acceleration of different materials or bodies.  Free-fall and 
E\"ot-Wash experiments originally performed 
to search for fifth force.  Blue band shows current 
bounds on $\eta$ for gravitating bodies 
from lunar laser ranging (LURE). }
\label{wepfig}
\end{figure}

Many high-precision E\"otv\"os-type experiments have been
performed, from the pendulum experiments of Newton, Bessel and
Potter, to the classic torsion-balance measurements of
E\"otv\"os \cite{eotvos}, Dicke \cite{dicke}, Braginsky \cite{braginsky} 
and their collaborators.  In the
modern torsion-balance experiments, two objects of different
composition are connected by a rod or placed on a tray
and suspended in a horizontal
orientation by a fine wire.  If the gravitational acceleration of
the bodies differs, there will be a torque induced on the
suspension wire, related to the angle between the wire and the
direction of the gravitational acceleration $\bf g$.  If the entire
apparatus is rotated about some direction with angular velocity
$\omega$, the torque will be modulated with period $2 \pi / \omega$.
In the experiments of E\"otv\"os and his collaborators, the wire
and $\bf g$ were not quite parallel because of the centripetal
acceleration on the apparatus due to the Earth's rotation; the
apparatus was rotated about the direction of the wire.  In the
Dicke and Braginsky
experiments, $\bf g$ was that of the Sun, and the rotation of the Earth
provided the modulation of the torque at a period of 24~hr
(TEGP 2.4(a)).  Beginning in the late 1980s, numerous
experiments were carried out
primarily to search for a ``fifth force'' (see Sec. \ref{newinteractions}), 
but their
null results also constituted tests of WEP.  In the ``free-fall Galileo
experiment'' performed at the University of Colorado, the
relative free-fall acceleration of two bodies made of uranium
and copper was measured using a laser interferometric technique.
The ``E\"ot-Wash'' experiments carried out at the
University of Washington used a sophisticated torsion balance
tray to compare the accelerations of various materials toward
local topography on Earth, movable laboratory masses, 
the Sun and the galaxy \cite{Su,baessler}, and have recently
reached levels of $4 \times 10^{-13}$.  The
resulting upper limits on $\eta$ are summarized in Figure \ref{wepfig} (TEGP
14.1; for a bibliography of experiments, see \cite{fischbach}).
 
\begin{figure}
\begin{center}
\leavevmode
\psfig{figure=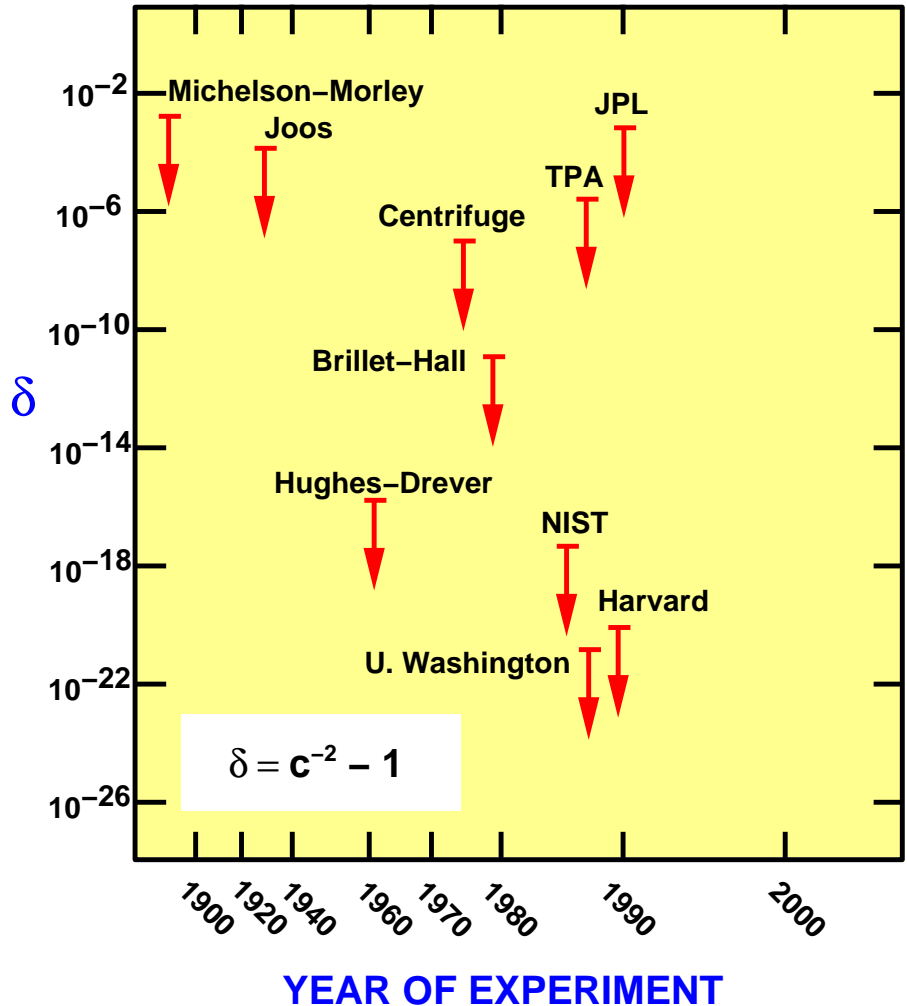,height=6.0in} 
\end{center}
\caption{ Selected tests of local Lorentz invariance showing 
bounds on parameter 
$\delta$, which measures degree of violation of Lorentz invariance 
in electromagnetism.  Michelson-Morley, Joos, and Brillet-Hall 
experiments test isotropy of round-trip speed of light,
the latter experiment using laser technology.  Centrifuge, 
two-photon absorption
(TPA) and JPL experiments test isotropy of light speed using one-way
propagation.
Remaining four
experiments test isotropy of nuclear energy levels.  Limits assume 
speed of Earth of 300\ km/s relative to the mean rest frame of the universe. 
} 
\label{llifig}
\end{figure}

The second ingredient of EEP, local Lorentz invariance, has been
tested to high-precision in the ``mass
anisotropy'' experiments\index{mass anisotropy experiments}:
the classic versions are the
Hughes-Drever experiments, performed in the period 1959-60
independently by Hughes and collaborators at Yale University, and
by Drever at Glasgow University (TEGP 2.4(b)).
Dramatically improved versions were carried out during the late 1980s
using laser-cooled trapped atom techniques (TEGP 14.1).
A simple and
useful way of interpreting these experiments is to suppose that
the electromagnetic interactions suffer a slight violation of
Lorentz invariance, through a change in the speed of
electromagnetic radiation $c$ relative to the limiting
speed of material test particles ($c_0$, chosen to be unity via a
choice of units), in other words, $c \ne 1$
(see Sec. \ref{c2formalism}).  Such a violation necessarily
selects a preferred universal rest frame, presumably that of the
cosmic background radiation, through which we are moving at about
300~km/s.  Such a Lorentz-non-invariant electromagnetic
interaction would cause shifts in the energy levels of atoms and
nuclei that depend on the orientation of the quantization axis of
the state relative to our universal velocity vector, and on the
quantum numbers of the state.  The presence or absence of such
energy shifts can be examined by measuring the energy of one such
state relative to another state that is either unaffected or is
affected differently by the supposed violation.  One way is to
look for a shifting of the energy levels of states that are
ordinarily equally spaced, such as the four $J{=}3/2$ ground states
of the $^7$Li nucleus in a magnetic field (Drever experiment);
another is to compare the levels of a complex nucleus with the
atomic hyperfine levels of a hydrogen maser clock.  These
experiments have all yielded extremely accurate results, quoted
as limits on the parameter $\delta \equiv c^{-2}-1$
in Figure \ref{llifig}.  Also included for comparison is the corresponding
limit obtained from Michelson-Morley type experiments (for a review, see
\cite{hauganwill}).

Recent advances in atomic spectroscopy and atomic timekeeping
have made it possible to test LLI by checking the isotropy of the
speed of light using one-way propagation (as opposed to round-trip 
propagation, as in the Michelson-Morley 
experiment).  In one experiment, for example, the
relative phases of two hydrogen maser clocks at two stations of
NASA 's Deep Space Tracking Network were compared over five
rotations of the Earth by propagating a light signal one-way
along an ultrastable fiberoptic link connecting them (see Sec.
\ref{c2formalism}).
Although the bounds from these experiments are not as tight as
those from mass-anisotropy experiments, they probe directly the
fundamental postulates of special relativity, and thereby of LLI.
(TEGP 14.1, \cite{Will92b}).

The principle of local position invariance, the third part of
EEP, can be tested by the gravitational redshift
experiment\index{gravitational redshift experiment},
 the
first experimental test of gravitation proposed by Einstein.
Despite the fact that Einstein regarded this as a crucial test of
GR, we now realize that it does not distinguish
between GR and any other metric theory of
gravity, instead is a test only of EEP.  A typical gravitational
redshift experiment measures the frequency or wavelength shift
$Z \equiv \Delta \nu / \nu = - \Delta \lambda / \lambda$ between two
identical frequency standards (clocks) placed at rest at
different heights in a static gravitational field.  If the
frequency of a given type of atomic clock is the same when
measured in a local, momentarily comoving freely falling frame
(Lorentz frame), independent of the location or velocity of that
frame, then the comparison of frequencies of two clocks at rest
at different locations boils down to a comparison of the
velocities of two local Lorentz frames, one at rest with respect
to one clock at the moment of emission of its signal, the other
at rest with respect to the other clock at the moment of
reception of the signal.  The frequency shift is then a
consequence of the first-order Doppler shift between the frames.
The structure of the clock plays no role whatsoever.  The result
is a shift
\begin{equation} \label{E3}
Z = \Delta U/ c^2 \,, 
\end{equation}
where $\Delta U$ is the difference in the Newtonian gravitational
potential between the receiver and the emitter.  If LPI is not
valid, then it turns out that the shift can be written
\begin{equation} \label{E4}
Z = (1+ \alpha ) \Delta U /c^2 \,, 
\end{equation}
where the parameter $\alpha$ may depend upon the nature of the
clock whose shift is being measured (see TEGP 2.4(c) for
details).

\begin{figure}
\begin{center}
\leavevmode
\psfig{figure=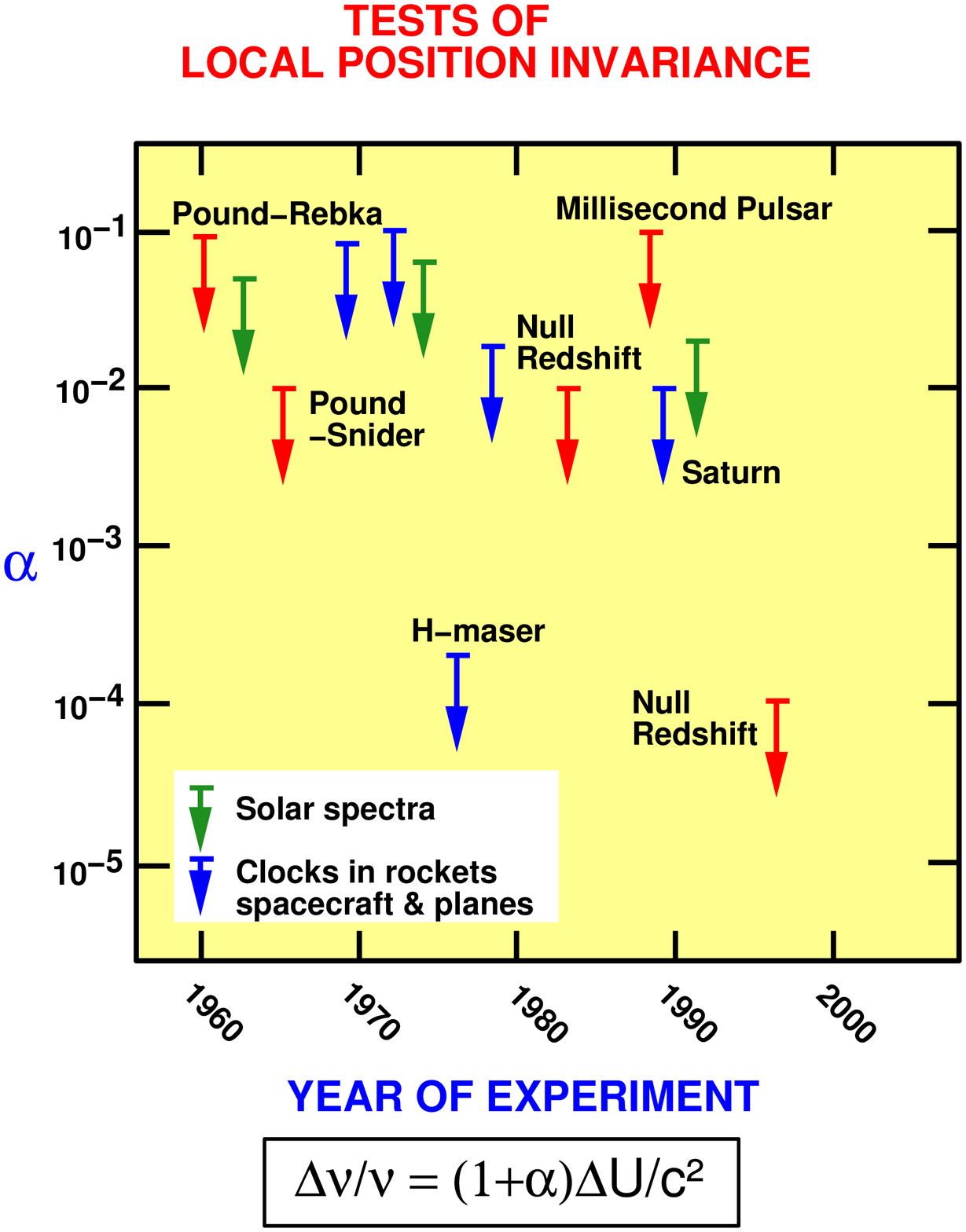,height=6.0in}
\end{center}
\caption{Selected tests of local position invariance via gravitational
redshift experiments, showing bounds on $\alpha$, which measures degree
of deviation of redshift from the formula 
$\Delta \nu / \nu = \Delta U/c^2$.
}  
\label{lpifig}
\end{figure}

The first successful, high-precision redshift measurement was the
series of Pound-Rebka-Snider experiments of 1960-1965, that
measured the frequency shift of gamma-ray photons from $^{57}$Fe
as they ascended or descended the Jefferson Physical Laboratory
tower at Harvard University.  The high accuracy 
achieved -- one percent -- was obtained by making
use of the M\"ossbauer effect to produce a narrow resonance line
whose shift could be accurately determined.  Other experiments
since 1960 measured the shift of spectral lines in the Sun's
gravitational field and the change in rate of atomic clocks
transported aloft on aircraft, rockets and satellites.  Figure \ref{lpifig}
summarizes the important redshift experiments that have been
performed since 1960 (TEGP 2.4(c)).

The most precise standard redshift
test to date was the Vessot-Levine rocket
experiment that took place in June 1976 \cite{vessot}.  A
hydrogen-maser clock was flown on a rocket to an altitude of
about 10,000~km and its frequency compared to a similar clock on
the ground.  The experiment took advantage of the masers'
frequency stability by monitoring the frequency shift as a
function of altitude.  A sophisticated data acquisition scheme
accurately eliminated all effects of the first-order Doppler
shift due to the rocket's motion, while tracking data were used
to determine the payload's location and the velocity (to evaluate
the potential difference $\Delta U$, and the special relativistic
time dilation).  Analysis of the data yielded a limit 
$| \alpha | < 2 \times 10^{-4}$.

A ``null'' redshift experiment performed in 1978 tested whether
the {\it relative} rates of two different clocks depended upon
position.  Two hydrogen maser clocks and an ensemble of three
superconducting-cavity stabilized oscillator ({\sc scso}) clocks were compared
over a 10-day period.  During the period of the experiment, 
the solar potential
$U/c^2$ changed sinusoidally with a 24-hour period by
$3 \times10^{-13}$ because of the Earth's rotation, and changed
linearly at $3 \times 10^{-12}$ per day because the Earth is 90 degrees
from perihelion in April.  However, analysis of the data
revealed no variations of either type within experimental errors,
leading to a limit on the LPI violation parameter 
$| \alpha^{\rm H} - \alpha^{\rm SCSO} | < 2 \times 10^{-2}$ 
\cite{turneaure}.  This
bound has been improved using more stable frequency standards
\cite{Godone,prestage}.  The
varying gravitational redshift of Earth-bound clocks relative to
the highly stable Millisecond Pulsar, caused by the Earth's
motion in the solar gravitational field around the Earth-Moon
center of mass (amplitude 4000~km), has been measured to
about 10 percent, and the redshift of stable oscillator
clocks on the Voyager spacecraft caused by Saturn's gravitational
field yielded a one percent test.  The solar gravitational
redshift has been tested to about two percent using infrared oxygen
triplet lines at the limb of the Sun, and to one percent using
oscillator clocks on the Galileo spacecraft (TEGP 2.4(c) and 14.1(a)).  

Modern advances in navigation using Earth-orbiting atomic clocks and
accurate time-transfer must routinely
take gravitational redshift and time-dilation effects into account.
For example, the Global Positioning System
(GPS) provides absolute accuracies of around 15 m (even better in its
military mode) anywhere on Earth, 
which corresponds to 50 nanoseconds in time accuracy
at all times.  Yet the difference in rate between satellite and ground
clocks as a result of special and general relativistic effects is a
whopping 39 {\it microseconds} per day ($46 \mu$s from the gravitational
redshift, and $-7 \, \mu$s from time dilation).
If these effects were not accurately accounted for, GPS
would fail to function at its stated accuracy.
This represents a welcome practical application of GR!
(For the role of GR in GPS, see
\cite{ashby}; for a popular essay, see
\cite{physicscentral}.)

Local position invariance also refers to position in time.  If LPI is
satisfied, the fundamental constants of non-gravitational physics
should be constants in time.  Table \ref{varyconstants} shows current bounds on
cosmological variations in selected dimensionless constants.  For
discussion and references to early work, see TEGP 2.4(c).

\begin{table} 
\begin{center}
\begin{tabular}{l l l}
\hline
\hline
&Limit on $\dot{k}/k$&\\
&per Hubble time& \\
Constant $k$&$1.2 \times 10^{10} \rm{yr}$ &Method \\ [0.5ex]
\hline
\hline
Fine structure constant&$4 \times 10^{-4}$&H-maser vs Hg ion clock
\cite{prestage}\\
\quad $\alpha=e^2/\hbar c$&$9 \times 10^{-5}$&$^{87}$Rb fountain vs Cs
clock \cite{salomon}\\
&$6 \times 10^{-7}$&Oklo Natural Reactor \cite{Dyson}
\\
&$6 \times 10^{-5}$&21-cm vs molecular absorption \\
&&\quad at $Z=0.7$ (\cite{drinkwater})\\ [0.5ex]
\hline
Weak interaction constant&1&$^{187}$Re, $^{40}$K decay rates\\
\quad $\beta=G_f m_p^2 c/\hbar^3$&0.1&Oklo Natural Reactor
\cite{Dyson} \\
&0.06&Big bang nucleosynthesis \cite{malaney,reeves}\\ [0.5ex]
\hline
e-p mass ratio&1&Mass shift in quasar\\
&&\quad spectra at $Z \sim 2$\\ [0.5ex]
\hline
Proton g-factor ($g_p$)&$10^{-5}$&21-cm vs molecular absorption \\
&&\quad at $Z=0.7$ (\cite{drinkwater})\\ [0.5ex]
\hline
\hline
\end{tabular}
\caption{Bounds on cosmological variation of 
fundamental constants of non-gravitational
physics.  For references to earlier work, see TEGP 2.4(c).}
\label{varyconstants}
\end{center}
\end{table}

%----------------------------------------------
% Subsection 2.2
%----------------------------------------------

\subsection{Theoretical Frameworks for Analyzing EEP}\label{EEPframeworks}

\subsubsection{Schiff's Conjecture}\label{Schiff}

Because the three parts of the Einstein equivalence principle
discussed above are so very different in their empirical
consequences, it is tempting to regard them as independent
theoretical principles.  On the other hand, any complete and
self-consistent gravitation theory must possess sufficient
mathematical machinery to make predictions for the outcomes of
experiments that test each principle, and because there are
limits to the number of ways that gravitation can be meshed with
the special relativistic laws of physics, one might not be
surprised if there were theoretical connections between the three
sub-principles.  For instance, the same mathematical formalism
that produces equations describing the free fall of a hydrogen
atom must also produce equations that determine the energy levels
of hydrogen in a gravitational field, and thereby the ticking
rate of a hydrogen maser clock.  Hence a violation of EEP in the
fundamental machinery of a theory that manifests itself as a
violation of WEP might also be expected to show up as a violation
of local position invariance.  Around 1960, Schiff conjectured
that this kind of connection was a necessary feature of any
self-consistent theory of gravity.  More precisely,
Schiff's conjecture states that {\it any complete, self-consistent
theory of gravity that embodies WEP necessarily embodies EEP}.
In other words, the validity of WEP alone guarantees the validity
of local Lorentz and position invariance, and thereby of EEP.

If Schiff's conjecture is correct, then E\"otv\"os experiments may
be seen as the direct empirical foundation for EEP, hence for the
interpretation of gravity as a curved-spacetime phenomenon.  Of course,
a rigorous proof of such a conjecture is impossible (indeed, some
special counter-examples are known) yet a number
of powerful ``plausibility'' arguments can be formulated.

The most general and elegant of these arguments is based upon the
assumption of energy conservation.  This assumption allows one to
perform very simple cyclic gedanken experiments in which the
energy at the end of the cycle must equal that at the beginning
of the cycle.  This approach was pioneered by Dicke, Nordtvedt and
Haugan (see, \eg \cite{haugan}).  
A system in a quantum state $A$ decays to state $B$,
emitting a quantum of frequency $\nu$.  The quantum falls a height
$H$ in an external gravitational field and is shifted to frequency
$\nu^\prime$, while the system in state $B$ falls with acceleration
$g_B$.  At the bottom, state $A$ is rebuilt out of state $B$, the
quantum of frequency $ \nu^\prime$, and the kinetic energy $m_B g_B H$
that state $B$ has gained during its fall.  The energy left over
must be exactly enough, $m_A g_A H$, to raise state $A$ to its original
location.  (Here an assumption of local Lorentz invariance
permits the inertial masses $m_A$  and $m_B$ to be identified with
the total energies of the bodies.)  If $g_A$ and $g_B$ depend on
that portion of the internal energy of the states that was
involved in the quantum transition from $A$ to $B$ according to
\begin{equation} \label{E5}
   g_A = g(1 + \alpha E_A / m_A c^2 ) \,, \qquad  
   g_B = g(1 + \alpha E_B / m_B c^2 ) \,, \qquad  
   E_A - E_B \equiv h \nu  
\end{equation}
(violation of WEP), then by conservation of energy, there must be
a corresponding violation of LPI in the frequency shift of the
form (to lowest order in $h \nu /mc^2$)
\begin{equation}\label{E6}
   Z = ( \nu^\prime - \nu )/ \nu^\prime
     = (1+\alpha ) gH/c^2 = (1+\alpha ) \Delta U/c^2 \,.
\end{equation}
Haugan generalized this approach to include violations of
LLI (\cite{haugan}, TEGP 2.5).

% Box 1 ------
\begin{table}
\medskip
\centerline{\bf Box 1. The $TH\epsilon\mu$ Formalism}
\medskip
\hrule\small
\medskip
\begin{enumerate}
\item{\bf Coordinate System and Conventions:}  

$x^0 =t=$ time coordinate associated with the
static nature of the static spherically symmetric ({\sc sss})
gravitational field; ${\bf x} =(x,y,z) =$ isotropic
quasi-Cartesian spatial coordinates; spatial vector and gradient
operations as in Cartesian space.

\item{\bf Matter and Field Variables:}
\begin{itemize}
\item
$m_{0a} =$ rest mass of particle $a$.
\item
$e_a =$ charge of particle $a$.
\item
$x_a^\mu (t) =$ world line of particle $a$.
\item
$v_a^\mu = dx_a^\mu/dt =$ coordinate velocity of particle $a$.
\item
$A_\mu =$ electromagnetic vector potential;
${\bf E}= \nabla A_0-\partial {\bf A}/\partial t \,,\,{\bf B}=\nabla \times
{\bf A}$ 
\end{itemize}

\item{\bf Gravitational Potential:}  $U( {\bf x} )$

\item{\bf Arbitrary Functions:}

$T(U)$, $H(U)$, $\epsilon (U)$, $\mu (U)$; 
EEP is satisfied iff 
$\epsilon = \mu = (H/T)^{1/2}$ for all $U$.

\item{\bf Action:}
\begin{eqnarray*}
  I &=& - \sum_a m_{0a} \int 
        (T-Hv_a^2)^{1/2} dt
    + \sum_a e_a \int 
        A_\mu (x_a^\nu ) v_a^\mu dt  \\
  && \quad + (8 \pi)^{-1} \int 
	(\epsilon E^2-\mu^{-1} B^2) d^4 x \,.
\end{eqnarray*}

\item{\bf Non-Metric Parameters:}
\begin{eqnarray*}
\Gamma_0 &=& -c_0^2(\partial /\partial U) \ln [\epsilon (T/H)^{1/2}]_0 \,,\\
\Lambda_0 &=& -c_0^2(\partial /\partial U) \ln [\mu (T/H)^{1/2}]_0 \,,\\
\Upsilon_0 &=& 1- (TH^{-1} \epsilon\mu)_0 \,,
\end{eqnarray*}
where $c_0 = (T_0 /H_0 )^{1/2}$ and subscript
``0'' refers to a chosen point in space.  
If EEP is satisfied, $\Gamma_0 \equiv \Lambda_0 \equiv \Upsilon_0 \equiv
0$.

\end{enumerate}
\hrule\normalsize
\medskip
\end{table}

% end of Box 1 ----
\subsubsection{The $TH\epsilon\mu$ Formalism}\label{theuformalism}

The first successful attempt to prove Schiff's conjecture more
formally was made by Lightman and Lee \cite{lightmanlee}.  They developed a
framework called the $TH\epsilon\mu$ formalism that encompasses
all metric theories of gravity and many non-metric theories
(Box 1).  It restricts attention to the behavior of charged
particles (electromagnetic interactions only) in an external
static spherically symmetric (SSS) gravitational field, described
by a potential $U$.  It characterizes the motion of the charged
particles in the external potential by two arbitrary functions
$T(U)$ and $H(U)$, and characterizes the response of
electromagnetic fields to the external potential (gravitationally
modified Maxwell equations) by two functions $\epsilon (U)$ and
$\mu (U)$.  The forms of $T$, $H$, $\epsilon$ and $\mu$ vary from theory
to theory, but every metric theory satisfies
\begin{equation}\label{E7}
    \epsilon = \mu = (H/T)^{1/2} \,,
\end{equation}
for all $U$.  This consequence follows from the action of
electrodynamics with a ``minimal'' or metric coupling:
\begin{eqnarray} \label{E8}
  I &=& - \sum_a m_{0a} \int 
        (g_{\mu\nu} v_a^\mu v_a^\nu )^{1/2} dt
    + \sum_a e_a \int 
        A_\mu (x_a^\nu ) v_a^\mu dt \nonumber\\
  &-& {1 \over {16 \pi}} \int \sqrt{-g}
          g^{\mu \alpha} g^{\nu \beta}
          F_{\mu \nu} F_{\alpha \beta}
          d^4 x  \,,
\end{eqnarray}
where the variables are defined in Box 1, and where 
$F_{\mu\nu} \equiv A_{\nu , \mu} - A_{\mu , \nu}$.  By identifying 
$g_{00} = T$ and $g_{ij} = H \delta_{ij}$ in a SSS field,
$F_{i0} = E_i$ and $F_{ij} = \epsilon_{ijk} B_k$,
one obtains Eq.~(\ref{E7}).
Conversely, every theory within this class that satisfies Eq. (\ref{E7})
can have its electrodynamic equations cast into ``metric'' form.
In a given non-metric theory, the functions $T$, $H$, $\epsilon$ and
$\mu$ will depend in general on the full gravitational environment,
including the potential of the Earth, Sun and Galaxy, as well as on
cosmological boundary conditions.  Which of these factors has the most
influence on a given experiment will depend on the nature of the
experiment.

Lightman and Lee then calculated explicitly the rate of
fall of a ``test'' body made up of interacting charged particles,
and found that the rate was independent of the internal
electromagnetic structure of the body (WEP) if and only if Eq. 
(\ref{E7})
was satisfied.  In other words WEP $\to$ EEP and Schiff's
conjecture was verified, at least within the restrictions built
into the formalism.

Certain combinations of the functions $T$, $H$, $\epsilon$ and $\mu$
reflect different aspects of EEP.  For instance, position or
$U$-dependence of either of the combinations $\epsilon (T/H)^{1/2}$ 
and $\mu (T/H)^{1/2}$ signals violations of LPI, the first
combination playing the role of the locally measured electric
charge or fine structure constant.  The ``non-metric parameters''
$\Gamma_0$ and $\Lambda_0$ (Box 1) are measures of such
violations of EEP.  Similarly, if the parameter 
$\Upsilon_0 \equiv 1-(TH^{-1} \epsilon \mu )_0$ is non-zero anywhere, 
then violations of LLI will occur.  This parameter is related to the
difference between the speed of light, $c$, and the
limiting speed of material test particles, $c_o$, given by
\begin{equation}\label{E9}
    c = ( \epsilon_0 \mu_0 )^{- 1/2} \,,\qquad   
    c_o ~=~ ( T_0 / H_0 )^{1/2} \,.  
\end{equation}
In many applications,
by suitable definition of units, $c_0$ can be set equal to unity.
If EEP is valid, $\Gamma_0 \equiv \Lambda_0 \equiv \Upsilon_0 = 0$ 
everywhere.

The rate of fall of a composite spherical test body of
electromagnetically interacting particles then has the form
\begin{eqnarray}
    {\bf a} &=& (m_P /m) \nabla U  \,,\label{E10}\\
          m_P /m 
 &=&1 + (E_B^{ES} /Mc_0^2 )
          [2 \Gamma_0 - {8 \over 3} \Upsilon_0 ]
      + (E_B^{MS} /Mc_0^2 ) 
          [2 \Lambda_0 - {4 \over 3} \Upsilon_0 ] 
      \nonumber \\
      && \quad + \dots ,\label{E11}
\end{eqnarray}
where $E_B^{ES}$ and $E_B^{MS}$ are the electrostatic
and magnetostatic binding energies of the body, given by
\begin{eqnarray}
	  E_B^{ES}
 &=&-{1 \over 4} T_0^{1/2} H_0^{-1} \epsilon_0^{-1}
\left < \sum_{ab} e_a e_b r_{ab}^{-1} \right >  \,, \label{E12} \\
       E_B^{MS} &=&
  - {1 \over 8} T_0^{1/2} H_0^{-1} \mu_0
  \left < \sum_{ab} e_a e_b r_{ab}^{-1}
       [ {\bf v}_a \cdot {\bf v}_b 
       + ( {\bf v}_a \cdot {\bf n}_{ab} )
	   ( {\bf v}_b \cdot {\bf n}_{ab} )]
\right >  \,, \label{E13}
\end{eqnarray}
where $r_{ab} = | {\bf x}_a  -  {\bf x}_b |$,
${\bf n}_{ab}  =  ( {\bf x}_a  -  {\bf x}_b )/r_{ab}$, and the
angle brackets denote an expectation value of the enclosed
operator for the system's internal state.  E\"otv\"os experiments
place limits on the WEP-violating terms in Eq. (\ref{E11}), and
ultimately place limits on the non-metric parameters 
$| \Gamma_0 | < 2 \times 10^{-10}$ and
$| \Lambda_0 |  <  3  \times  10^{-6}$. 
(We set 
$\Upsilon_0 = 0$ because of very tight constraints on it from
tests of LLI .)  These limits are
sufficiently tight to rule out a number of non-metric theories of
gravity thought previously to be viable (TEGP 2.6(f)).

The $TH \epsilon\mu$ formalism also yields a
gravitationally modified Dirac equation that can be used to
determine the gravitational redshift experienced by a variety of
atomic clocks.  For the redshift parameter $\alpha$ 
[Eq. (\ref{E4})], the results are (TEGP 2.6(c)):
\begin{equation}\label{E14}
\alpha= \left \{ \begin{array}{ll}
  -3\Gamma_0 + \Lambda_0  &{\rm hydrogen~hyperfine~transition,~ H-Maser~ clock} \\
-{1 \over 2} (3\Gamma_0 + \Lambda_0)  &{\rm electromagnetic~mode~in
~cavity,~SCSO~clock} \\
-2\Gamma_0  &{\rm phonon~mode~in~solid,~principal~transition~in} \\
& \quad {\rm~hydrogen}.
\end{array} \right.
\end{equation}

The redshift is the standard one $( \alpha = 0)$, independently
of the nature of the clock if and only if 
$\Gamma_0 \equiv \Lambda_0 \equiv 0$.  Thus the Vessot-Levine
rocket redshift experiment sets a limit on the parameter
combination $| 3 \Gamma_0 - \Lambda_0 |$ (Figure \ref{lpifig}); the
null-redshift experiment comparing hydrogen-maser and {\sc scso} clocks
sets a limit on 
$| \alpha_H - \alpha_{SCSO} | = {3 \over 2} | \Gamma_0 - \Lambda_0 |$.
Alvarez and Mann
\cite{AlvarezMann96a,AlvarezMann96b,AlvarezMann97a,
AlvarezMann97b,AlvarezMann97c} extended the $TH\epsilon\mu$ formalism to
permit analysis of such effects as the Lamb shift, anomalous magnetic moments
and non-baryonic effects, and placed interesting bounds on EEP violations.

\subsubsection{The ${c^2}$ Formalism}\label{c2formalism}

The $TH \epsilon \mu$ formalism can also be applied to tests of
local Lorentz invariance, but in this context it can be
simplified.  Since most such tests do not concern themselves with
the spatial variation of the functions $T$, $H$, $\epsilon$, and $\mu$,
but rather with observations made in moving frames, we can treat
them as spatial constants.  Then by rescaling the time and space
coordinates, the charges and the electromagnetic fields, we can
put the action in Box 1 into the form (TEGP 2.6(a)).
\begin{equation}\label{E15}
  I = - \sum_a m_{0a} \int 
        (1-v_a^2)^{1/2} dt
    + \sum_a e_a \int 
        A_\mu (x_a^\nu ) v_a^\mu dt 
  + (8 \pi)^{-1} \int 
	(E^2-c^2B^2) d^4 x   \,,  
\end{equation}
where $c^2 \equiv H_0 /T_0 \epsilon_0 \mu_0=(1-\Upsilon_0)^{-1}$.
This amounts to using units in which the limiting speed $c_o$
of massive test particles is unity, and the speed of light is c.
If $c \ne 1$, LLI is violated; furthermore, the form of the
action above must be assumed to be valid only in some preferred
universal rest frame.  The natural candidate for such a frame is
the rest frame of the microwave background.

The electrodynamical equations which follow from Eq. (\ref{E15})
yield the behavior of rods and clocks, just as in the full $TH
\epsilon \mu$ 
formalism.  For example, the length of a rod moving
through the rest frame in a direction parallel to its length
will be observed by a rest observer to be contracted relative to
an identical rod perpendicular to the motion by a factor 
$1 - V^2 /2 +  O(V^4 )$.  Notice that $c$ does not appear in this
expression.  The energy and momentum of an electromagnetically
bound body which moves with velocity $ \bf V$ relative to the rest frame
are given by
\begin{eqnarray}\label{E16}
  E &=& M_R  +  {1 \over 2}  M_R V^2
          +  {1 \over 2}  \delta M_I^{ij} V^i V^j \,,  \nonumber \\
  P^i  &=&  M_R V^i +  \delta M_I^{ij} V^j  \,,  
\end{eqnarray}
where $M_R  =  M_0  - E_B^{ES}$, $M_0$ is the
sum of the particle rest masses, $E_B^{ES}$ is the
electrostatic binding energy of the system ([Eq. (\ref{E12})] with
$T_0^{1/2}H_0 \epsilon_0^{-1}=1$), and
\begin{equation}\label{E17}
    \delta M_I^{ij}  =  -  2
( {1 \over {c^2}} -  1 )
[ {4 \over 3} E_B^{ES} \delta^{ij} 
    +   \tilde E_B^{ESij} ]  \,,  
\end{equation}
where
\begin{equation} \label{E18}
       \tilde E_B^{ESij} 
 = -{1 \over 4} \left < \sum_{ab} e_a e_b r_{ab}^{-1} \left (
       (n_{ab}^i n_{ab}^j
   - {1 \over 3} \delta^{ij} \right ) \right > \,.
\end{equation}
Note that $(c^{-2} - 1)$ corresponds to the parameter $\delta$
plotted in Figure \ref{llifig}.

The electrodynamics given by Eq. (\ref{E15}) can also be
quantized, so that we may treat the interaction of photons with
atoms via perturbation theory.  The energy of a photon is $\hbar$
times its frequency $\omega$, while its momentum is $\hbar \omega /c$.  
Using this approach, one finds that the difference in round trip
travel times of light along the two arms of the interferometer in the
Michelson-Morley experiment is given by 
$L_0 (v^2 /c)(c^{-2}  - 1)$.  
The experimental null result then leads to the bound on 
$(c^{-2} - 1)$ shown on Figure \ref{llifig}.  Similarly the anisotropy in
energy levels is clearly illustrated by the tensorial terms
in Eqs. (\ref{E16}) and (\ref{E18}); by
evaluating $\tilde E_B^{ESij}$ for each nucleus in the
various Hughes-Drever-type experiments and comparing with the
experimental limits on energy differences, one obtains the extremely
tight bounds also shown on Figure \ref{llifig}.  

The behavior of moving
atomic clocks can also be analysed in detail, and bounds
on $(c^{-2} - 1)$ can be placed using results from tests of
time dilation and of the propagation of light.  In some cases, it is
advantageous to combine the $c^2$ framework with a ``kinematical''
viewpoint that treats a general class of boost transformations between
moving frames.  Such kinematical approaches have been discussed by
Robertson, Mansouri and Sexl, and Will (see \cite{Will92b}).

For example, 
in the ``JPL'' experiment, in which 
the phases of two hydrogen masers connected by a fiberoptic link were
compared as a function of the Earth's orientation, 
the predicted phase difference as a function of
direction is, to first order in $\bf V$, the velocity of the Earth
through the cosmic background,
\begin{equation}\label{E19}
    \Delta \phi / \tilde \phi \approx - {4 \over 3} (1-c^2)
    ( {\bf V} \cdot {\bf n} ~-~ {\bf V} \cdot {\bf n}_0 ) \,,  
\end{equation}
where $\tilde \phi  = 2 \pi \nu L$, $\nu$ is the maser frequency,
$L=21$ km is the baseline, and where ${\bf n}$ and ${\bf n}_0$ are
unit vectors along the direction of propagation of the light, at
a given time, and at the initial time of the experiment, respectively.  The
observed limit on a diurnal variation in the relative phase
resulted in the bound 
$| c^{-2}-1 | < 3 \times 10^{-4}$.
Tighter bounds were obtained from a ``two-photon absorption'' (TPA)
experiment, and a 1960s series of 
``M\"ossbauer-rotor'' experiments, which tested the isotropy of
time dilation between a gamma ray emitter on the rim of a rotating
disk and an absorber placed at the center \cite{Will92b}.

\subsection{EEP, Particle Physics, and the Search for New
Interactions}\label{newinteractions}

In 1986, as a result 
of a detailed reanalysis of E\"otv\"os' original data, 
Fischbach \etal \cite{fischbach5} suggested the existence of a fifth 
force of nature, with a strength of about a percent that of 
gravity, but with a range (as defined by the range $\lambda$ of a Yukawa 
potential, $e^{-r/\lambda} /r$) of a few hundred meters.
This proposal dovetailed with earlier hints of a deviation from the 
inverse-square law of Newtonian gravitation derived from measurements of 
the gravity profile down deep mines in Australia, 
and with ideas from particle physics suggesting the possible 
presence of very low-mass particles with gravitational-strength couplings. 
During the next four years 
numerous experiments looked for evidence of the fifth force by searching 
for composition-dependent differences in acceleration, with variants of 
the E\"otv\"os experiment or with free-fall Galileo-type experiments.  
Although two early experiments reported positive evidence, the others 
all yielded null results.  Over the range between one and $10^4$ meters, 
the null experiments produced upper limits on the strength of a postulated 
fifth force between $10^{-3}$ and $10^{-6}$ of the strength of gravity. 
Interpreted as tests of WEP (corresponding to the limit of 
infinite-range forces), the results of two representative experiments from
this period: the free-fall Galileo experiment, 
and the early E\"ot-Wash experiment, are shown in 
Figure \ref{wepfig}.  At the same time, tests of the inverse-square 
law of gravity were carried out by comparing variations in gravity 
measurements up tall towers or down mines or boreholes with gravity 
variations predicted using the inverse square law together with Earth 
models and surface gravity data mathematically ``continued'' up 
the tower or down the hole.  Despite early reports of anomalies, 
independent tower, borehole and seawater measurements now show no evidence of a 
deviation.  Analyses of orbital data from planetary range
measurements, lunar laser ranging, and laser tracking of the LAGEOS
satellite verified the inverse-square law to parts in $10^8$ over
scales of $10^3$ to $10^5$ km, and to parts in $10^9$ over planetary
scales of several astronomical units \cite{talmadge}.
The consensus at present is that there is no 
credible experimental evidence for a fifth force of nature.
For reviews and bibliographies, 
see \cite{fischbach,FischbachTalmadge,FischbachTalmadge2,Adelberger,WillSky}.

Nevertheless,
theoretical evidence continues to mount that EEP is {\it
likely} to be violated at some level, whether by quantum gravity
effects, by effects arising from string theory, or by hitherto
undetected interactions, albeit at levels well below those that motivated
the fifth-force searches.  
Roughly speaking, in addition to the pure Einsteinian
gravitational interaction, which respects EEP, theories such as string
theory predict
other interactions which do not.  In string theory, for example, the
existence of such EEP-violating fields is assured, but the theory is not
yet mature enough to enable calculation of their strength (relative to
gravity), or their range (whether they are long range, like gravity, or
short range, like the nuclear and weak interactions, 
and thus too short-range to be
detectable).  

In one simple example, one can write the Lagrangian for the low-energy
limit of string theory in the so-called ``Einstein frame'', in which
the gravitational Lagrangian is purely general relativistic:
\begin{eqnarray}
\tilde {\cal L} =& \sqrt{- \tilde g} \biggl (
{\tilde g}^{\mu\nu} \biggl [ {1 \over {2\kappa}} {\tilde
R}_{\mu\nu} - {1 \over 2} \tilde G (\varphi) \partial_\mu \varphi
\partial_\nu \varphi  \biggr ] 
-  U(\varphi) {\tilde g}^{\mu\nu}{\tilde g}^{\alpha\beta}
F_{\mu\alpha} F_{\nu\beta} \nonumber \\
&+ \bar {\tilde \psi} \biggl [ i {\tilde e}^\mu_a \gamma^a
(\partial_\mu + {\tilde \Omega}_\mu +qA_\mu ) - \tilde M (\varphi) \biggr
]
\tilde \psi \biggr ) \,,
\label{stringlagrangian1}
\end{eqnarray}
where ${\tilde g}_{\mu\nu}$ is the non-physical metric, 
${\tilde R}_{\mu\nu}$ is the Ricci tensor derived from
it, 
$\varphi$ is a
dilaton field, and $\tilde G$, $U$ and $\tilde M$ are functions of
$\varphi$.  The Lagrangian includes that for the electromagnetic field
$F_{\mu\nu}$, and that for 
particles, written in terms of Dirac spinors $\tilde \psi$.  This is
not a metric representation because of the coupling of $\varphi$ to
matter via $\tilde M (\varphi)$ and $U(\varphi)$.   
A conformal transformation ${\tilde g}_{\mu\nu} =
F(\varphi) g_{\mu\nu}$, $\tilde \psi = F(\varphi)^{-3/4} \psi$, 
puts the Lagrangian in the form (``Jordan'' frame)
\begin{eqnarray}
{\cal L} &=& \sqrt{- g} \biggl ( {g}^{\mu\nu}
\biggl [ {1 \over {
2\kappa}} F(\varphi) {R}_{\mu\nu} 
- {1 \over 2} F(\varphi)
\tilde G (\varphi) \partial_\mu \varphi
\partial_\nu \varphi
+{3 \over {4\kappa F(\varphi)}} \partial_\mu F
\partial_\nu F \biggr ] \nonumber \\
&&- U(\varphi) {g}^{\mu\nu}{g}^{\alpha\beta}
F_{\mu\alpha} F_{\nu\beta} 
\nonumber \\
&&
+ \bar {\psi} \biggl [ i {e}^\mu_a
\gamma^a
(\partial_\mu + {\Omega}_\mu +qA_\mu ) - \tilde M (\varphi)
F^{1/2} \biggr ]
\psi \biggr ) \,.
\label{stringlagrangian2}
\end{eqnarray}
One may choose $F(\varphi)= {\rm const}/\tilde M (\varphi)^2$ 
so that the particle Lagrangian takes the
metric form (no explicit 
coupling to $\varphi$), but the electromagnetic Lagrangian
will still couple non-metrically to $U(\varphi)$.  The gravitational
Lagrangian here takes the form of a scalar-tensor theory (Sec.
\ref{scalartensor}).  But the non-metric electromagnetic term will, in
general, produce violations of EEP.  For examples of specific models,
see \cite{TaylorVeneziano,DamourPolyakov}.
 
Thus, EEP and related tests are now viewed as ways to discover or place
constraints on new physical interactions, or as a branch of
``non-accelerator particle physics'', searching for the possible imprints
of high-energy particle effects in the low-energy realm of gravity. 
Whether current or proposed experiments 
can actually probe these phenomena meaningfully is an open
question at the moment, largely because of a dearth of firm
theoretical predictions.  Despite this uncertainty, a number of experimental
possibilities are being explored.

Concepts for an equivalence principle experiment in space have been developed. 
The project MICROSCOPE, designed to test WEP to $10^{-15}$ has been
approved by the French space agency CNES for a possible 2004 launch.
Another, known as 
Satellite Test of the Equivalence Principle (STEP), 
is under consideration as a possible
joint effort of NASA  and the European Space Agency (ESA), with the
goal of a $10^{-18}$ test.  
The gravitational redshift could be improved to the $10^{-10}$
level 
using atomic clocks on board 
a spacecraft which would travel to 
within four solar radii of the Sun.  
Laboratory tests of the gravitational inverse square law at
sub-millimeter scales are being developed as ways to search for new
short-range interactions or for the existence of large extra dimensions; 
the challenge of these experiments is to
distinguish gravitation-like interactions from electromagnetic and
quantum mechanical (Casimir) effects \cite{price}.

%===========================================================
% Section 3
%===========================================================

\section{Tests of Post-Newtonian Gravity}\label{S3}
\index{post-Newtonian gravity, tests} 

%---------------------------------------------------------------
% Subsection 3.1
%---------------------------------------------------------------

\subsection{Metric Theories of Gravity and the Strong Equivalence Principle}
\label{metrictheories}

\subsubsection{Universal Coupling and the Metric Postulates}\label{universal}

The overwhelming empirical evidence supporting the Einstein
equivalence principle, discussed in the previous section, 
supports the conclusion that the only theories of
gravity that have a hope of being viable are metric
theories, or possibly theories that are metric apart from possible weak
or short-range non-metric couplings (as in string theory).  Therefore for
the remainder of this article, we shall turn our attention
exclusively to metric theories of gravity, which assume that
(i)~there exists a
symmetric metric, (ii)~test bodies follow geodesics of the
metric, and (iii)~in local Lorentz frames, the non-gravitational
laws of physics are those of special relativity.

The property that all non-gravitational fields should couple in
the same manner to a single gravitational field is sometimes
called ``universal coupling''.  Because of it, one can discuss the
metric as a property of spacetime itself rather than as a field
over spacetime.  This is because its properties may be measured
and studied using a variety of different experimental devices,
composed of different non-gravitational fields and particles,
and, because of universal coupling, the results will be
independent of the device.  Thus, for instance, the proper time
between two events is a characteristic of spacetime and of the
location of the events, not of the clocks used to measure it.

Consequently, if EEP is valid, the non-gravitational laws of
physics may be formulated by taking their special relativistic
forms in terms of the Min\-kowski metric $\mbox{\boldmath$\eta$}$ and simply
``going over'' to new forms in terms of the curved spacetime
metric $\bf g$, using the mathematics of differential geometry.
The details of this ``going over'' can be found in standard
textbooks (\cite{MTW,Weinberg}; TEGP 3.2)

\subsubsection{The Strong Equivalence Principle}
\label{sep}

In any metric theory of gravity, matter and non-gravitational
fields respond only to the spacetime metric $\bf g$.  In
principle, however, there could exist other gravitational fields
besides the metric, such as scalar fields, vector fields, and so
on.  If, by our strict definition of metric theory, 
matter does not couple to these fields, what can their role
in gravitation theory be?  Their role must be that of mediating
the manner in which matter and non-gravitational fields generate
gravitational fields and produce the metric; once determined,
however, the metric alone acts back on the matter in the manner
prescribed by EEP.

What distinguishes one metric theory from another, therefore, is
the number and kind of gravitational fields it contains in
addition to the metric, and the equations that determine the
structure and evolution of these fields.  From this viewpoint,
one can divide all metric theories of gravity into two fundamental
classes:  ``purely dynamical'' and ``prior-geometric''.

By ``purely dynamical metric theory''\index{purely dynamic metric theory}
 we mean any metric theory
whose gravitational fields have their structure and evolution
determined by coupled partial differential field equations.  In
other words, the behavior of each field is influenced to some
extent by a coupling to at least one of the other fields in the
theory.  By ``prior geometric''\index{prior geometric theory}
 theory, we mean any metric theory
that contains ``absolute elements'', fields or equations whose
structure and evolution are given {\it a priori}, and are
independent of the structure and evolution of the other fields of
the theory.  These ``absolute elements'' typically include flat
background metrics $\mbox{\boldmath$\eta$}$, cosmic time coordinates $t$,
algebraic relationships among otherwise dynamical fields, such as
$g_{\mu \nu} =  h_{\mu \nu} + k_\mu k_\nu$, where
$h_{\mu \nu}$ and $k_\mu$ may be dynamical fields.

General relativity is a purely dynamical theory since it contains
only one gravitational field, the metric itself, and its
structure and evolution are governed by partial differential
equations (Einstein's equations).  Brans-Dicke theory and its
generalizations are purely
dynamical theories; the field equation for the metric involves the
scalar field (as well as the matter as source), and that for the
scalar field involves the metric.  Rosen's bimetric theory is a
prior-geometric theory:  it has a non-dynamical, Riemann-flat 
background metric, 
$\mbox{\boldmath$\eta$}$, and the field equations for the physical metric 
$\bf g$ involve $\mbox{\boldmath$\eta$}$. 

By discussing metric theories of gravity from this broad point of
view, it is possible to draw some general conclusions about the
nature of gravity in different metric theories, conclusions that
are reminiscent of the Einstein equivalence principle, but that
are subsumed under the name ``strong equivalence principle''.

Consider a local, freely falling frame in any metric theory of
gravity.  Let this frame be small enough that inhomogeneities in
the external gravitational fields can be neglected throughout its
volume.  On the other hand, let the frame be large enough to
encompass a system of gravitating matter and its associated
gravitational fields.  The system could be a star, a black hole,
the solar system or a Cavendish experiment.  Call this frame a
``quasi-local Lorentz frame'' .  To determine the behavior
of the system we must calculate the metric.  The computation
proceeds in two stages.  First we determine the external
behavior of the metric and gravitational fields, thereby
establishing boundary values for the fields generated by the
local system, at a boundary of the quasi-local frame ``far'' from
the local system.  Second, we solve for the fields generated by
the local system.  But because the metric is coupled directly or
indirectly to the other fields of the theory, its structure and
evolution will be influenced by those fields, and in particular
by the boundary values taken on by those fields far from the
local system.  This will be true even if we work in a coordinate
system in which the asymptotic form of $g_{\mu\nu}$ in the
boundary region between the local system and the external world
is that of the Minkowski metric.  Thus the gravitational
environment in which the local gravitating system resides can
influence the metric generated by the local system via the
boundary values of the auxiliary fields.  Consequently, the
results of local gravitational experiments may depend on the
location and velocity of the frame relative to the external
environment.  Of course, local {\it non}-gravitational
experiments are unaffected since the gravitational fields they
generate are assumed to be negligible, and since those
experiments couple only to the metric, whose form can always be
made locally Minkowskian at a given spacetime event.  
Local gravitational experiments might
include Cavendish experiments, measurement of the acceleration of
massive self-gravitating bodies, 
studies of the structure of stars and planets, or
analyses of the periods of ``gravitational clocks''.  We
can now make several statements about different kinds of metric
theories.

(i)   A theory which contains only the metric $\bf g$ yields
local gravitational physics which is independent of the location
and velocity of the local system.  This follows from the fact
that the only field coupling the local system to the environment
is $\bf g$, and it is always possible to find a coordinate
system in which $\bf g$ takes the Minkowski form at the boundary
between the local system and the external environment.  Thus the
asymptotic values of $g_{\mu\nu}$ are constants independent of
location, and are asymptotically Lorentz invariant, thus
independent of velocity.  General relativity is an example of
such a theory.  

(ii)  A theory which contains the metric $\bf g$ and dynamical
scalar fields $\varphi_A$ yields local gravitational physics
which may depend on the location of the frame but which is
independent of the velocity of the frame.  This follows from the
asymptotic Lorentz invariance of the Minkowski metric and of the
scalar fields, but now the asymptotic values of the scalar fields
may depend on the location of the frame.  An example is
Brans-Dicke theory, where the asymptotic scalar field determines
the effective value of the gravitational constant, which can thus vary as
$\varphi$ varies.  On the other hand, a form of velocity dependence in
local physics can enter indirectly if the asymptotic values of the
scalar field vary with time cosmologically.  Then the {\it rate} of
variation of the gravitational constant could 
depend on the velocity of the frame.

(iii) A theory which contains the metric $\bf g$ and additional
dynamical vector or tensor fields or prior-geometric fields
yields local gravitational physics which may have both location
and velocity-dependent effects.

These ideas can be summarized in 
the strong equivalence principle (SEP), which states that 
\begin{enumerate}
\item 
WEP is valid for self-gravitating bodies as well as for test bodies. 
\item
The outcome of any local test experiment is
independent of the velocity of the (freely falling) apparatus. 
\item
The outcome of any local test experiment is
independent of where and when in the universe it is performed.
\end{enumerate}
The distinction between SEP and EEP is the inclusion of bodies
with self-gravitational interactions (planets, stars) and of
experiments involving gravitational forces (Cavendish
experiments, gravimeter measurements).  Note that SEP contains
EEP as the special case in which local gravitational forces are
ignored.

The above discussion of the coupling of auxiliary fields to local
gravitating systems indicates that if SEP is strictly valid, there must be
one and only one gravitational field in the universe, the metric
$\bf g$.  These arguments are only suggestive however, and no
rigorous proof of this statement is available at present.
Empirically it has been found that every metric theory other than
GR introduces auxiliary gravitational fields,
either dynamical or prior geometric, and thus predicts violations
of SEP at some level (here we ignore quantum-theory inspired
modifications to GR involving ``$R^2$'' terms).  
General relativity seems to be the only
metric theory that embodies SEP completely.  This lends some
credence to the conjecture SEP $\to$ General Relativity.  In
Sec. \ref{septests}, we shall discuss experimental evidence for the
validity of SEP.

%----------------------------------------------------
% Subsection 3.2
%----------------------------------------------------

\subsection{The Parametrized Post-Newtonian Formalism}\label{ppn}

Despite the possible existence of long-range gravitational fields
in addition to the metric in various metric theories of gravity,
the postulates of those theories demand that matter and
non-gravitational fields be completely oblivious to them.  The
only gravitational field that enters the equations of motion is
the metric $\bf g$.  The role of the other fields that a theory may
contain can only be that of helping to generate the spacetime
curvature associated with the metric.  Matter may create these
fields, and they plus the matter may generate the metric, but they
cannot act back directly on the matter.  Matter responds only to
the metric.

Thus the metric and the equations of motion for matter become the
primary entities for calculating observable effects, 
and all that distinguishes one
metric theory from another is the particular way in which matter
and possibly other gravitational fields generate the metric.

The comparison of metric theories of gravity with each other and
with experiment becomes particularly simple when one takes the
slow-motion, weak-field limit.  This approximation, known as the
post-Newtonian limit, is sufficiently accurate to encompass most
solar-system tests that can be performed in the foreseeable
future.  It turns out that, in this limit, the spacetime metric
$\bf g$ predicted by nearly every metric theory of gravity has the
same structure.  It can be written as an expansion about the 
Minkowski metric ($ \eta_{\mu\nu} = {\rm diag}(-1,1,1,1)$) in terms
of dimensionless gravitational potentials of varying degrees of
smallness.
These potentials are constructed from the matter
variables (Box 2) in imitation of the Newtonian gravitational
potential
\begin{equation}\label{E20}
    U ( {\bf x} ,t) \equiv \int \rho ( {\bf x}^\prime , t) 
   | {\bf x} - {\bf x}^\prime |^{-1}
    d^3 x^\prime \,.
\end{equation}
The ``order of smallness'' is determined according to the rules
$U \sim~v^2 \sim \Pi \sim p/ \rho \sim \epsilon$, 
$v^i \sim | d/dt | / | d/dx | \sim \epsilon^{1/2}$, and so on (we use units
in which $G=c=1$; see Box 2).

\begin{table} 
\begin{center}
\begin{tabular}{c l c c c}
\hline
\hline
&&&Value&Value \\
&&&in semi-&in fully- \\
&What it measures &Value&conservative &conservative \\
Parameter&relative to GR&in GR&theories&theories \\[0.5ex]
\hline
\hline
$\gamma$&How much space-curvature&1&$\gamma$&$\gamma$ \\
&produced by unit rest mass?&&& \\[0.5ex]
\hline
$\beta$&How much ``nonlinearity''&1&$\beta$&$\beta$ \\
&in the superposition&&& \\
&law for gravity?&&& \\[0.5ex]
\hline
$\xi$&Preferred-location effects?&0&$\xi$&$\xi$ \\[0.5ex]
\hline
$\alpha_1$&Preferred-frame effects?&0&$\alpha_1$&0 \\
$\alpha_2$&&0&$\alpha_2$&0 \\
$\alpha_3$&&0&0&0 \\[0.5ex]
\hline
$\alpha_3$&Violation of conservation&0&0&0 \\
$\zeta_1$&of total momentum?&0&0&0 \\
$\zeta_2$&&0&0&0 \\
$\zeta_3$&&0&0&0 \\
$\zeta_4$&&0&0&0 \\[0.5ex]
\hline
\hline
\end{tabular}
\caption{The PPN Parameters and their significance (note that
$\alpha_3$ has been shown twice to indicate that it is a measure of
two effects)}
\label{ppnmeaning}
\end{center}
\end{table}

A consistent post-Newtonian limit requires determination of $g_{00}$ 
correct through $O(\epsilon^2)$, 
$g_{0i}$ through $O(\epsilon^{3/2})$ and $g_{ij}$
through $O(\epsilon)$ (for details see TEGP 4.1).  The only way that
one metric theory differs from another is in the numerical values
of the coefficients that appear in front of the metric
potentials.  The parametrized post-Newtonian (PPN ) formalism
inserts parameters in place of these coefficients, parameters
whose values depend on the theory under  study.  In the current
version of the PPN  formalism, summarized in Box 2, ten
parameters are used, chosen in such a manner that they measure or
indicate general properties of metric theories of gravity
(Table \ref{ppnmeaning}).  Under reasonable assumptions about the
kinds of potentials that can be present at post-Newtonian order
(basically only Poisson-like potentials), one finds that ten PPN
parameters exhaust the possibilities.

The parameters $\gamma$ and $\beta$ are the usual
Eddington-Robertson-Schiff parameters used to describe the
``classical'' tests of GR, and are in some sense the most important; they
are the only non-zero parameters in GR and scalar-tensor gravity.
The parameter $\xi$ is non-zero in
any theory of gravity that predicts preferred-location effects
such as a galaxy-induced anisotropy in the local gravitational
constant $G_L$ (also called ``Whitehead'' effects);
$\alpha_1$, $\alpha_2$, $\alpha_3$ measure whether or
not the theory predicts post-Newtonian preferred-frame
effects; $\alpha_3$, $\zeta_1$, $\zeta_2$,
$\zeta_3$, $\zeta_4$ measure whether or not the theory
predicts violations of global conservation laws for total
momentum.  Next to $\gamma$ and $\beta$, the parameters $\alpha_1$ and
$\alpha_2$ occur most frequently with non-trivial null values.
In Table \ref{ppnmeaning} we show the values these
parameters take (i)~in GR, (ii)~in any theory
of gravity that possesses conservation laws for total momentum,
called ``semi-conservative'' (any theory that is based on an
invariant action principle is semi-conservative), and
(iii)~in any theory that in addition possesses six global
conservation laws for angular momentum, called ``fully
conservative'' (such theories automatically predict no
post-Newtonian preferred-frame effects).  Semi-conservative
theories have five free PPN  parameters ($\gamma$, $\beta$, $\xi$,
$\alpha_1$, $\alpha_2$) while fully conservative theories
have three ($\gamma$, $\beta$ , $\xi$).

The PPN  formalism was pioneered by
Kenneth Nordtvedt \cite{nordtvedt2}, who studied the post-Newtonian
metric of a system of gravitating point masses, extending earlier
work by Eddington, Robertson and Schiff (TEGP 4.2).  A
general and unified version of the PPN  formalism was developed by
Will and Nordtvedt.  The canonical version, with
conventions altered to be more in accord with standard textbooks
such as \cite{MTW}, is discussed in detail in TEGP, Chapter 4.  Other versions
of the PPN  formalism have been developed to deal with point
masses with charge, fluid with anisotropic stresses,
bodies with strong internal gravity, and
post-post-Newtonian effects (TEGP 4.2, 14.2).  

%-- Box 2
\begin{table}
\medskip
\centerline{\bf Box 2.   The Parametrized Post-Newtonian Formalism}
\medskip
\hrule\small
\medskip
\begin{enumerate}
\item{\bf Coordinate System:}  
The framework uses a nearly
globally Lorentz coordinate system in which the coordinates are
$(t , x^1 , x^2 , x^3 )$.  Three-dimensional,
Euclidean vector notation is used throughout.  All coordinate
arbitrariness (``gauge freedom'') has been removed by
specialization of the coordinates to the standard PPN  gauge
(TEGP 4.2).  Units are chosen so that $G = c = 1$, where $G$ is
the physically measured Newtonian constant far from the solar system.

\item{\bf Matter Variables:}

\begin{itemize}
\item{$\rho =$}
density of rest mass as measured in a local freely falling 
frame momentarily comoving with the gravitating matter.
\item{$v^i=$}
$(dx^i /dt)=$ coordinate velocity of the matter.
\item{$w^i=$}
coordinate velocity of PPN  coordinate system relative 
to the mean rest-frame of the universe.
\item{$p=$}
pressure as measured in a local freely falling frame 
momentarily comoving with the matter.
\item{$\Pi=$}
internal energy per unit rest mass.  It includes all forms 
of non-rest-mass, non-gravitational energy, \eg energy 
of compression and thermal energy.
\end{itemize}

\item{\bf PPN  Parameters:}

$\gamma \,,\, \beta \,,\, \xi \,,\, \alpha_1 \,,\, \alpha_2 \,,\,
\alpha_3 \,,\, \zeta_1 \,,\,\zeta_2 \,,\,\zeta_3 \,,\,\zeta_4 \,. $

\item{\bf Metric:}
\begin{eqnarray*}
          g_{00} &=&
-1 + 2U - 2 \beta U^2 - 2 \xi \Phi_W 
        + (2 \gamma +2+ \alpha_3 + \zeta_1 - 2 \xi ) \Phi_1 \\
        &&+ 2(3 \gamma - 2 \beta + 1 + \zeta_2 + \xi ) \Phi_2 
+ 2(1 + \zeta_3 ) \Phi_3 
      + 2(3 \gamma + 3 \zeta_4 - 2 \xi ) \Phi_4 \\
      &&- ( \zeta_1 - 2 \xi ) {\cal A}
- ( \alpha_1 - \alpha_2 - \alpha_3 ) w^2 U
      - \alpha_2 w^i w^j U_{ij}
      + (2 \alpha_3 - \alpha_1 ) w^i V_i \\
     && +  O(\epsilon^3) \\
          g_{0i} 
&=& - {1 \over 2 }
          (4 \gamma + 3 + \alpha_1 - \alpha_2 
           + \zeta_1 - 2 \xi ) V_i 
       - {1 \over 2} 
          (1 + \alpha_2 - \zeta_1 + 2 \xi )W_i \\
       &&- {1 \over 2} ( \alpha_1 - 2 \alpha_2 ) w^i U
       - \alpha_2 w^j U_{ij} + O(\epsilon^{5/2})\\ 
           g_{ij} 
&=& (1 + 2 \gamma U + O(\epsilon^2)) \delta_{ij} 
\end{eqnarray*}

\end{enumerate}
\hrule\normalsize
\medskip
\end{table}

\begin{table}
\medskip
\centerline{\bf Box 2. (continued)}
\medskip
\hrule\small
\medskip
\begin{enumerate}

\item{\bf Metric Potentials:}
\begin{eqnarray*}
U&=&\int {{\rho^\prime } \over {| {\bf x}-{\bf x}^\prime |}}
d^3x^\prime \,,\qquad U_{ij}=
\int {{\rho^\prime (x-x^\prime)_i (x-x^\prime)_j} 
\over {| {\bf x}-{\bf x}^\prime |^3}} d^3x^\prime \\
\Phi_W &=& 
\int {{\rho^\prime \rho^{\prime\prime} ({\bf x}-{\bf x}^\prime)} 
\over {| {\bf x}-{\bf x}^\prime |^3}} \cdot \left (
{{{\bf x}^\prime -{\bf x}^{\prime\prime} } 
\over {| {\bf x}-{\bf x}^{\prime\prime} |}}-
{{{\bf x}-{\bf x}^{\prime\prime}} 
\over {| {\bf x}^\prime-{\bf x}^{\prime\prime} |}} 
\right ) d^3x^\prime d^3x^{\prime\prime} \\
{\cal A} &=& \int {{\rho^\prime [{\bf v}^\prime 
\cdot ({\bf x}-{\bf x}^\prime)]^2 } 
\over {| {\bf x}-{\bf x}^\prime |^3}}
d^3x^\prime \,, \qquad
\Phi_1 = 
\int {{\rho^\prime v^{\prime 2}} 
\over {| {\bf x}-{\bf x}^\prime |}} d^3x^\prime \\
\Phi_2 &=& 
\int {{\rho^\prime U^\prime} \over {| {\bf x}-{\bf x}^\prime |}} d^3x^\prime
\,,\quad
\Phi_3 =
\int {{\rho^\prime \Pi^\prime} \over {| {\bf x}-{\bf x}^\prime |}} d^3x^\prime
\,, \quad
\Phi_4=
\int {{p^\prime } \over {| {\bf x}-{\bf x}^\prime |}} d^3x^\prime \\
V_i &=&
\int {{\rho^\prime v_i^\prime} \over {| {\bf x}-{\bf x}^\prime |}} d^3x^\prime
\,, \qquad
W_i=
\int {{\rho^\prime [{\bf v}^\prime \cdot 
({\bf x}-{\bf x}^\prime)](x-x^\prime)_i} 
\over {| {\bf x}-{\bf x}^\prime |^3}} d^3x^\prime
\end{eqnarray*}

\item{\bf Stress-Energy Tensor} (perfect fluid)
\begin{eqnarray*}
T^{00} &=& \rho (1+ \Pi + v^2 +2U) \\
T^{0i} &=& \rho v^i (1+ \Pi + v^2 +2U + p/\rho) \\
T^{ij} &=& \rho v^iv^j (1+ \Pi + v^2 +2U + p/\rho) +
p\delta^{ij}(1-2\gamma U) 
\end{eqnarray*}

\item{\bf Equations of Motion}

\begin{itemize}
\item Stressed Matter, \quad
${T^{\mu\nu}}_{;\nu}= 0$
\item Test Bodies, \quad
$d^2 x^\mu /d \lambda^2 + {\Gamma^\mu}_{\nu \lambda}
(dx^\nu /d \lambda )( dx^\lambda / d \lambda ) = 0$
\item Maxwell's Equations, \quad
${F^{\mu \nu}}_{; \nu} = 4 \pi J^\mu$ \qquad
$F_{\mu \nu} = A_{\nu ; \mu} - A_{\mu ; \nu }$
\end{itemize}
\end{enumerate}
\hrule\normalsize
\medskip
\end{table}

%-- end of Box 2

%-----------------------------------------------------
% subsection 3.3
%-----------------------------------------------------

\subsection{Competing Theories of Gravity}\label{theories}

One of the important applications of the PPN  formalism is the
comparison and classification of alternative metric theories of
gravity.  The population of viable theories has fluctuated over
the years as new effects and tests have been discovered, largely
through the use of the PPN  framework, which eliminated many 
theories thought previously to be viable.  The theory population
has also fluctuated as new, potentially viable theories have been
invented.

In this article, we shall focus on general relativity and the
general class of scalar-tensor modifications of it, of which the
Jordan-Fierz-Brans-Dicke theory (Brans-Dicke, for short)
is the classic example.  The reasons are several-fold:

\begin{itemize}
\item A full compendium of alternative theories is given in Chapter 5 of
TEGP.
\item Many alternative metric theories developed during the 1970s and
1980s could be viewed as ``straw-man'' theories, invented to prove
that such theories exist or to illustrate particular properties.  Few
of these could be regarded as well-motivated theories from the point
of view, say, of field theory or particle physics.  Examples are the
vector-tensor theories studied by Will, Nordtvedt and Hellings.
\item A number of theories fall into the class of ``prior-geometric''
theories, with absolute elements such as a flat background metric in
addition to the physical metric.  Most of these theories predict
``preferred-frame'' effects, that have been tightly constrained by
observations (see Sec. \ref{preferred}).  
An example is Rosen's bimetric theory.  
\item A large number of alternative theories of gravity predict
gravitational-wave emission substantially different from that of general
relativity,
in strong disagreement with observations of the binary pulsar (see
Sec. \ref{S5}).
\item Scalar-tensor modifications of GR have recently become very
popular in unification schemes
such as string theory, and in cosmological model building.
Because the scalar fields are generally massive, the potentials in the
post-Newtonian limit will be modified by Yukawa-like terms.
\end{itemize}

\subsubsection{General Relativity}\label{generalrelativity}

The metric $\bf g$ is the sole
dynamical field and the theory contains no arbitrary functions or
parameters, apart from the value of the Newtonian coupling constant
$G$, which is measurable in laboratory experiments.  Throughout this
article, we ignore the cosmological constant $\lambda$.  Although
$\lambda$ has significance for quantum field theory, quantum
gravity, and cosmology, on the scale of the solar-system or of stellar
systems, its effects are negligible, for values of $\lambda$
corresponding to a cosmological closure density.

The field equations of GR are derivable from an invariant action
principle $\delta I=0$, where
\begin{equation}\label{E21}
I=(16\pi G)^{-1} \int R (-g)^{1/2} d^4x + I_m(\psi_m, g_{\mu\nu})\,,
\end{equation}
where $R$ is the Ricci scalar, and $I_m$ is the matter action, which
depends on matter fields $\psi_m$ universally coupled to the metric
$\bf g$.  By varying the action with respect to $g_{\mu\nu}$, we
obtain the field equations
\begin{equation}\label{E22}
G_{\mu\nu} \equiv R_{\mu\nu} - {1 \over 2} g_{\mu\nu} R = 8\pi G T_{\mu\nu}  
\,,
\end{equation}
where $T_{\mu\nu}$ is the matter energy-momentum tensor.  General
covariance of the matter action implies the equations of motion
${T^{\mu\nu}}_{;\nu}=0$; varying $I_m$ with respect to $\psi_M$ yields
the matter field equations.  By virtue of the {\it absence} of
prior-geometric elements, the equations of motion are also a
consequence of the field equations via the Bianchi identities
${G^{\mu\nu}}_{;\nu}=0$.

The general procedure for deriving the post-Newtonian limit is spelled
out in TEGP 5.1, and is described in detail for GR in TEGP 5.2.  The PPN 
parameter values are listed in Table \ref{ppnvalues}.

\begin{table} 
\begin{center}
\begin{tabular}{l c c c c c c c}
\hline
\hline
\noalign{\smallskip}
&Arbitrary&Cosmic&\multispan 5 \hfil PPN 
Parameters \hfil \\
&Functions&Matching&\multispan 5 \hrulefill \\
Theory&or
Constants&Parameters&$\gamma$&$\beta$&$\xi$&$\alpha_1$&$\alpha_2$\\
[0.5ex]
\hline
\hline
General Relativity &none&none&1&1&0&0&0 \\
Scalar-Tensor &&&&&&& \\
\quad Brans-Dicke&$\omega$&$\phi_0$&${{(1+\omega)}
\over {(2+\omega)}}$
&1&0&0&0 \\
\quad General&$A(\varphi) ,\, V(\varphi)$&$\varphi_0$
&${{(1+\omega)} \over {(2+\omega)}}$&$1+\Lambda$&0&0&0
\\
Rosen's Bimetric&none&$c_0,\,c_1$&1&1&0&0&${c_0 \over c_1}-1$ \\
[0.5ex]
\hline
\hline
\end{tabular}
\caption{Metric Theories and Their PPN  Parameter Values ($\alpha_3
= \zeta_i=0$ for all cases)}
\label{ppnvalues}
\end{center}
\end{table}

\subsubsection{Scalar-Tensor Theories}\label{scalartensor}

These theories contain the metric $\bf g$, a
scalar field $\varphi$, a potential  function $V(\varphi)$, and a
coupling function $A(\varphi)$ (generalizations to more than one scalar 
field have also been carried out \cite{DamourEspo92}).
For some purposes, the action is conveniently written in a non-metric
representation, sometimes denoted the ``Einstein frame'', in which the
gravitational action looks exactly like that of GR:
\begin{equation}\label{E23}
\tilde I=(16\pi G)^{-1} \int [\tilde R -2\tilde g^{\mu\nu} \partial_\mu \varphi 
\partial_\nu
\varphi -V(\varphi)] (-\tilde g)^{1/2} d^4x + I_m(\psi_m, A^2(\varphi) 
\tilde g_{\mu\nu})  
\,,
\end{equation}
where $\tilde R \equiv \tilde g^{\mu\nu} \tilde R_{\mu\nu}$ is the 
Ricci scalar of the
``Einstein'' metric $\tilde g_{\mu\nu}$.  (Apart from the scalar potential term
$V(\varphi)$, this corresponds to Eq. (\ref{stringlagrangian1})
with $\tilde G(\varphi) \equiv (4\pi G)^{-1}$, $U(\varphi) \equiv 1$,
and $\tilde M(\varphi) \propto A(\varphi)$.)  This representation is a
``non-metric'' one because the matter fields $\psi_m$ couple to a
combination of $\varphi$ and $\tilde g_{\mu\nu}$.  
Despite appearances, however, 
it is a metric theory, because it can be put
into a metric representation by identifying the ``physical metric''
\begin{equation}\label{E24}
g_{\mu\nu} \equiv A^2(\varphi) \tilde g_{\mu\nu} \,.
\end{equation}
The action can then be rewritten in the metric form
\begin{equation}\label{E25}
I=(16\pi G)^{-1} \int [\phi R - \phi^{-1} \omega(\phi) g^{\mu\nu}\partial_\mu 
\phi \partial_\nu \phi - \phi^2 V] (-g)^{1/2} d^4x + I_m(\psi_m,  
g_{\mu\nu})  
\,,
\end{equation}
where 
\begin{eqnarray}\label{E26}
\phi &\equiv& A(\varphi)^{-2} \,, \nonumber \\
3+2\omega(\phi) &\equiv& \alpha(\varphi)^{-2} \,, \nonumber \\
 \alpha(\varphi) &\equiv& d (\ln A(\varphi))/d\varphi \,.
\end{eqnarray}
The Einstein frame is useful for discussing general characteristics of
such theories, and for some cosmological applications, while the metric
representation is most useful for calculating observable effects.  
The field equations, post-Newtonian limit and PPN  parameters are
discussed in TEGP 5.3, and the values of the PPN  parameters are
listed in Table \ref{ppnvalues}.

The
parameters that enter the post-Newtonian limit are
\begin{equation}\label{E27}
\omega \equiv \omega(\phi_0)   \qquad 
\Lambda \equiv [(d\omega/d\phi)(3+2\omega)^{-2}(4+2\omega)^{-1}]_{\phi_0}  
\,,
\end{equation}
where $\phi_0$ is the value of $\phi$ today far from the 
system being studied, as determined by appropriate cosmological boundary
conditions.  
The following formula is also useful:
$1/(2+\omega)=2\alpha_0^2/(1+\alpha_0^2)$.
In Brans-Dicke theory ($\omega(\phi)=$ constant), 
the larger the value of
$\omega$, the smaller the effects of the scalar field, and in the
limit $\omega \to \infty$ ($\alpha_0 \to 0$), 
the theory becomes indistinguishable from
GR in all its predictions.  In more general
theories, the function $\omega ( \phi )$ could have the
property that, at the present epoch, and in weak-field situations, 
the value of the scalar field $\phi_0$ is such that,
$\omega$ is very large and $\Lambda$ is very small (theory almost
identical to GR today), but that for past or
future values of $\phi$, or in strong-field regions such as the
interiors of neutron stars, $\omega$ and $\Lambda$ could take on values
that would lead to significant differences from GR.  
Indeed, Damour and Nordtvedt have shown that in such general
scalar-tensor theories, GR is a natural ``attractor'': regardless of
how different the theory may be from GR in the early universe (apart
from special cases), cosmological
evolution naturally drives the fields toward small values  of the
function $\alpha$, thence to large $\omega$.  Estimates of the
expected relic deviations from GR today in such theories depend on the
cosmological model, but range from $10^{-5}$ to a few times $10^{-7}$
for $1{-}\gamma$ (\cite{DamourNord93a,DamourNord93b}). 

Scalar fields coupled to gravity or matter are also
ubiquitous in particle-physics-inspired models of unification, such as
string theory.  
In some models, the coupling to matter may lead to
violations of WEP, which are tested by E\"otv\"os-type experiments.  In
many models the scalar field is massive; if the Compton wavelength is
of macroscopic scale, its effects are those of a ``fifth force''.
Only if the theory can be cast as a metric theory with a
scalar field of infinite range or of range long compared to the scale
of the system in question (solar system) can the PPN  framework be
strictly
applied.  If the mass of the scalar field is sufficiently large that its
range is microscopic, then, on solar-system scales, the scalar field is
suppressed, and the theory is essentially equivalent to general
relativity.  This is the case, for example in the ``oscillating-G''
models of Accetta, Steinhardt and Will (see \cite{Steinhardt}), 
in which the
potential function $V(\varphi)$ contains both quadratic (mass)
and quartic (self-interaction) terms, causing the scalar field to
oscillate (the initial amplitude of oscillation is provided by an
inflationary epoch); high-frequency 
oscillations in the ``effective'' Newtonian
constant $G_{\sl eff} \equiv G/\phi = GA(\varphi)^2$ then result.      
The energy density in the oscillating scalar field can be enough to
provide a cosmological closure density without resorting to dark 
matter, yet the
value of $\omega$ today is so large that the theory's local predictions are
experimentally
indistinguishable from GR.   In other models, explored by Damour and
Esposito-Far\`ese \cite{DamourEspo96}, 
non-linear scalar-field couplings can lead
to ``spontaneous scalarization'' inside strong-field objects such as
neutron stars, leading to large deviations from GR, even in the limit
of very large $\omega$.

%------------------------------------------------
% subsection 3.4
%------------------------------------------------

\subsection{Tests of the Parameter $\gamma$} \label{gamma}

With the PPN  formalism in hand, we are now ready to confront
gravitation theories with the results of solar-system
experiments.  In this section we focus on tests of the parameter
$\gamma$, consisting of the deflection of light and the time delay
of light.

\subsubsection{The Deflection of Light}\label{deflection}

\begin{figure}
\begin{center}
\leavevmode
\psfig{figure=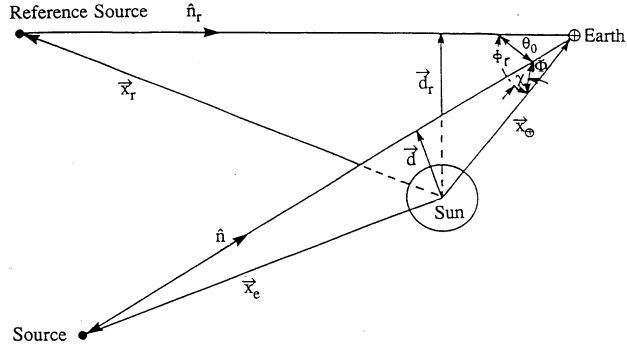,height=2.0in}
\end{center}
\caption{Geometry of light deflection measurements.} 
\label{deflectiongeom}
\end{figure}

A light ray
(or photon) which passes the Sun at a distance $d$ is deflected by
an angle
\begin{equation}\label{E28}
    \delta \theta = {1 \over 2}(1+ \gamma ) (4m_\odot/d)
    [(1+\cos\Phi )/2]  
\end{equation}
(TEGP 7.1), where $m_\odot$ is the mass of the Sun
and $\Phi$ is the angle between the Earth-Sun line and the
incoming direction of the photon (Figure \ref{deflectiongeom}).  
For a grazing ray, 
$d \approx d_\odot$, $\Phi \approx 0$, and
\begin{equation}\label{E29}
     \delta \theta ~\approx~{1 \over 2} (1+ \gamma ) 1.^{\prime\prime}75  
\,,
\end{equation}
independent of the frequency of light.  Another, more useful
expression gives the change in the relative angular separation
between an observed source of light and a nearby reference source
as both rays pass near the Sun:
\begin{equation}\label{E30}
          \delta \theta = {1 \over 2}(1+ \gamma )
\left [ - {{4m_\odot} \over d} \cos\chi
      + {{4m_\odot} \over d_r} 
\left ( {{1+ \cos\Phi_r} \over 2} \right ) \right ]  
\,,
\end{equation}
where $d$ and $d_r$ are the distances of closest approach of
the source and reference rays respectively, $\Phi_r$ is the
angular separation between the Sun and the reference source, and
$\chi$ is the angle between the Sun-source and the Sun-reference
directions, projected on the plane of the sky (Figure \ref{deflectiongeom}).
Thus, for example, the relative angular separation between the
two sources may vary if the line of sight of one of them passes
near the Sun ($d \sim R_\odot$, $d_r \gg d$,
$\chi$ varying with time).

It is interesting to note that the classic derivations of the
deflection of light that use only the principle of equivalence
or the corpuscular theory of light
yield only the ``$1/2$'' part of the coefficient in front of the
expression in Eq. (\ref{E28}).  But the result of these calculations
is the deflection of light relative to local straight lines, as
established for example by rigid rods; however, because of space
curvature around the Sun, determined by the PPN  parameter
$\gamma$, local straight lines are bent relative to asymptotic
straight lines far from the Sun by just enough to yield the
remaining factor ``$\gamma /2$''.  The first factor ``$1/2$''
holds in any metric theory, the second ``$\gamma /2$'' varies
from theory to theory.  Thus, calculations that purport to derive
the full deflection using the equivalence principle alone are
incorrect.

\begin{figure}
\begin{center}
\leavevmode
\psfig{figure=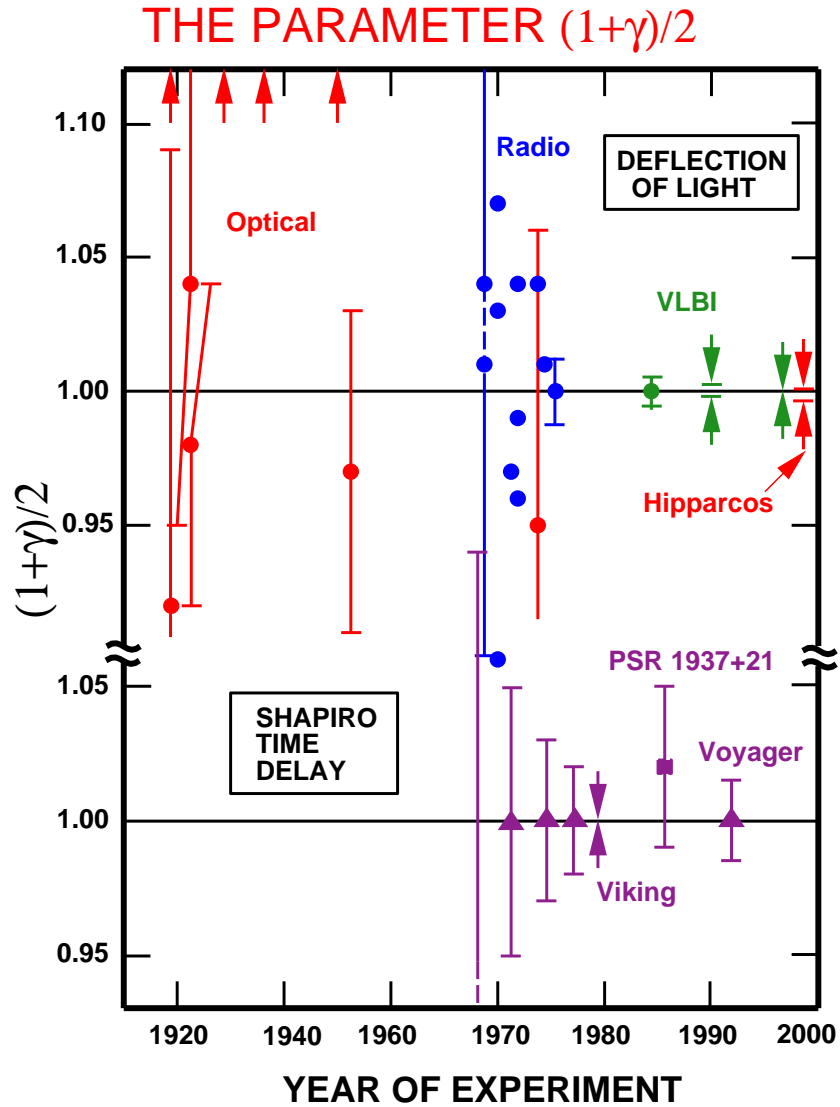,height=5.8in}
\end{center}
\caption{Measurements of the coefficient 
$(1 + \gamma )/2$ from light deflection and time
delay measurements.  General relativity value is unity.  
Arrows denote anomalously large
values from early eclipse expeditions.  Shapiro time-delay
measurements using Viking spacecraft yielded agreement with GR to 0.1 percent,
and VLBI light deflection
measurements have reached 0.02 percent.
Hipparcos denotes the optical astrometry satellite, which has reached 0.1
percent.
} 
\label{gammavalues}
\end{figure}

The prediction of the full bending of light by the Sun was one of
the great successes of Einstein's GR.
Eddington's confirmation of the bending of optical starlight
observed during a solar eclipse in the first days following World
War I helped make Einstein famous.  However, the experiments of
Eddington and his co-workers had only 30~percent accuracy, and
succeeding experiments were not much better:  the results were
scattered between one half and twice the Einstein value 
(Figure \ref{gammavalues}),
and the accuracies were low.
 
However, the development of VLBI, very-long-baseline radio
interfer\-om\-etry, produced greatly improved determinations of
the deflection of light.  These techniques now have the capability
of measuring angular separations and changes in angles
as small as 100 microarcseconds.  Early measurements took advantage 
of a series of heavenly coincidences:
each year, groups of strong quasistellar radio sources pass
very close to the Sun (as seen from the Earth), including the
group 3C273, 3C279, and 3C48, and the group 0111+02, 0119+11
and 0116+08.  As the Earth moves in its orbit, changing the
lines of sight of the quasars relative to the Sun, the angular
separation $\delta \theta$ between pairs of quasars
varies [Eq. (\ref{E30})].  The time variation in the
quantities $d$, $d_r$, $\chi$ and $\Phi_r$ in Eq. (\ref{E30}) is
determined using an accurate ephemeris for the Earth and initial
directions for the quasars, and the resulting prediction for 
$\delta \theta$  as a function of time is used as a basis for a
least-squares fit of the measured $\delta \theta$,  with one of
the fitted parameters being the coefficient ${1 \over 2}(1+ \gamma )$.  
A number of measurements of this kind over the period 1969--1975 yielded 
an accurate determination of the coefficient ${1 \over 2}(1+ \gamma)$  
which has the value unity in GR.  A 1995 VLBI measurement using 3C273 and
3C279 yielded $(1+\gamma)/2=0.9996 \pm 0.0017$ \cite{lebach}. 

A recent series of transcontinental and intercontinental VLBI  quasar and 
radio galaxy observations made primarily to monitor the Earth's rotation
(``VLBI '' in Figure \ref{gammavalues}) 
was sensitive to the deflection of light over 
almost the entire celestial sphere (at $90 ^\circ$ from the Sun, the
deflection is still 4 milli\-arcseconds). 
A recent analysis of  over 2 million VLBI observations
yielded $(1+\gamma)/2=0.99992 \pm 0.00014$ \cite{eubanks}.  
Analysis of observations made by the Hipparcos optical astrometry
satellite yielded a test at the level of 0.3
percent \cite{hipparcos}.
A VLBI 
measurement of the deflection of light by Jupiter was
reported; the predicted deflection of about 300
microarcseconds was seen with about 50 percent accuracy \cite{treuhaft}.
The results of light-deflection measurements are summarized in
Figure \ref{gammavalues}.

\subsubsection{The Time Delay of Light}\label{timedelay}

A radar signal sent across the solar system past the Sun to a
planet or satellite and returned to the Earth suffers an
additional non-Newtonian delay in its round-trip travel time,
given by (see Figure \ref{deflectiongeom})
\begin{equation}\label{E31}
      \delta t = 2(1+ \gamma ) m_\odot
      \ln [(r_\oplus
  ~+~ {\bf x}_\oplus \cdot {\bf n} )
      (r_e - {\bf x}_e \cdot {\bf n} )/d^2 ]  
\end{equation}
(TEGP 7.2).  For a ray which passes close to the Sun,
\begin{equation} \label{E32}
    \delta t \approx {1 \over 2} (1+ \gamma )
    [240-20 \ln (d^2 /r)]\; \mu {\rm s}  
\,,
\end{equation}
where $d$ is the distance of closest approach of the ray in solar
radii, and $r$ is the distance of the planet or satellite from the
Sun, in astronomical units.

In the two decades following Irwin Shapiro's 1964 discovery of
this effect as a theoretical consequence of general
relativity, several high-precision measurements were made
using radar ranging to targets passing through superior
conjunction.  Since one does not have access to a ``Newtonian''
signal against which to compare the round-trip travel time of the
observed signal, it is necessary to do a differential measurement
of the variations in round-trip travel times as the target passes
through superior conjunction, and to look for the logarithmic
behavior of Eq. (\ref{E32}).  In order to do this accurately however,
one must take into account the variations in round-trip travel
time due to the orbital motion of the target relative to the
Earth. This is done by using radar-ranging (and possibly other)
data on the target taken when it is far from superior conjunction
({\it i.e.}, when the time-delay term is negligible) to determine
an accurate ephemeris for the target, using the ephemeris to
predict the PPN  coordinate trajectory ${\bf x}_e (t)$ near
superior conjunction, then combining that trajectory with the
trajectory of the Earth ${\bf x}_\oplus (t)$ to determine the
Newtonian round-trip time and the logarithmic term in Eq. (\ref{E32}).
The resulting predicted round-trip travel times in terms of the
unknown coefficient ${1 \over 2}(1+ \gamma)$
are then fit to the measured travel times using the method 
of least-squares, and an estimate obtained for
${1 \over 2}(1+ \gamma)$.

The targets employed included
planets, such as Mercury or Venus, used as a passive reflectors of
the radar signals (``passive radar''); and
artificial satellites, such as Mariners~6 and 7, Voyager 2,
and the Viking
Mars landers and orbiters,
used as
active retransmitters of the radar signals (``active radar').

The results for the coefficient ${1 \over 2}(1+ \gamma)$
of all radar time-delay measurements
performed to date (including a measurement of the one-way time delay
of signals from the millisecond pulsar PSR 1937+21) 
are shown in Figure \ref{gammavalues} (see TEGP
7.2 for discussion and references).  The Viking experiment resulted in a
0.1 percent measurement \cite{reasenberg}.

  From the results of VLBI light-deflection experiments, we
can conclude that the coefficient 
${1 \over 2}(1+ \gamma)$
must be within at most 0.014~percent of unity.  
Scalar-tensor theories must have $\omega > 3500$ to be compatible with
this constraint.

\begin{table}
\begin{center}
\begin{tabular}{c l c l}
\hline
\hline
Parameter&Effect&Limit&Remarks \\ [0.5ex]
\hline
\hline
$\gamma-1$&time delay&$2 \times 10^{-3}$&Viking ranging \\
&light deflection&$3 \times 10^{-4}$&VLBI\\
$\beta-1$&perihelion shift&$3 \times 10^{-3}$&$J_2=10^{-7}$ from
helioseismology \\
&Nordtvedt effect&$6 \times 10^{-4}$&$\eta=4\beta-\gamma-3$
assumed \\
$\xi$&Earth tides&$10^{-3}$&gravimeter data \\
$\alpha_1$&orbital polarization&$10^{-4}$&Lunar laser
ranging \\
&&$2 \times 10^{-4}$&PSR J2317+1439\\
$\alpha_2$&spin precession&$4 \times 10^{-7}$&solar alignment 
with ecliptic \\
$\alpha_3$&pulsar acceleration&$2 \times 10^{-20}$
&pulsar $\dot P$ statistics \\
$\eta^1$&Nordtvedt effect&$10^{-3}$&lunar laser ranging \\
$\zeta_1$&-- &$2 \times 10^{-2}$&combined PPN  bounds \\
$\zeta_2$&binary acceleration&$4 \times 10^{-5}$&$\ddot P_p$
for PSR 1913+16 \\
$\zeta_3$&Newton's 3rd law&$10^{-8}$&Lunar acceleration \\
$\zeta_4$&-- &-- &not independent\\ [0.5ex]
\hline
\hline
\noalign{\smallskip}
\multispan 4 $^1$Here $\eta = 4\beta -\gamma -3 - 10 \xi /3 -\alpha_1
-2
\alpha_2 /3 - 2\zeta_1 /3 - \zeta_2 /3 $\hfill \\
\end{tabular}
\caption{Current Limits on the PPN  Parameters}
\index{limits on PPN  parameters}
\label{ppnlimits}
\end{center} 
\end{table}

%---------------------------------------------------------
% subsection 3.5
%----------------------------------------------------------

\subsection{The Perihelion Shift of Mercury}\label{perihelion}

The explanation of the anomalous perihelion shift of Mercury's
orbit was another of the triumphs of GR.  This had
been an unsolved problem in celestial mechanics for over half a
century, since the announcement by Le Verrier in 1859 that, after
the perturbing effects of the planets on Mercury's orbit had been
accounted for, and after the effect of the precession of the
equinoxes on the astronomical coordinate system had been
subtracted, there remained in the data an unexplained advance
in the perihelion of Mercury.  The modern value for this
discrepancy is 43 arcseconds per century.  A number of {\it ad
hoc} proposals were made in an attempt to account for this
excess, including, among others, the existence of new planet
Vulcan near the Sun, a ring of planetoids, a solar quadrupole
moment and a deviation from the inverse-square law of gravitation,
but none was successful.  General relativity accounted
for the anomalous shift in a natural way without disturbing the
agreement with other planetary observations.

The predicted advance, $\Delta \tilde \omega$ , per orbit, including
both relativistic PPN  contributions and the Newtonian
contribution resulting from a
possible solar quadrupole moment, is given by
\begin{equation}\label{E33}
 \Delta \tilde \omega
 = (6 \pi m/p)[ {1 \over 3} (2+2 \gamma - \beta )
 + {1 \over 6}~
     (2 \alpha_1 - \alpha_2 + \alpha_3 +2 \zeta_2 ) \mu /m
 + J_2 (R^2 /2mp)]  \,,
\end{equation}
where $m \equiv m_1 + m_2$ and $\mu  \equiv m_1 m_2 /m$
are the total mass and reduced mass of the two-body system
respectively; $p \equiv a(1-e^2 )$ is the semi-latus rectum of
the orbit, with a the semi-major axis and $e$ the eccentricity; $R$ is
the mean radius of the oblate body; and $J_2$ is a
dimensionless measure of its quadrupole moment, given by 
$J_2 = (C-A)/m_1 R^2$, where $C$ and $A$ are the
moments of inertia about the body's rotation and equatorial
axes, respectively (for details of the derivation see TEGP 7.3).  
We have ignored preferred-frame and galaxy-induced
contributions to  $\Delta \tilde \omega$; these are discussed in
TEGP 8.3.

The first term in Eq. (\ref{E33}) is the classical relativistic
perihelion shift, which depends upon the PPN  parameters $\gamma$
and $\beta$.  The second term depends upon the ratio of the masses
of the two bodies; it is zero in any fully conservative theory of
gravity 
($ \alpha_1 \equiv \alpha_2 \equiv \alpha_3 \equiv \zeta_2 \equiv 0$);
it is also negligible for Mercury, since 
$\mu /m \approx m_{\rm Merc} /m_\odot \approx 2 \times 10^{-7}$. 
We shall drop this term henceforth.  The third term depends upon
the solar quadrupole moment $J_2$.  For a Sun that rotates
uniformly with its observed surface angular velocity,  so that
the quadrupole moment is produced by centrifugal flattening, one
may estimate $J_2$ to be 
$\sim 1 \times 10^{-7}$.  This actually agrees reasonably well with
values inferred from rotating solar models that are in accord with
observations of the normal modes of solar oscillations
(helioseismology). 
Substituting standard orbital elements and physical
constants for Mercury and the Sun we obtain the rate of
perihelion shift $\dot {\tilde \omega}$, in seconds of arc per century,
\begin{equation}\label{E34}
\dot {\tilde \omega}= 42.^{\prime\prime}98 
\left [ {1 \over 3} (2+2 \gamma - \beta ) 
                 + 3 \times 10^{-4} (J_2 /10^{-7} ) \right ]  \,.
\end{equation}
Now, the measured perihelion shift of Mercury is known 
accurately:  after the perturbing effects of the other planets have been
accounted for, the excess shift is known to about 0.1 percent
from radar observations of Mercury between 1966 and 1990 \cite{shapiroGR12}.  
Analysis of data taken since 1990 could improve the accuracy.
The solar oblateness
effect is smaller than the observational error, so we obtain the PPN 
bound $| 2\gamma -\beta-1 | < 3 \times 10^{-3}$.  

%---------------------------------------------------
% subsection 3.6
%----------------------------------------------------

\subsection{Tests of the Strong Equivalence Principle}\label{septests}

The next class of solar-system experiments that test relativistic 
gravitational effects may be called tests of the strong
equivalence principle (SEP).  In Sec.\ 3.1.2, we pointed out that many metric
theories of gravity (perhaps all except GR) can be
expected to violate one or more aspects of SEP.  Among the
testable violations of SEP are a
violation of the weak equivalence principle for gravitating bodies
that leads to perturbations in the Earth-Moon
orbit; preferred-location and preferred-frame effects in the
locally measured gravitational constant that could produce
observable geophysical effects; and possible variations in the
gravitational constant over cosmological timescales.

\subsubsection{The Nordtvedt Effect and the Lunar E\"otv\"os Experiment}
\label{Nordtvedteffect}

In a pioneering calculation using his early form of the PPN 
formalism, Nord\-tvedt \cite{nordtvedt1} showed that many metric theories
of gravity predict that massive bodies violate the weak
equivalence principle -- that is, fall with different
accelerations depending on their gravitational self-energy.  Dicke
\cite{Dicke} linked such an effect to the possibility of a spatially
varying gravitational constant, in theories such as scalar-tensor
gravity.  For a
spherically symmetric body, the acceleration from rest in an
external gravitational potential $U$ has the form
\begin{eqnarray}\label{E35}
{\bf a} &=& (m_p/m) \nabla U \,, \nonumber \\
m_p/m &=& 1- \eta (E_g/m)  \,, \nonumber \\
    \eta 
 &=& 4 \beta - \gamma -3 - {10 \over 3} \xi
        - \alpha_1 +  {2 \over 3} \alpha_2
        -  {2 \over 3} \zeta_1
        -  {1 \over 3} \zeta_2  
\,,
\end{eqnarray}
where $E_g$ is the negative of the gravitational self-energy
of the body ($E_g >0$).  This violation of the massive-body
equivalence principle is known as the ``Nordtvedt effect''.  The
effect is absent in GR ($ \eta = 0$) but present
in scalar-tensor theory ($ \eta = 1/(2+ \omega )+4\Lambda$).  The existence
of the Nordtvedt effect does not violate the results of laboratory
E\"otv\"os experiments, since for laboratory-sized objects,
$E_g /m \le 10^{-27}$, far below the sensitivity of
current or future experiments.  However, for astronomical bodies, 
$E_g /m$ may be significant ($10^{-5}$ for the Sun, $10^{-8}$
for Jupiter, 
$4.6 \times 10^{-10}$  for the Earth, $0.2 \times 10^{-10}$
for the Moon).  If the Nordtvedt effect is present 
($\eta \ne 0$) then the Earth should fall toward the Sun with a
slightly different acceleration than the Moon.  This perturbation
in the Earth-Moon orbit leads to a polarization of the orbit that
is directed toward the Sun as it moves around the Earth-Moon
system, as seen from Earth.  This polarization represents a
perturbation in the Earth-Moon distance of the form
\begin{equation}\label{E36}
    \delta r = 13.1 \eta \cos( \omega_0 -\omega_s )t \quad {\rm m}
\,,
\end{equation}
where $\omega_0$ and $\omega_s$ are the angular frequencies
of the orbits of the Moon and Sun around the Earth (see TEGP
8.1, for detailed derivations and references; for improved
calculations of the numerical coefficient, see 
\cite{Nordtvedt95,DamourVokrou96}).

Since August 1969, when the first successful acquisition was made
of a laser signal reflected from the Apollo\ 11 retroreflector on
the Moon, the lunar laser-ranging experiment (LURE) has made
regular measurements of the round-trip travel times of laser
pulses between a network of observatories
and the lunar retroreflectors, with accuracies that are
approaching 50\ ps (1\ cm).  These measurements are fit
using the method of least-squares to a theoretical model for the
lunar motion that takes into account perturbations due to the Sun
and the other planets, tidal interactions, and post-Newtonian
gravitational effects.  The predicted round-trip travel times
between retroreflector and telescope also take into account the
librations of the Moon, the orientation of the Earth, the
location of the observatory, and atmospheric effects on the
signal propagation.  The ``Nordtvedt'' parameter $\eta$ along with
several other important parameters of the model are then
estimated in the least-squares method.

Several independent analyses of the data found no evidence, within
experimental uncertainty, for the Nordtvedt effect (for recent results
see \cite{Dickey,Williams,MullerMG}).  Their results
can be summarized by the bound $| \eta | < 0.001$.
These results represent a limit on a possible violation of WEP for
massive bodies of 5 parts in $10^{13}$ (compare Figure \ref{wepfig}).  For
Brans-Dicke theory, these results force a lower limit on the
coupling constant $\omega$ of 1000.   Note that,
at this level of precision, one cannot regard the results of lunar laser 
ranging as a ``clean'' test of SEP until one eliminates the
possibility of a compensating violation of WEP for the two bodies,
because the chemical compositions of the Earth 
and Moon differ in the relative fractions of iron and silicates.  To
this end, the E{\"o}t-Wash group carried out an improved test of WEP
for laboratory bodies whose chemical compositions mimic that of the
Earth and Moon.  The resulting bound of four parts in $10^{13}$
\cite{baessler} reduces the ambiguity in the Lunar laser ranging
bound, and establishes the firm limit on the universality of
acceleration of gravitational binding energy at the level of $\eta < 1.3
\times 10^{-3}$.

In GR, the Nordtvedt effect vanishes; at the level of
several centimeters and below, 
a number of non-null general relativistic effects
should be present \cite{Nordtvedt95}.

\subsubsection{Preferred-Frame and Preferred-Location Effects}\label{preferred}

Some theories of gravity violate SEP by predicting that the
outcomes of local gravitational experiments may depend on the
velocity of the laboratory relative to the mean rest frame of the
universe (preferred-frame effects) or on the location of the
laboratory relative to a nearby gravitating body
(preferred-location effects).  In the post-Newtonian limit,
preferred-frame effects are governed by the values of the PPN 
parameters $\alpha_1$, $\alpha_2$, and $\alpha_3$, and some
preferred-location effects are governed by $\xi$ (see Table \ref{ppnmeaning}). 

The most important such effects are variations and anisotropies
in the locally-measured value of the gravitational constant, which
lead to anomalous Earth tides and variations in the Earth's
rotation rate; anomalous contributions to the 
orbital dynamics of planets and the Moon; self-accelerations of 
pulsars, and anomalous torques on the Sun that
would cause its spin axis to be randomly oriented relative to the
ecliptic (see TEGP
8.2, 8.3, 9.3 and 14.3(c)).  An improved bound on $\alpha_3$ of $2
\times 10^{-20}$ from the period derivatives of 20 millisecond pulsars
was reported in \cite{Bell};  improved bounds on 
$\alpha_1$ were achieved using lunar laser ranging data \cite{MullerPRD}, 
and using
observations of the circular binary
orbit of the pulsar J2317+1439 (\cite{Bell2}).  
Negative searches for these effects have
produced strong constraints on the PPN  parameters (Table \ref{ppnlimits}).

\subsubsection{Constancy of the Newtonian Gravitational Constant}
\label{bigG}

Most theories of gravity that violate SEP predict that the locally 
measured Newtonian gravitational constant may vary with time as 
the universe evolves.  For the scalar-tensor theories listed in 
Table \ref{ppnvalues}, 
the predictions for $\dot G/G$ can be written in terms of time
derivatives of the asymptotic scalar field.
Where $G$
does change with cosmic evolution, its rate of variation should
be of the order of the expansion rate of the universe,
{\it i.e.,}
$\dot G/G \sim H_0$, where $H_0$ is the Hubble expansion parameter 
and is given by 
$H_0 = 100 h\; {\rm km~s^{-1}~Mpc^{-1}}=h \times 10^{-10}\, {\rm yr^{-1}}$,
where current observations of the expansion of the
universe give $h \approx 0.7$.

Several observational constraints can be placed on $\dot G/G$
using methods that include studies of the evolution of the Sun,
observations of lunar occultations (including analyses of ancient
eclipse data), lunar laser-ranging measurements,
planetary
radar-ranging measurements, and pulsar timing data.
Laboratory experiments may one day lead to interesting limits (for
review and references to past work see TEGP 8.4 and 14.3(c)).  Recent
results are shown in Table \ref{Gdottable}. 

\begin{table}
\begin{center}
\begin{tabular}{l c}
\hline
\hline
Method&$\dot G/G (10^{-12}\; {\rm yr}^{-1})$ \\[0.5ex]
\hline
\hline
Lunar Laser Ranging&$ 0  \pm  8$ \\[0.5ex]
&$3 \pm 5$ \\[0.5ex]
Viking Radar&$ 2  \pm  4$ \\
&$ -2  \pm  10$ \\[0.5ex]
Binary Pulsar$^1$&$ 11  \pm  11$ \\[0.5ex]
Pulsar PSR 0655+64$^1$&$ <  55$ \\[0.5ex]
\hline
\hline
\noalign{\medskip}
\multispan 2 
 ${}^1$ Bounds dependent
upon theory of gravity in \hfill\\
\multispan 2 
strong-field regime and on neutron star equation \hfill \\
\multispan 2 
of state. \hfill \\
\end{tabular}
\caption{Constancy of the Gravitational Constant}
\label{Gdottable}
\end{center} 
\end{table}

The best limits on $\dot G/G$ still
come from ranging measurements to the Viking landers and
Lunar laser ranging measurements \cite{Dickey,Williams,MullerMG}.
It has
been suggested that radar observations of a Mercury orbiter over a
two-year mission (30\ cm accuracy in range) could yield
$\Delta (\dot G/G) \sim 10^{-14}\; {\rm yr}^{-1}$. 
 
Although bounds on $\dot G/G$ from solar-system measurements can be
correctly
obtained in a phenomenological manner through the simple expedient of
replacing $G$ by $G_0 + {\dot G}_0 (t - t_0 )$ in
Newton's equations of motion, the same does not hold true for pulsar
and binary pulsar timing measurements.  The reason is that, in theories 
of gravity that violate SEP, such as scalar-tensor theories,
the ``mass'' and moment of inertia of a
gravitationally bound body may vary with variation in $G$.  Because
neutron stars are highly relativistic, the fractional variation in
these quantities can be comparable to $\Delta G/G$, the precise
variation depending both on the equation of state of neutron star
matter and on the theory of gravity in the strong-field regime.  The
variation in the moment of inertia affects the spin rate of the pulsar,
while the variation in the mass can affect the orbital period in a
manner that can subtract from the direct effect of a variation in G,
given by $\dot P_b /P_b =-{2} \dot G/G$ \cite{nordtvedt3}.  Thus,
the bounds quoted in Table \ref{Gdottable} 
for the binary pulsar PSR 1913+16 \cite{DamourTaylor91} and the
pulsar PSR 0655+64 \cite{goldman} 
are theory-dependent and must be treated as merely
suggestive.

%---------------------------------------------------
% subsection 3.7
%----------------------------------------------------

\subsection{Other Tests of Post-Newtonian Gravity} \label{othertests}

\subsubsection{Tests of Post-Newtonian Conservation Laws}
\label{conservation}

Of the five ``conservation law'' PPN  parameters 
$\zeta_1$, $\zeta_2$, $\zeta_3$, $\zeta_4$, 
and $\alpha_3$, only three, $\zeta_2$, $\zeta_3$ and $\alpha_3$,  
have been constrained directly with any precision;
$\zeta_1$ is constrained
indirectly through its appearance in the Nordtvedt effect parameter
$\eta$, Eq. (\ref{E35}).  There is strong theoretical
evidence that $\zeta_4$, which is related to the gravity generated by
fluid pressure, is not really an independent parameter -- in any
reasonable theory of gravity there should be a connection between the
gravity produced by kinetic energy ($\rho v^2$), internal energy ($\rho
\Pi$), and pressure ($p$).
From such considerations, there follows\cite{Will76}  
the additional theoretical
constraint
\begin{equation}\label{E37}
6\zeta_4= 3\alpha_3+2\zeta_1-3\zeta_3  \,.
\end{equation}

A non-zero value for any of these parameters would result in a
violation of conservation of momentum, or of Newton's third law
in gravitating systems.  An alternative statement of Newton's 
third law for gravitating systems is that the ``active gravitational mass'', 
that is the mass that determines the gravitational potential exhibited 
by a body, should equal the ``passive gravitational mass'', 
the mass that determines the force on a body in a gravitational field.   
Such an equality guarantees the equality of action and reaction and 
of conservation of momentum, at least in the Newtonian limit.

A classic test of Newton's third law for gravitating systems was
carried out in 1968 by Kreuzer, in which the gravitational attraction
of fluorine and bromine were compared to a precision of 
5\ parts in $10^5$.

A remarkable planetary test was reported by
Bartlett and van Buren \cite{bartlett}.  They noted that current understanding
of the structure of the Moon involves an iron-rich, aluminum-poor
mantle whose center of mass is offset about 10 km from the center of 
mass of an aluminum-rich, iron-poor crust.  The direction of offset 
is toward the Earth, about $14 ^\circ$ to the east of the Earth-Moon line.  
Such a model accounts for the basaltic maria which face the Earth, 
and the aluminum-rich highlands on the Moon's far side, and for a 2\ km 
offset between the observed center of mass and center of figure for 
the Moon.  Because of this asymmetry, a violation of Newton's third 
law for aluminum and iron would result in a momentum non-conserving 
self-force on the Moon, whose component along the orbital direction
would contribute to the secular acceleration of the lunar orbit.
Improved knowledge of the lunar orbit through lunar laser ranging, and
a better understanding of tidal effects in the Earth-Moon system
(which also contribute to the secular acceleration) through satellite 
data, severely limit any anomalous secular acceleration, with 
the resulting limit
\begin{equation}\label{E38}
\left | {{(m_A / m_P )_{\rm Al} - (m_A /m_P )_{\rm Fe}}
       \over
       {(m_A /m_P )_{\rm Fe}}}
\right | <4 \times 10^{-12} \,.
\end{equation}
According to the PPN  formalism, in a theory of gravity that violates
conservation of momentum, but that obeys the constraint of Eq. (\ref{E37}), 
the electrostatic binding energy $E_e$
of an atomic nucleus could make a contribution to the ratio of active to
passive mass of the form
\begin{equation}\label{E39}
     m_A = m_P 
 + {1 \over 2} \zeta_3 E_e /c^2 \,.
\end{equation}
The resulting limit on $\zeta_3$ from the lunar experiment 
is $\zeta_3 < 1 \times 10^{-8}$ (TEGP 9.2, 14.3(d)).

Another consequence of a violation of conservation of momentum is a
self-accel\-er\-at\-ion of the center of mass of a binary stellar system,
given by
\begin{equation}\label{E40}
{\bf a}_{\rm CM} = {1 \over 2} (\zeta_2+\alpha_3) {m \over a^2}{\mu
\over a}{{\delta m} \over m} {e \over {(1-e^2)^{3/2}}} {\bf n}_P \,, 
\end{equation}
where $\delta m = m_1-m_2$, $a$ is the semi-major axis, and ${\bf n}_P$
is a unit vector directed from the center of mass to the point of
periastron of $m_1$ (TEGP 9.3).
A consequence of this acceleration would be non-vanishing values for
$d^2 P/dt^2$, where $P$ denotes the period of any intrinsic process in
the system (orbit, spectra,
pulsar periods).  The observed upper limit on $d^2 P_p
/dt^2$ of the binary pulsar PSR 1913+16 places a
strong constraint on such an effect, resulting in the bound $|
\alpha_3 + \zeta_2 |<4 \times 10^{-5}$.  Since $\alpha_3$ has already
been constrained to be much less than this (Table \ref{ppnlimits}), 
we obtain a strong
bound on $\zeta_2$ alone \cite{Will92c}.

\subsubsection{Geodetic Precession}\label{geodeticprecession}

A gyroscope moving through curved spacetime suffers a precession of
its axis given by 
\begin{equation}\label{E41}
d {\bf S} /d \tau = \mbox{\boldmath$\Omega$}_G \times {\bf S} \,, \qquad  
\mbox{\boldmath$\Omega$}_G = (\gamma + {1 \over 2} ) {\bf v} 
           \times \nabla U  
\,,
\end{equation}
where $\bf v$ is the velocity of the gyroscope, and $U$ is the
Newtonian gravitational potential of the source (TEGP 9.1).  
The Earth-Moon system can be considered as a ``gyroscope'', with its axis
perpendicular to the orbital plane.  The predicted precession is about
2 arcseconds per century, an effect first calculated by de Sitter.
This effect has been measured to about 0.7 percent using Lunar laser
ranging data \cite{Dickey,Williams}.

For a gyroscope orbiting the Earth, the precession is about 8 arcseconds 
per year.  The Stanford Gyroscope Experiment has as one of its goals 
the measurement of this effect to $5 \times 10^{-5}$ (see below); if
achieved, this would substantially improve the accuracy of the parameter
$\gamma$.

\subsubsection{Search for Gravitomagnetism}\label{gravitomagnetism}

According to GR, moving or rotating matter should 
produce a contribution to the gravitational field that is the analogue 
of the magnetic field of a moving charge or a magnetic dipole.  
Although gravitomagnetism plays a role in a variety of measured 
relativistic effects, it has not been seen to date, isolated from 
other post-Newtonian effects (for a discussion of the evidence for
gravitomagnetism in solar system measurements and the binary pulsar, see
\cite{nordtvedt88a,nordtvedt88b}).  The Relativity 
Gyroscope Experiment (Gravity Probe B or GP-B) 
at Stanford University, in collaboration with
NASA  and Lockheed-Martin Corporation, is in the advanced stage 
of developing a space mission to detect this phenomenon 
directly \cite{gpbwebsite}.  A set of four superconducting-niobium-coated, 
spherical quartz gyroscopes will be flown in a low polar Earth orbit, 
and the precession of the gyroscopes relative to the distant stars 
will be measured.  In the PPN  formalism, the predicted effect 
of gravitomagnetism is a precession (also known as the Lense-Thirring
effect, or the dragging of inertial frames), given by
\begin{equation}\label{E42}
d {\bf S} /d \tau = \mbox{\boldmath$\Omega$}_{\rm LT} \times {\bf S} \,,
  \qquad
\mbox{\boldmath$\Omega$}_{\rm LT} = -{1 \over 2}(1+\gamma + {1 \over 4}\alpha_1 ) 
          [{\bf J}-3{\bf n}({\bf n} \cdot {\bf J})]/r^3  
\,,
\end{equation}
where $\bf J$ is the angular momentum of the Earth, $\bf n$ is a
unit radial vector, and $r$ is the distance from the center of the
Earth (TEGP 9.1).  For a polar orbit at about 650\ km altitude, 
this leads to a secular angular precession at a rate
${1 \over 2}(1+\gamma + {1 \over 4}\alpha_1 ) 
42 \times 10^{-3}$ arcsec/yr.
The accuracy goal of the experiment is about 0.5\ milliarcseconds 
per year.  The science instrument 
package and the spacecraft are
in the final phases of assembly, with launch scheduled for 
2002.

Another proposal to look for an effect of gravitomagnetism is to 
measure the relative precession of the line of nodes of a pair 
of laser-ranged geodynamics satellites (LAGEOS), ideally with supplementary 
inclination angles; the inclinations must be supplementary in order 
to cancel the dominant nodal precession caused by the Earth's 
Newtonian gravitational multipole moments.  Unfortunately, the two
existing LAGEOS satellites are not in appropriately inclined orbits,
and no plans exist at present to launch a third satellite in a
supplementary orbit.  Nevertheless, by combing nodal precession data
from LAGEOS I and II with perigee advance data from the slightly
eccentric orbit of LAGEOS II, Ciufolini \etal reported a partial
cancellation of multipole effects, and a resulting 20 percent
confirmation of GR \cite{ciufolini}.

\subsubsection{Improved PPN  Parameter Values}\label{improvedPPN}  

A number of advanced space missions have been proposed in which spacecraft 
orbiters or landers and improved tracking capabilities could lead to 
significant improvements in values of the PPN  parameters, 
of $J_2$ of the Sun, and of $\dot G/G$.  Doppler tracking of the Cassini
spacecraft (launched to orbit and study Saturn in 1997) 
during its 2003 superior conjunction could measure $\gamma$ to a few parts in
$10^5$, by measuring the time variation of the Shapiro delay \cite{iess}.
A Mercury orbiter, 
in a two-year experiment, with 3\ cm range capability, could yield 
improvements in the perihelion shift to a part in $10^4$, in $\gamma$ 
to $4 \times 10^{-5}$, in $\dot G/G$ to $10^{-14}\; {\rm yr}^{-1}$, and in 
$J_2$ to a few parts in $10^8$. 
Proposals are being developed, primarily in Europe, for advanced space
missions which will have tests of PPN  parameters as key components,
including GAIA, a high-precision astrometric telescope (successor to
Hipparcos), which could
measure light-deflection and 
$\gamma$ to the $10^{-6}$ level \cite{gaia}.  
Nordtvedt \cite{Nordtvedt00} has argued that ``grand fits'' of large solar
system range data sets, including ranging to Mercury, Mars and the Moon,
could yield substantially improved measurements of PPN parameters.

%=======================================================
% Section 4
%=======================================================

\section{Strong Gravity and Gravitational Waves: A New Testing
Ground}\label{S4}

%------------------------------------------------
% subsection 4.1
%------------------------------------------------

\subsection{Strong-field systems in general relativity}
\label{strong}

\subsubsection{Defining weak and strong gravity}
\label{strongvweak}

In the solar system, gravity is weak, in the sense that the  Newtonian
gravitational
potential and related variables ($ U ( {\bf x} ,t) \sim v^2 \sim p/\rho \sim
\epsilon$ )
are everywhere much smaller than unity everywhere.  
This is the basis for the post-Newtonian expansion and for the
``parametrized
post-Newtonian'' framework 
described in Sec. \ref{ppn}.
``Strong-field'' systems are those for which 
the simple 1PN approximation of the PPN framework
is no longer
appropriate.  This can occur in a number of situations:

\begin{itemize}

\item
The system may contain strongly relativistic objects, such as neutron
stars or black holes, near and inside which $\epsilon \sim 1$, and the
post-Newtonian approximation breaks down.  Nevertheless, under 
some circumstances,
the orbital motion may be such that the interbody potential and
orbital velocities still satisfy $\epsilon \ll 1$ so that a kind of
post-Newtonian approximation for the orbital motion 
might work; however, the strong-field
internal gravity of the bodies could (especially in alternative theories
of gravity) leave imprints on the orbital motion.

\item
The evolution of the system may be affected by the emission of
gravitational radiation.  The 1PN approximation 
does not contain the effects of gravitational radiation back-reaction.
In the expression for the metric given in Box 2,
radiation back-reaction effects do not occur until $O(\epsilon^{7/2})$
in $g_{00}$, $O(\epsilon^{3})$ 
in $g_{0i}$, and $O(\epsilon^{5/2})$ 
in $g_{ij}$.
Consequently, in order to describe such systems, one must carry out a
solution of the equations substantially beyond 1PN order,
sufficient to incorporate the leading
radiation damping terms at 2.5PN order.

\item
The system may be highly relativistic in its orbital motion, so that
$U \sim v^2 \sim 1$ even for the interbody field and orbital velocity.
Systems like this include the late stage of the inspiral of binary
systems of neutron stars or black holes, driven by gravitational
radiation damping, prior to a merger and collapse to a final
stationary state.  Binary inspiral is one of the leading candidate
sources for detection by a world-wide network of laser interferometric
gravitational-wave observatories nearing completion.  A proper
description of such systems requires not only equations for the motion
of the binary carried to extraordinarily high PN orders (at least
3.5PN), but also requires equations for the far-zone
gravitational waveform measured at the detector, that are equally
accurate to high PN orders beyond the leading ``quadrupole''
approximation.

\end{itemize}

Of course, some systems cannot be properly described by any post-Newt\-onian
approximation because their behavior is fundamentally controlled by
strong gravity.  These include the imploding cores of supernovae, the
final merger of two compact objects, the quasinormal-mode vibrations
of neutron stars and black holes, the structure of rapidly rotating
neutron stars, and so on.  Phenomena such as these must be analysed
using different techniques.  Chief among these is the full solution of
Einstein's equations via numerical methods.  This field of ``numerical
relativity'' is a rapidly growing and maturing branch of gravitational
physics, whose description is beyond the scope of this article.
Another is black hole perturbation theory (see \cite{msstt97} for a review).

\subsubsection{Compact bodies and the Strong Equivalence Principle}
\label{compact-SEP}

When dealing with the motion and gravitational-wave generation by
orbiting bodies, one finds a remarkable simplification within general
relativity.  As long as the bodies are sufficiently well-separated
that one can ignore
tidal interactions and other effects that depend upon the finite extent of
the bodies (such as their quadrupole and higher multipole moments), 
then all aspects of their orbital behavior and gravitational wave
generation 
can be characterized by just two parameters:
mass and angular momentum.
Whether their internal structure is highly relativistic, as in black
holes or neutron stars, or non-relativistic as in the Earth and Sun,
only the mass and angular momentum are needed.  Furthermore, both
quantities are measurable in principle by examining the external
gravitational field of the bodies, and make no reference whatsoever to
their interiors.  

Damour
\cite{Damour300} calls this the ``effacement'' of the bodies' internal
structure.
It is a consequence of the Strong Equivalence Principle (SEP), described in
Section \ref{sep}.

General relativity satisfies SEP because it contains one and only one
gravitational field, the spacetime metric $g_{\mu\nu}$.  Consider the
motion of a body in a binary system, whose size is small compared to
the binary separation.  Surround the body by a region that is large
compared to the size of the body, yet small compared to the
separation.  Because of the general covariance of the theory, one can
choose a freely-falling coordinate system which comoves with the body,
whose spacetime metric 
takes the Minkowski form at its outer boundary (ignoring tidal
effects generated by the companion).  
There is thus no evidence of the presence of the companion body,
and the structure of the chosen body can be obtained using the field
equations of GR in this coordinate system.  Far from the chosen body,
the metric is characterized by the mass and angular momentum (assuming
that one ignores quadrupole and higher multipole moments of the body) as
measured far from the body using orbiting test particles and gyroscopes.
These asymptotically measured quantities are oblivious to 
the body's internal structure.  A black hole of mass $m$ and a planet
of mass $m$ would produce identical spacetimes in this outer region.

The geometry of this region surrounding the one body must be matched to
the geometry provided by the companion body.  Einstein's equations
provide consistency conditions for this matching that yield
constraints on the motion of the bodies.  These are the equations of
motion.  As a result the motion of two planets of mass and angular
momentum $m_1$, $m_2$, ${\bf J}_1$ and  ${\bf J}_2$ is 
identical to that of two black holes of the same mass and angular
momentum (again, ignoring tidal effects).

This effacement does not occur in an alternative gravitional theory
like scalar-tensor gravity.  There, in addition to the spacetime
metric, a scalar field $\phi$ is generated by the masses of
the bodies, and controls the local value of the gravitational coupling
constant (\ie $G$ is a function of $\phi$).   Now, in the local frame
surrounding one of the bodies in our binary system, while the metric
can still be made Minkowskian far away, the scalar field will take on
a value $\phi_0$ determined by the companion body.  This can affect
the value of $G$ inside the chosen body, alter its 
internal structure (specifically its gravitational binding energy) 
and hence alter its mass.  
Effectively, each mass becomes several functions $m_A(\phi)$
of
the value of the scalar field at its location, and several distinct masses
come into play, inertial mass, gravitational mass, ``radiation'' mass,
\etc  The precise nature of the
functions will depend on the body, specifically on its gravitational
binding energy, and as a result, the motion and
gravitational radiation may depend on the internal structure of each
body.  For compact bodies such as neutron stars, and black holes these internal
structure effects could be large; for example, the gravitational binding energy
of a neutron star can be 40 percent of its total mass.  
At 1PN order, the leading manifestation of this effect is 
the Nordtvedt effect. 

This is how the study of orbiting systems containing
compact objects provides strong-field tests of general relativity.
Even though the strong-field nature of the bodies is effaced in GR, it
is not in other theories, thus any result in agreement with the
predictions of GR constitutes a
kind of ``null'' test of strong-field gravity.
 
\subsection{Motion and gravitational
radiation in general relativity}
\label{eomgw}

The motion of bodies and the
generation of gravitational radiation are long-standing problems
that date back to the first years following the publication of
GR, when Einstein calculated the gravitational
radiation emitted by a laboratory-scale object using the linearized
version of GR, and de Sitter calculated N-body equations of motion for
bodies in the 1PN approximation to GR.
It has at times been controversial, with disputes over such issues as
whether Einstein's equations alone imply equations of motion for
bodies (Einstein, Infeld and Hoffman demonstrated explicitly that they
do, using a matching procedure similar to the one described 
above), whether gravitational waves are real or are artifacts of general
covariance (Einstein waffled; Bondi and colleagues proved their
reality rigorously in the 1950s), and even over algebraic errors
(Einstein erred by a factor of 2 in his first radiation calculation;
Eddington found the mistake).  
Shortly after the discovery of the binary pulsar PSR
1913+16 in 1974, questions were raised about the foundations of the
``quadrupole formula'' for gravitational radiation damping
(and in
some quarters, even about its quantitative validity).
These questions were answered in part by theoretical
work designed to shore up the
foundations of the quadrupole approximation, and in
part
(perhaps mostly) by the
agreement between the predictions of the
quadrupole formula and the {\it observed}
rate of damping of the pulsar's orbit (see Section \ref{binarypulsars}).
Damour \cite{Damour300} gives a
thorough review of this subject. 

The problem of motion and radiation has received renewed interest
since 1990, with proposals for construction of 
large-scale laser interferometric
grav\-it\-at\-ional-wave observatories, such as the LIGO project in the US,
VIRGO and GEO600 in Europe, and TAMA300 in Japan,
and the realization that a leading
candidate source of detectable waves would be the inspiral, driven 
by grav\-it\-at\-ional radiation damping,
of a binary system of compact objects (neutron stars
or black holes) \cite{LIGO,snowmass}.  The
analysis of signals from such systems 
will require theoretical predictions from GR that are
extremely accurate, well beyond the leading-order prediction of
Newtonian or even post-Newtonian gravity for the orbits, and well beyond
the leading-order formulae for gravitational waves.

This presented a major theoretical challenge: to calculate the motion
and radiation of systems of compact objects
to very high PN order, a formidable algebraic task,
while addressing a number of issues of principle that have
historically plagued this subject, sufficiently well
to ensure that the results were physically meaningful.  This
challenge is in the process of being met, so that we
may soon see a remarkable convergence between observational
data and accurate predictions of gravitational theory that could
provide new, strong-field tests of GR. 

Here we give a brief overview of the problem of motion
and gravitational radiation.  

\subsection{Einstein's equations in ``relaxed'' form}
\label{EErelaxed}

The Einstein equations $G_{\mu\nu} = 8\pi T_{\mu\nu}$ are elegant and
deceptively simple, showing geometry (in the form of the Einstein
tensor $G_{\mu\nu}$, which is a function of spacetime curvature)
being generated by matter (in the form of the material stress-energy tensor
$T_{\mu\nu}$.  However, this is not the most useful form for actual
calculations.  For post-Newtonian calculations, a far more useful form
is the so-called ``relaxed'' Einstein equations:
\begin{eqnarray}
\Box h^{ \alpha \beta } = -16 \pi {\tau}^{ \alpha \beta } \; ,
\label{relaxed}
\end{eqnarray}
where $\Box \equiv  -{\partial}^2 / \partial t^2 + {\nabla}^2 $
is the flat-spacetime wave operator,
$h^{ \alpha \beta }$ is a ``gravitational tensor potential''
related to the deviation of the spacetime metric from its Minkowski
form by the formula 
$h^{\alpha \beta} \equiv \eta^{\alpha \beta} - (-g)^{1/2} g^{\alpha
\beta}$, $g$ is the determinant of $g_{\alpha
\beta}$, and a particular coordinate system has been specified 
by the deDonder
or harmonic gauge condition
$\partial h^{\alpha \beta} /\partial x^\beta =0$ (summation on
repeated indices is assumed). 
This form of Einstein's equations bears a striking similarity to Maxwell's
equations for the vector potential $A^\alpha$ in Lorentz gauge: $\Box
A^\alpha = -4\pi J^\alpha$, $\partial A^\alpha /\partial
x^\alpha =0$.   There is a key difference, however: the source on the right
hand side of Eq. (\ref{relaxed}) is given by the ``effective''
stress-energy pseudotensor
\begin{eqnarray}
\tau^{\alpha\beta} = (-g)T^{\alpha\beta} + (16\pi)^{-1}
\Lambda^{\alpha\beta} \;,
\label{effective}
\end{eqnarray}
where $\Lambda^{\alpha\beta}$ is the non-linear ``field'' contribution
given by terms quadratic (and higher) in $h^{\alpha \beta}$ and its 
derivatives (see \cite{MTW}, Eqs. (20.20) - (20.21) for formulae).
In general relativity, the gravitational field itself generates
gravity, a reflection of the nonlinearity of Einstein's equations, and
in
contrast to the linearity of Maxwell's equations.  

Equation (\ref{relaxed}) is exact, and depends only on the assumption
that spacetime can be covered by harmonic coordinates.  It is called
``relaxed'' because it
can be solved formally as a functional of source variables without
specifying the motion of the source, in the form
\begin{eqnarray}
h^{\alpha \beta} (t,{\bf x})  &=& 4 \int_{\cal C}
{ \tau^{\alpha \beta} (t -| {\bf x} - {\bf x^\prime} |, {\bf x^\prime}
)
\over | {\bf x} - {\bf x^\prime} | } d^3x^\prime \;,
\label{nearintegral}
\end{eqnarray}
where the integration is over the past flat-spacetime null cone $\cal
C$ of the field point $(t,{\bf x})$.  
The motion of the source is then determined either by the equation 
$\partial {\tau}^{\alpha \beta} /\partial x^\beta =0$ (which follows
from the harmonic gauge condition), or from the usual covariant
equation of motion ${T^{\alpha\beta}}_{;\beta}=0$, where the subscript
$;\beta$ denotes a covariant divergence.
This formal solution can then be iterated in a slow motion ($v<1$)
weak-field ($||h^{\alpha \beta}||<1$) approximation.  One begins by
substituting
$h_0^{\alpha \beta} =0$ into the source $\tau^{\alpha \beta}$ in Eq.
(\ref{nearintegral}), and
solving for the first iterate $h_1^{\alpha \beta}$, and then repeating the
procedure sufficiently many times to achieve a solution of the desired
accuracy.  For example, to obtain the 1PN equations of motion, {\it
two} iterations are needed (\ie $h_2^{\alpha \beta}$ must be
calculated); likewise, to obtain the leading gravitational waveform
for a binary system, two iterations are needed. 

At the same time, just as in electromagnetism, the formal integral
(\ref{nearintegral}) must be handled differently, depending on whether
the field point is in the far zone or the near zone.
For 
field points in the far zone or radiation zone, $|{\bf x}| >
{\lambda\!\!\!{\scriptscriptstyle{{}^{-}}}} > |{\bf x}^\prime|$ 
(${\lambda\!\!\!{\scriptscriptstyle{{}^{-}}}}$ is the gravitational
wavelength$/2\pi$), the field can be expanded in inverse powers of
$R=|{\bf x}|$ in a multipole expansion, evaluated at the ``retarded
time'' $t-R$.  The leading term in $1/R$ is
the gravitational waveform.  For field points in the near zone or
induction zone, $|{\bf x}| \sim |{\bf x}^\prime| <
{\lambda\!\!\!{\scriptscriptstyle{{}^{-}}}}$, the field is expanded in
powers of $|{\bf x}-{\bf x}^\prime|$ about the local time $t$,
yielding instantaneous potentials that go into the equations of
motion.  

However, because the source ${\tau}^{\alpha \beta}$ contains
$h^{\alpha \beta}$ itself, it is not confined to a compact region, but
extends over all spacetime.  As a result, there is a danger that the
integrals involved in the various expansions will diverge or be
ill-defined.  This consequence of the non-linearity of Einstein's
equations has bedeviled the subject of gravitational radiation for
decades.  Numerous approaches have been developed to try to
handle this difficulty.  The ``post-Minkowskian'' method of Blanchet,
Damour and Iyer \cite{bd86,bd88,bd89,di91,bd92,blanchet95} 
solves Einstein's equations by two
different techniques, one in the near zone and one in the far zone,
and uses the method of singular asymptotic matching to join the
solutions in an overlap region.  The method provides a natural
``regularization'' technique to control potentially divergent
integrals.  The ``Direct Integration of the Relaxed Einstein
Equations'' (DIRE) approach of Will, Wiseman and Pati
\cite{opus,DIRE}, retains
Eq. (\ref{nearintegral}) as the global solution, but splits the
integration into one over the near zone and another over the far zone,
and uses different
integration variables to carry out the explicit integrals over the two
zones.  In the DIRE method, all integrals are finite and convergent.

These methods assume from the outset that gravity is
sufficiently weak that $||h^{\alpha\beta}||<1$ and harmonic
coordinates exists everywhere, including inside the bodies.  Thus, in
order to apply the results to cases where the bodies may be neutron
stars or black holes, one relies upon the Strong Equivalence Principle
to argue that, if tidal forces are ignored, and equations are
expressed in terms of masses and spins,  one can simply 
extrapolate the results unchanged 
to the situation where the bodies are ultrarelativistic.
While no general proof of this exists, it has been shown to be valid
in specific circumstances, such as at 2PN order in the equations of
motion, and for black holes moving in a Newtonian background field
\cite{Damour300}.

Methods such as these 
have resolved most of the issues that led to criticism of the
foundations of gravitational radiation theory during the 1970s. 

\subsection{Equations of motion and gravitational waveform}
\label{eomwaveform}

Among the results of these approaches are formulae for the equations of
motion and gravitational waveform of binary systems of compact
objects, carried out to high orders in an PN expansion.  Here we shall
only state the key formulae that will be needed for this article.
For example,
the relative two-body equation of motion has the form
\begin{equation}
{\bf a} = {{d{\bf v}} \over dt} = {m \over r^2} \left \{- {\bf {\hat n}} + {\bf A}_{1PN} + {\bf
A}_{2PN} + {\bf A}_{2.5PN} + {\bf A}_{3PN} + {\bf A}_{3.5PN} + \dots
\right \} \,,
\label{EOM}
\end{equation}
where $m=m_1+m_2$ is the total mass, $r= |{\bf x}_1 -{\bf x}_2|$,
${\bf v}={\bf v}_1-{\bf v}_2$, and
${\bf {\hat n}} = ({\bf x}_1 -{\bf x}_2)/r$.
The notation ${\bf A}_{nPN}$ indicates that the term is
$O(\epsilon^n)$ relative to the Newtonian term $-{\bf {\hat n}}$.
Explicit formulae for terms through various orders have been
calculated by
various authors: non-radiative terms through 2PN order
\cite{DD81,Damour82,GK86,Damour300,bfp98},
radiation reaction terms at 2.5PN and
3.5PN order \cite{iyerwill,iyerwill2,blanchet97},
and non-radiative 3PN terms
\cite{jaraschaefer98,jaranowski,damjaraschaefer,blanchetfaye1,blanchetfaye2}.
Here we quote only the first PN corrections and the
leading radiation-reaction terms at 2.5PN order:
\begin{eqnarray}
 {\bf A}_{1PN} &=& \left \{ (4+2\eta){m \over r} - (1+3\eta)v^2 +
{3 \over 2} \eta {\dot r}^2 \right \}{\bf {\hat n}} + (4-2\eta) \dot r
{\bf v} \,, \label{APN} \\
{\bf A}_{2.5PN} &=& -{8 \over 15} \eta {m \over r} \left \{ \left ( 9v^2 +17{m
\over r} \right ) \dot r {\bf {\hat n}} - \left ( 3v^2 +9{m \over r}
\right ) {\bf v} \right \} \,, \label{A2.5PN}
\end{eqnarray}
where $\eta = m_1m_2/(m_1+m_2)^2$.
These terms are sufficient to analyse the orbit and evolution of the
binary pulsar (Sec. \ref{binarypulsars}).  For example, the 1PN terms are
responsible for
the periastron advance of an eccentric orbit, given by $\dot \omega =
6\pi f_b m/a(1-e^2)$, where $a$ and $e$ are the semi-major axis and
eccentricity, respectively, of the orbit, and $f_b$ is the orbital
frequency, given to the needed order by Kepler's third law 
$2 \pi f_b = (m/a^3)^{1/2}$.

Another product is a formula for the gravitational field 
far from the system, written schematically in the form
\begin{equation}
h^{ij} = {2m \over R} \left \{ Q^{ij} + Q_{0.5PN}^{ij} +
Q_{1PN}^{ij} + Q_{1.5PN}^{ij} + Q_{2PN}^{ij} + Q_{2.5PN}^{ij} +
\dots \right \} \,, 
\label{waveform}
\end{equation}
where $R$ is the distance from the source, and the variables
are to be evaluated at retarded time $t-R$.  The leading term
is the so-called quadrupole formula 
\begin{equation}
h^{ij}(t,{\bf x}) =  {2 \over R}{\ddot I}^{ij}(t-R) \,,
\label{waveformquad}
\end{equation}
where $I^{ij}$ is the quadrupole moment of the source, and overdots
denote time derivatives.  For a binary system this leads to
\begin{equation}
Q^{ij} = 2\eta (v^iv^j - m{\hat n}^i{\hat n}^j/r) \,.
\label{Qij} 
\end{equation}
For binary systems, explicit 
formulae for all the terms through 2.5PN order have been
derived by various authors
\cite{wagwill,magnum,poisson93,bdi2pn,bdiww,opus,blanchet96,blanchet98}.  
Given the gravitational waveform, one can
compute the rate at which energy is carried off by the radiation
(schematically $\int \dot h \dot h d\Omega$,
the gravitational analog of the Poynting
flux).
The lowest-order quadrupole formula leads to the
gravitational-wave energy flux
\begin{equation}
\dot E = {8 \over 15} \eta^2 {m^4 \over r^4} (12v^2-11 {\dot r}^2)\,.
\label{EdotGR}
\end{equation}
Formulae for fluxes of angular and linear momentum can also be
derived.
The 2.5PN radiation-reaction terms in the equation of motion (\ref{EOM})
result in
a damping of the orbital energy that precisely balances the energy
flux (\ref{EdotGR}) 
determined from the waveform.  Averaged over one orbit, this
results in a rate of increase of the binary's orbital frequency  given by 
\begin{equation}
\dot f_b = {192\pi \over 5} f_b^2 (2\pi{\cal M}f_b)^{5/3} F(e)
\,,
\label{fdotGR}
\end{equation} 
where ${\cal M}$ is the so-called ``chirp'' mass, given by ${\cal
M}=\eta^{3/5} m$, and $F(e)=(1+ 73e^2/24+37e^4/96)/(1-e^2)^{7/2}$.
Notice that by making precise measurements of the phase $\Phi (t) = 2\pi
\int^t
f(t^\prime) dt^\prime$ of either the orbit or the gravitational waves
(for which $f =2f_b$ for the dominant component) as a function of
the frequency, one in effect measures the ``chirp'' mass of the
system.  

These formalisms have also been generalized to include the leading effects of
spin-orbit and spin-spin coupling between the bodies \cite{kww93,kidder95}.

Another approach to gravitational radiation is applicable to the special
limit in which one mass is much smaller than the other.
This is the method of black-hole perturbation theory.  One begins with an
exact background spacetime of a black hole, either the non-rotating
Schwarzschild or the rotating Kerr solution, and perturbs it according to 
$g_{\mu\nu}=g^{(0)}_{\mu\nu} + h_{\mu\nu}$.  The particle moves on a
geodesic of the background spacetime, and a suitably defined source
stress-energy tensor for the particle acts as a source for the gravitational
perturbation and wave field $h_{\mu\nu}$.  This method provides 
numerical results that are exact in $v$, as well as analytical results 
expressed as series in powers of $v$, both for
non-rotating and for rotating black holes.  For
non-rotating holes, the analytical expansions have been carried to
{\it 5.5} PN order, or $\epsilon^{5.5}$ beyond the 
quadrupole approximation.  All results of black hole
perturbation agree precisely with the $m_1 \to 0$ limit of the PN results,
up to the highest PN order where they can be compared (for a detailed review
see \cite{msstt97}).  

\subsection{Gravitational-wave detection}
\label{gwdetection}

A gravitational-wave detector can be modelled as a body of mass $M$ at
a
distance $L$ from a fiducial laboratory point, connected to the point
by a spring of resonant frequency $\omega_0$ and quality factor $Q$.
From the equation of geodesic deviation, the infinitesimal
displacement $\xi$ of the mass along the line of separation from its
equilibrium position satisfies the equation of motion
\begin{equation}
\ddot \xi + 2{\omega_0 \over Q} \dot \xi + \omega_0^2 \xi
= {L \over 2} \left ( F_+ (\theta,\phi,\psi) {\ddot h}_+ (t) + F_\times
(\theta,\phi,\psi) {\ddot h}_\times (t) \right ) \,,
\label{detector}
\end{equation}
where $F_+ (\theta,\phi,\psi)$ and $F_\times
(\theta,\phi,\psi)$ are ``beam-pattern'' factors, that depend on the
direction of the source $(\theta,\phi)$, and on a polarization  angle
$\psi$, and $h_+(t)$ and $h_\times (t)$ are gravitational waveforms
corresponding to the two polarizations of
the gravitational wave (for a review, see \cite{Thorne300}).  In 
a source coordinate system in which the $x-y$
plane is the plane of the sky and the $z$-direction points toward the
detector, these two modes are given by 
\begin{equation}
h_+ (t) = {1 \over 2} (h^{xx} (t) - h^{yy} (t) ) \,, \quad
h_\times  (t) = h^{xy} (t) \,,
\label{modes}
\end{equation}
where $h^{ij}$ represent transverse-traceless (TT) projections of the
calculated waveform of Eq. (\ref{waveform}), given by 
\begin{equation}
h^{ij}_{TT} = h^{kl} [(\delta{ik}-{\hat N}^i{\hat N}^k)
(\delta{jl}-{\hat N}^j{\hat N}^l ) - {1 \over 2} (\delta{ij}-{\hat
N}^i{\hat N}^j)
(\delta{kl}-{\hat N}^k{\hat N}^l ) ] \,,
\label{TTprojection}
\end{equation}
where ${\hat N}^j$ is a unit vector pointing toward the detector.
The beam pattern factors depend on the orientation and nature of the
detector.  For a wave approaching along the laboratory
$z$-direction, and for a mass whose location on the $x-y$ plane
makes an angle $\phi$
with the $x$ axis, the beam pattern factors are given by $F_+ =
\cos 2 \phi$ and $F_\times = \sin 2 \phi$.
For a resonant cylinder oriented along the laboratory $z$
axis, and for source direction $(\theta,\phi)$, 
they are given by $F_+ = \sin^2 \theta \cos 2 \psi $, $F_\times
= \sin^2 \theta \sin 2 \psi $ (the angle $\psi$ measures the relative
orientation of the laboratory and source $x$-axes).  For a laser interferometer with one
arm along the laboratory $x$-axis, the other along the $y$-axis, and
with $\xi$ defined as the {\it differential} displacement 
along the two arms, the beam pattern functions are
$F_+ = {1 \over 2} (1+\cos^2 \theta )\cos 2 \phi \cos 2 \psi - \cos
\theta \sin 2 \phi \sin 2 \psi $ and
$F_\times = {1 \over 2} (1+\cos^2 \theta )\cos 2 \phi \sin 2 \psi + \cos
\theta \sin 2 \phi \cos 2 \psi $.  

The waveforms $h_+ (t)$ and $h_\times (t)$ depend on the nature and
evolution of the source.  For example, for a binary system in a
circular orbit, with an inclination $i$ relative to the plane of the
sky, and the $x$-axis oriented along the major axis of the projected
orbit, the quadrupole approximation of Eq. (\ref{Qij}) gives 
\begin{eqnarray}
h_+ (t) &=& - {{2{\cal M}} \over R} (2\pi {\cal M} f_b)^{2/3} (1+ \cos^2 i) \cos
2 \Phi_b (t) \,, \nonumber \\
h_\times (t) &=& - {{2{\cal M}} \over R} (2\pi {\cal M} f_b)^{2/3} 2 \cos i
\cos 2 \Phi_b (t) \,,
\end{eqnarray}
where $\Phi_b (t) = 2\pi \int^t f_b (t^\prime) dt^\prime$ is the orbital
phase.

%=======================================================
% Section 5
%=======================================================

\section{Stellar system tests of gravitational theory}\label{stellar}
%------------------------------------------------
% subsection 5.1
%------------------------------------------------

\subsection{The binary pulsar and general relativity}\label{binarypulsars}

The 1974 discovery of the binary pulsar PSR 1913+16 by Joseph Taylor
and Russell Hulse during a routine search for new pulsars
provided the first possibility of probing new aspects of gravitational 
theory:  the effects of strong relativistic internal gravitational fields 
on orbital dynamics, and the effects of gravitational radiation reaction.
For reviews of the discovery and current status, see the published
Nobel Prize lectures by Hulse and Taylor \cite{Hulse,Taylor94}.
For a thorough review of pulsars, including binary and millisecond
pulsars, see \cite{lorimer}.

The system consists of a pulsar of nominal period 59 ms in a close binary 
orbit with an as yet unseen companion.  The orbital period is about
7.75 hours, and the eccentricity is 0.617.  From detailed analyses of the 
arrival times of pulses (which amounts to an integrated version of the 
Doppler-shift methods used in spectroscopic binary systems), extremely 
accurate orbital and physical parameters for the system have been obtained 
(Table \ref{bpdata}).  Because the orbit is so close 
($ \approx 1 R_\odot$) 
and because there is no evidence of an eclipse of the pulsar signal or of 
mass transfer from the companion, it is generally believed that the companion 
is compact:  evolutionary arguments suggest that it is most likely 
a dead pulsar.  Thus the orbital motion is 
very clean, free from tidal or other 
complicating effects.  Furthermore, the data acquisition is ``clean'' in 
the sense that by exploiting the intrinsic stability of the pulsar
clock combined with the ability to maintain and transfer atomic time
accurately using such devices as the Global Positioning System,
the observers can keep track of the pulsar phase with 
an accuracy of $15 \mu$s, despite extended gaps between 
observing sessions (including a several-year gap during the middle 1990s
upgrade of
the Arecibo radio telecsope).  The pulsar has shown no evidence of ``glitches'' 
in its pulse period.

\begin{table}
\begin{center}
\begin{tabular}{l l l}
\hline
\hline
\noalign{\smallskip}
&Symbol&\\
Parameter&(units)&Value$^1$ \\
\hline
\hline
\noalign{\smallskip}
\multispan 3 {(i)   {\bf ``Physical'' Parameters} \hfil}\\
\quad Right Ascension&$\alpha$&$19^h 15^m 28.^s 00018(15)$ \\
\quad Declination&$\delta$&$16 ^\circ 06 ^\prime 27. ^{\prime \prime}
4043(3)$
\\
\quad  Pulsar Period&$P_p$ (ms)&$59.029997929613(7)$ \\
\quad  Derivative of Period&$\dot P_p$&$8.62713(8) \times 10^{-18}$
\\
\noalign{\smallskip}
\multispan 3 {(ii)  {\bf ``Keplerian'' Parameters} \hfil}\\
\quad  Projected semimajor axis&$a_p \sin i$ (s)&$2.3417592(19)$ \\
\quad  Eccentricity&$e$&$0.6171308(4)$ \\
\quad  Orbital Period&$P_b$ (day)&$0.322997462736(7)$ \\
\quad  Longitude of periastron&$\omega_0$ ($^\circ$)&$226.57528(6)$
\\
\quad Julian date of periastron&$T_0$
(MJD)&$46443.99588319(3)$ \\
\noalign{\smallskip}
\multispan 3 {(iii) {\bf ``Post-Keplerian'' Parameters} \hfil} \\
\quad  Mean rate of periastron advance&$\langle \dot\omega \rangle\,
( ^\circ \, {\rm
yr}^{-1} )$&$4.226621(11)$ \\
\quad  Redshift/time dilation&$\gamma^\prime$
(ms)&$4.295(2)$ \\
\quad  Orbital period derivative&$\dot P_b \, (10^{-12} )$&$-
2.422(6)$ \\
\hline
\hline
\noalign{\smallskip}
\multispan 3 {$^1$Numbers in parentheses denote errors in last
digit. \hfil }\\
\multispan 3 {\quad  Data from http://puppsr8.princeton.edu/psrcat.html\hfill
}\\
\end{tabular}
\caption{Parameters of the Binary Pulsar PSR 1913+16}
\label{bpdata}
\end{center}
\end{table}

Three factors make this system an arena where relativistic celestial
mechanics must be used: the relatively large size of relativistic
effects
[$ v_{\rm orbit} \approx (m/r)^{1/2} \approx 10^{-3}$], a factor of 10
larger than the corresponding values for solar-system orbits;
the short orbital period, allowing secular effects to
build up rapidly; and the cleanliness of the system, allowing
accurate determinations of small effects.  Because the orbital
separation is large compared to the neutron stars' compact size, tidal
effects can be ignored.  Just as Newtonian gravity
is used as a tool for measuring astrophysical parameters of ordinary binary
systems, so GR is used as a tool for measuring
astrophysical parameters in the binary pulsar.

The observational parameters that are obtained from a least-squares
solution of the arrival-time data fall into three groups:  (i)\ non-orbital 
parameters, such as the pulsar period and its rate of change (defined
at a given epoch), and the 
position of the pulsar on the sky; (ii)\ five ``Keplerian'' parameters, 
most closely related to those appropriate for standard Newtonian binary
systems, 
such as the eccentricity $e$ and the orbital period $P_b$;
and (iii)\ five ``post-Keplerian'' parameters.   The five post-Keplerian
parameters are $\langle \dot \omega \rangle$, the average rate of periastron advance; 
$\gamma^\prime$, the amplitude of delays in arrival of pulses caused by the
varying effects of the gravitational redshift and time dilation as the
pulsar moves in its elliptical orbit at varying distances from the
companion and with varying speeds;
$\dot P_b$, the rate of change of orbital period, caused
predominantly by gravitational radiation damping; and  $r$ and 
$s = \sin i$, respectively the ``range'' and ``shape'' of the Shapiro
time delay of the pulsar signal as it propagates through the curved
spacetime region near the companion, where $i$ is the angle of 
inclination of the orbit relative 
to the plane of the sky.

In GR, these post-Keplerian parameters can be related 
to the masses of the two bodies and to measured Keplerian parameters 
by the equations (TEGP 12.1, 14.6(a))
%\begin{mathletters}
\begin{eqnarray}
        \langle \dot \omega \rangle 
&=& 6\pi f_b (2\pi m f_b )^{2/3} (1-e^2 )^{-1}   \,, \nonumber \\
          \gamma^\prime 
 &=&e (2 \pi f_b )^{-1} (2\pi m f_b)^{2/3}
          (m_2 /m) (1+m_2 /m )   \,, \nonumber \\
          \dot P_b 
 &=&-(192 \pi /5)(2 \pi {\cal M}f_b )^{5/3} F(e)\,,  \nonumber \\
          s &=& \sin i  \,,  \nonumber \\
          r &=& m_2  \,, 
\label{pkparameters}
\end{eqnarray}
%\end{mathletters}
where 
$m_1$ and $m_2$ denote the pulsar and companion masses, respectively.  
The formula for $\langle \dot \omega \rangle$ ignores
possible non-relativistic contributions to the periastron shift, 
such as tidally or rotationally induced effects caused by the companion 
(for discussion of these effects, see TEGP 12.1(c)).   The formula 
for $\dot P_b$ includes only quadrupole
gravitational radiation; 
it ignores other sources of energy loss, such as tidal
dissipation (TEGP 12.1(f)).  Notice that, by virtue of Kepler's third
law, $(2\pi f_b)^2 = m/a^3$,  $(2\pi m f_b)^{2/3} \sim m/a \sim
\epsilon$, thus the first two post-Keplerian parameters can be seen
as $O(\epsilon)$, or 1PN corrections to the underlying variable, while the
third is an $O(\epsilon^{5/2})$, or 2.5PN correction.
The current observed values for the Keplerian 
and post-Keplerian parameters are shown in Table \ref{bpdata}.
The parameters $r$ and $s$ are not separately
measurable with interesting accuracy for PSR 1913+16 because the
orbit's $47 ^\circ$ inclination does not lead to a substantial Shapiro
delay.  

Because $f_b$ and $e$ are separately measured parameters, the
measurement of the three post-Keplerian parameters provide three
constraints on the two unknown masses.  The periastron shift measures
the total mass of the system, $\dot P_b$ measures the chirp mass, and
$\gamma^\prime$ measures a complicated function of the masses.
GR passes the test if it provides a consistent solution to these
constraints, within the measurement errors.
  
\begin{figure}
\begin{center}
\leavevmode
\psfig{figure=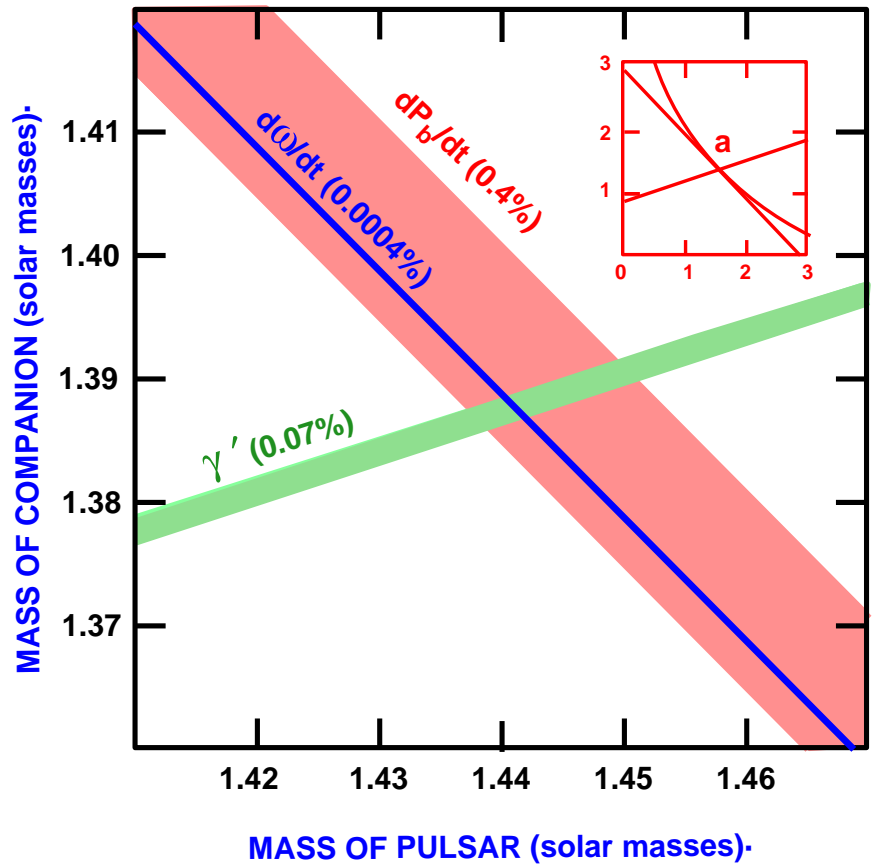,height=4.5in}
\end{center}
\caption{Constraints on masses of pulsar and companion from data on
PSR 1913+16, assuming GR to be valid.  Width of each
strip in the plane reflects observational accuracy, shown as a
percentage.  Inset shows the three constraints on the full mass plane;
intersection region (a) has been magnified 400 times for the full
figure.
} 
\label{bpfigure1}
\end{figure}

From the intersection of the  $\langle \dot \omega \rangle$ 
and $\gamma^\prime $ constraints we obtain the values
$m_1 = 1.4411 \pm 0.0007 M_\odot$ and 
$m_2 = 1.3873 \pm 0.0007 M_\odot$.  The third of Eqs. \ref{pkparameters} 
then predicts the value 
$\dot P_b = -2.40243 \pm 0.00005 \times 10^{-12}$.
In order to compare the predicted
value for $\dot P_b$ with the observed value of Table \ref{bpdata}, it 
is necessary to
take into account the small effect of a relative acceleration between the
binary pulsar system and the solar system caused by the differential
rotation of the galaxy.  This effect was previously considered unimportant 
when $\dot P_b$ was known only to 10 percent accuracy.  Damour and 
Taylor \cite{DamourTaylor92} 
carried out a careful estimate of this effect using data 
on the location and proper motion of the pulsar, combined with the best
information available on galactic rotation, and found
\begin{equation} 
 \dot P_b^{ \rm GAL} \simeq -(1.7 \pm 0.5) \times 10^{-14}  \,.
\label{Pdotgal}
\end{equation}
Subtracting this from the observed $\dot P_b$ (Table \ref{bpdata})
gives the residual
\begin{equation}
\dot P_b^{\rm CORR} = -(2.408 \pm 0.010[{\rm OBS}] \pm 0.005[{\rm GAL}])
\times 10^{-12} \,, 
\label{Pdotcorr}
\end{equation}
which agrees with the prediction, within the errors.  In other words,
\begin{equation}
     {{\dot P_b^{\rm GR}} \over {\dot P_b^{\rm CORR}}} 
 = 1.0023 \pm 0.0041[{\rm OBS}] \pm 0.0021[{\rm GAL}] \,.
\label{Pdotcompare}
\end{equation}
The consistency among the measurements is displayed in Figure \ref{bpfigure1},
in which the regions allowed by the three most precise constraints
have a single common overlap.  

\begin{figure}
\begin{center}
\leavevmode
\psfig{figure=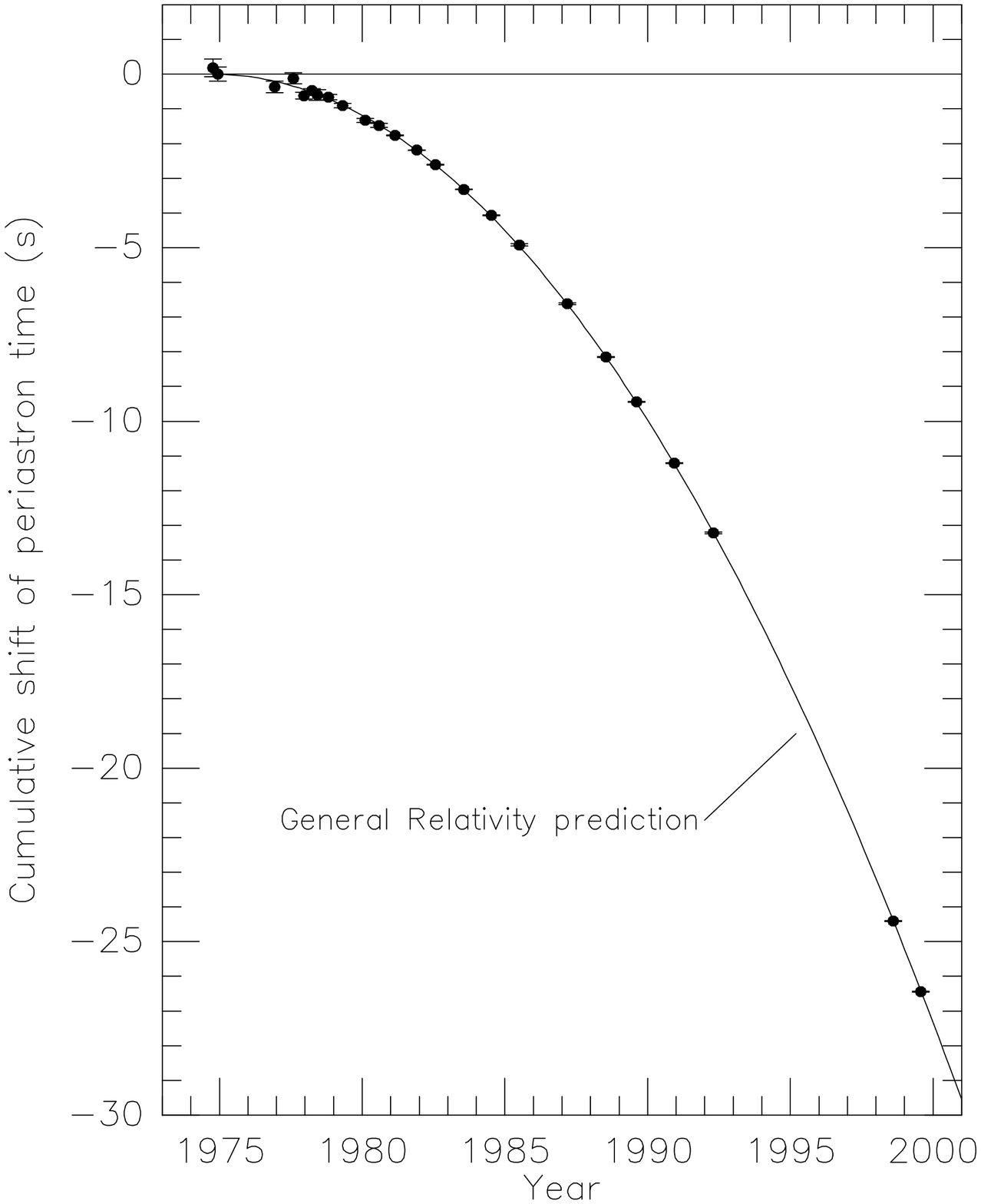,angle=0,height=4.5in}
\end{center}
\caption{Plot of cumulative shift of the periastron time from 1975 -- 2000.
Points are data, curve is the GR prediction.  Gap during the middle
1990s was caused by closure of Arecibo for upgrading. 
[J. H. Taylor and J. M.  Weisberg, 2000, private communication].
}
\label{bpfigure2}
\end{figure}

A third way to display the agreement with general relativity is by
comparing the observed phase of the orbit with a theoretical template
phase as a function of time.  If $f_b$ varies slowly in time, then to
first order in a Taylor expansion, the orbital phase is given by
$\Phi_b (t) = 2\pi f_{b0} t + \pi {\dot f}_{b0} t^2$.  The time of
periastron passage $t_P$ is given by $\Phi (t_P)=2\pi N$, where $N$ is
an integer, and consequently, the periastron time will not grow
linearly with $N$.  Thus the cumulative difference between periastron
time $t_P$ and $N/f_{b0}$, the quantities actually measured in
practice,  should vary according to 
$t_P - N/f_{b0} = -{\dot f}_{b0} N^2/2 f_{b0}^3 \approx - ({\dot
f}_{b0}/2f_{b0}) t^2$.  Figure \ref{bpfigure2} shows the results: the
dots are the data points, while the curve is the predicted difference
using the measured masses and the quadrupole formula for ${\dot
f}_{b0}$ \cite{weisberg}.   

The consistency among the constraints 
provides a test of the assumption that the two bodies 
behave as ``point'' masses, without complicated tidal effects, obeying 
the general relativistic equations of motion including
gravitational radiation.  It is also a test of strong gravity,
in that the highly relativistic internal structure of the
neutron stars does not influence their orbital motion, as predicted by
the Strong Equivalence Principle of GR.

Recent observations \cite{kramer,weisberg2} indicate
variations in the pulse profile, which suggests that the pulsar is
undergoing precession as it moves through the curved spacetime
generated by its companion, an effect known as geodetic precession. 
The amount is consistent with GR, assuming that the pulsar's
spin is suitably misaligned with the orbital angular momentum.  
Unfortunately, the evidence suggests that the pulsar beam may precess
out of our line of sight by 2020. 

%-----------------------------------------
% subsection 4.2
%-----------------------------------------

\subsection{A population of binary pulsars?}\label{population}

Since 1990, several new massive binary pulsars similar to PSR 1913+16 were
discovered, leading to the possibility of new or improved tests of
GR.

\begin{table}
\begin{center}
\begin{tabular}{l l l l l}
\hline
\hline
\noalign{\smallskip}
Parameter&B1534+12&B2127+11C&B1855+09&B0655+64 \\
\hline
\hline
\noalign{\smallskip}
\multispan 4 {\bf (i)  ``Keplerian'' Parameters \hfil}\\
\quad  $a_p \sin i$ (s) &3.729464(3)&2.520(3)&9.2307802(4)&4.125612(5)\\
\quad  $e$&0.2736775(5)&0.68141(2)&0.00002168(5)&0.0000075(11)\\
\quad  $P_b$(day)&0.42073729933(3)&0.335282052(6)&12.3271711905(6)&1.028669703(1)\\
\noalign{\smallskip}
\multispan 4 {\bf (ii) ``Post-Keplerian'' Parameters$^1$ \hfil} \\
\quad  $\langle \dot \omega \rangle \, ( ^\circ \, {\rm yr}^{-1} )$
&1.755794(19)&4.457(12)&\\
\quad  $\gamma^\prime$ (ms)&2.071(6)&4.9(1.1)&\\
\quad  ${\dot P}_b \,(10^{-12})$&$-$0.131(9)&&&$<$ 0.5\\
\quad  $r(\mu {\rm s})$&6.3(1.3)&&1.27(10)\\
\quad  $s= \sin i$&0.983(8)&&0.9992(5)\\
\hline
\hline
\noalign{\smallskip}
\multispan 5 {$^1$ From \cite{wolszczan,stairs,stairs2} \hfil }\\
\multispan 5 {\quad and http://puppsr8.princeton.edu/psrcat.html \hfil}\\
\end{tabular}
\caption{Parameters of Other Binary Pulsars}
\label{bpdata2}
\end{center}
\end{table}

{\it PSR 1534+12}.  \quad This is a binary pulsar system in our
galaxy.  Its pulses are significantly stronger and narrower
than those of PSR 1913+16, so timing measurements are more precise,
reaching $3 \mu s$ accuracy.  Its parameters are listed in 
Table \ref{bpdata2} \cite{stairs,stairs2}.  
The orbital plane appears to be almost edge-on relative to the line 
of sight ($i \simeq 80 ^\circ$); as a result the Shapiro
delay is substantial, and separate values of the parameters $r$ and
$s$ have been obtained with interesting accuracy.  Assuming general
relativity, one infers that the two masses are 
$m_1=1.335 \pm 0.002 M_\odot$ and $m_2=1.344 \pm 0.002 M_\odot$.  The rate of orbit decay $\dot P_b$ agrees
with GR to about 15 percent, the precision limited by
the poorly known distance to the pulsar, which introduces a
significant uncertainty into the subtraction of galactic
acceleration.  Independently of $\dot P_b$, measurement of the four
other post-Keplerian parameters gives two tests of strong-field
gravity in the non-radiative regime \cite{TWDW92}.

{\it PSR 2127+11C.}
\quad This system appears to be a clone of the Hulse-Taylor binary 
pulsar, with very similar values for orbital period and eccentricity
(see Table \ref{bpdata2}).  The inferred
total mass of the system is $2.706 \pm 0.011 M_\odot$.
Because the system is in the globular cluster M15 (NGC 7078), 
it suffers Doppler shifts resulting from local accelerations, 
either by the mean cluster gravitational field or by nearby stars, 
that are more difficult to estimate than was the case with the galactic 
system PSR 1913+16.  This may make a separate, precision measurement of the 
relativistic contribution to $\dot P_b$ impossible.

{\it PSR 1855+09}.  \quad This binary pulsar system is not
particularly relativistic, with a long period (12 days) and highly
circular orbit.  However, because we observe the orbit nearly edge on,
the Shapiro delay is large and measurable, as reflected in the
post-Keplerian parameters $r$ and $s$.

{\it PSR 0655+64}. \quad This system consists of a pulsar and a white
dwarf companion in a nearly circular orbit.  Only an upper limit on
$\dot P_b$ has been placed.

%------------------------------------------
% subsection 4.3
%-------------------------------------------

\subsection{Binary pulsars and alternative theories} \label{binarypulsarsalt}

Soon after the discovery of the binary pulsar it was widely hailed as 
a new testing ground for relativistic gravitational effects.  
As we have seen in the case of GR, in most respects, 
the system has lived up to, indeed exceeded, the early expectations.  

In another respect, however, the system has only partially lived up to its
promise, namely as a direct testing ground for alternative theories of
gravity.  The origin of this promise was the discovery
that alternative theories of gravity generically predict the emission 
of dipole gravitational radiation from binary star systems.  
In general relativity, there is no dipole radiation because the
``dipole moment'' (center of mass) of isolated systems is 
uniform in time (conservation
of momentum), and because the ``inertial mass'' that determines the
dipole moment is the same as the mass that generates gravitational
waves (SEP).  In other theories, while the
inertial dipole moment may remain uniform, the ``gravity-wave'' dipole
moment need not, because the mass that generates gravitational waves
depends differently on the internal
gravitational binding energy of each body than does the inertial mass
(violation of SEP).  
Schematically, in a coordinate system in which the center of inertial
mass is at the origin, so that $m_{I,1} {\bf x}_1 + m_{I,2} {\bf x}_2 =0$,
the dipole part of the retarded gravitational field would be given by
\begin{equation}
h \sim {1 \over R} {d \over dt} (m_{GW,1} {\bf x}_1 + m_{GW,2} {\bf x}_2 )
\sim {{\eta m} \over R} {\bf v} \left ( {m_{GW,1} \over m_{I,1}} -
{m_{GW,2} \over m_{I,2}} \right ) \,,
\label{hdipole} 
\end{equation}
where ${\bf v} = {\bf v}_1 -{\bf v}_2$ and $\eta$ and $m$ are defined
using inertial masses.  In theories that violate SEP,
the difference between gravitational-wave mass and inertial mass is a
function of the internal gravitational binding energy of the bodies.
This additional form of gravitational radiation damping could, 
at least in principle, be significantly stronger than the usual quadrupole 
damping, because it depends on fewer powers of the orbital velocity $v$,
and it depends on the gravitational binding energy per unit mass of
the bodies, which, for neutron stars, could be as large as 40 percent
(see TEGP 10 for further details).
As one fulfillment of this promise, Will and Eardley worked out in 
detail the effects of dipole gravitational radiation in the bimetric theory 
of Rosen, and, when the first observation of the decrease of 
the orbital period was announced in 1979, the Rosen theory
suffered a terminal blow.  A wide
class of alternative theories also fail the binary pulsar test because
of dipole gravitational radiation (TEGP 12.3).

On the other hand, the early observations of PSR 1913+16 
already indicated that, in
GR, the masses of the two bodies were nearly equal, so
that, in theories of gravity that are in some sense ``close'' to
GR, dipole gravitational radiation would not be a
strong effect, because of the apparent symmetry of the system.  
The Rosen theory, and others like it, are not ``close'' to general 
relativity, except in their predictions for the weak-field, slow-motion 
regime of the solar system.  When relativistic neutron stars are present, 
theories like these can predict strong effects on the motion of the bodies 
resulting from their internal highly relativistic gravitational structure
(violations of SEP).  As a consequence,
the masses inferred from observations of the periastron shift 
and $\gamma^\prime$ may
be significantly different from those inferred using general
relativity, and may be different from each other, leading to strong
dipole gravitational radiation damping.  By contrast, the Brans-Dicke
theory is ``close'' to GR, roughly
speaking within $1/ \omega_{\rm BD}$ of the predictions of the latter, 
for large values of the coupling constant $\omega_{\rm BD}$  (here we use
the subscript BD to distinguish the coupling constant from the periastron
advance $\dot \omega$).  
Thus, despite the presence of dipole gravitational radiation, 
the binary pulsar provides at present only a weak test of Brans-Dicke
theory, not yet competitive with solar-system tests.  

%------------------------------------------------------
% subsection 4.4
%------------------------------------------------------

\subsection{Binary pulsars and scalar-tensor gravity} 
\label{binarypulsarsscalar}

Making the usual assumption that both members of the system are neutron 
stars, and using the methods summarized in TEGP Chapters 10--12, 
one can obtain 
formulas for the periastron shift, the gravitational redshift/second-order 
Doppler shift parameter, and the rate of change of orbital period, 
analogous to Eqs. (\ref{pkparameters}).  These formulas depend on the
masses of the two neutron stars, on their self-gravitational binding
energy, represented by ``sensitivities'' $s$ and $\kappa^*$ and on 
the Brans-Dicke coupling constant $\omega_{\rm BD}$.  First, there is 
a modification of Kepler's third law, given by
\begin{equation}
    2 \pi f_b = ( {\cal G} m /a^3)^{1/2} \,.
\label{KeplerBD}
\end{equation}
Then, the predictions for $\langle \dot \omega \rangle$, $\gamma^\prime$ 
and $\dot P_b$
are
%\begin{mathletters}
\begin{eqnarray}
       \langle \dot \omega \rangle 
&=& 6 \pi f_b (2\pi m f_b)^{2/3} (1-e^2 )^{-1} {\cal P}{\cal G}^{-4/3}
 \,, \label{periastronBD}\\
          \gamma^\prime 
&=& e (2 \pi f_b )^{-1} (2\pi m f_b)^{2/3}
          (m_2 /m) {\cal G}^{-1/3}
          ( \alpha_2^* + {\cal G} m_2 /m
      + \kappa_1^* \eta_2^* )  \,,\label{gammaBD} \\
          \dot P_b 
&=& -(192 \pi /5)(2 \pi {\cal M} f_b )^{5/3} F^\prime (e) 
     - 4 \pi (2 \pi \mu f_b ) \xi {\cal S}^2 G(e)  \,, \label{PdotBD}
\end{eqnarray}
%\end{mathletters}
where ${\cal M} \equiv \chi^{3/5} {\cal G}^{-4/5} \eta^{3/5} m$, and, 
to first order in $\xi \equiv (2 + \omega_{\rm BD} )^{-1}$, we
have 
%\begin{mathletters}
\begin{eqnarray} 
          F^\prime (e)  
&=& F(e) + {5 \over 144} \xi (\Gamma + 3 \Gamma^\prime )^2
       (  {1 \over 2} e^2 
       +   {1 \over 8} e^4 ) (1-e^2 )^{-7/2}  \,, \nonumber \\
          G(e)  
&=& (1-e^2 )^{-5/2} (1+ {1 \over 2} e^2 )
    \,,\nonumber \\
          {\cal S}  
&=& s_1  -  s_2  \,, \nonumber \\
          {\cal G}  
&=&  1  -  \xi (s_1 + s_2 - 2s_1 s_2 )  \,, \nonumber \\
          {\cal P}  
&=& {\cal G} [1 -  {2 \over 3} \xi
       +  {1 \over 3} \xi  
          (s_1 + s_2 -2s_1 s_2 )]  \,, \nonumber \\
          \alpha_2^*  
&=& 1 -  \xi s_2  \,, \qquad
          \eta_2^*  
= (1-2s_2 ) \xi  \,, \nonumber \\
          \chi
&=& {\cal G}^2 
          [1 -   {1 \over 2} \xi  +  {1 \over 12} \xi \Gamma^2 ]  \,, 
\nonumber \\
          \Gamma  
&=& 1 - 2(m_1 s_2  +  m_2 s_1 )/m  \,, \quad
          \Gamma^\prime  
= 1 - s_1 - s_2 \,. 
\label{BDcoefficients}
\end{eqnarray}
%\end{mathletters}
The quantities $s_a$ and $\kappa_a^*$ are defined by
\begin{equation} 
       s_a = - 
\left ( {{\partial (\ln m_a )} \over {\partial (\ln G)}} \right )_N 
    \,, \qquad
       \kappa_a^*  =  - 
\left ( {{\partial (\ln I_a )} \over {\partial (\ln G)}} \right )_N 
\,,\label{sensitivities}
\end{equation}
and measure the ``sensitivity'' of the mass $m_a$ and moment
of inertia $I_a$ of each body to changes in the scalar field
(reflected in changes in $G$) for a fixed baryon number $N$ (see TEGP
11, 12 and 14.6(c) for further details).  The quantity $s_a$ is
related to the gravitational binding energy.  Notice how the violation of
SEP in Brans-Dicke theory introduces complex structure-dependent
effects in everything from the Newtonian limit (modification of the
effective coupling constant in Kepler's third law) to
gravitational-radiation.  In the limit $\xi \to 0$, we recover GR, and
all structure dependence disappears.
The first term in $\dot P_b$ [Eq. (\ref{PdotBD})]
is the effect of quadrupole and
monopole gravitational radiation, while the second term is the
effect of dipole radiation.  

In order to estimate the sensitivities $s_a$  
and $\kappa^*_a$, one must adopt an equation of state for
the neutron stars.  It is sufficient to restrict
attention to relatively stiff neutron 
star equations of state in order to guarantee neutron stars of sufficient 
mass, approximately $1.4 M_\odot$.  The lower limit 
on $\omega_{\rm BD}$ required to give consistency among the constraints on 
$\langle \dot \omega \rangle$, $\gamma$ and $\dot P_b$ 
as in Figure \ref{bpfigure1} is several hundred \cite{zaglauer}.  
The combination of $\langle \dot \omega \rangle$ and $\gamma$ give a constraint on 
the masses that is relatively weakly dependent on $\xi$, thus the constraint 
on $\xi$ is dominated by $\dot P_b$ and is directly proportional to 
the measurement error in $\dot P_b$; in order to achieve a constraint 
comparable to the solar system value of $3 \times 10^{-4}$, the error in
$\dot P_b^{\rm OBS}$ would have to be reduced by more than a factor of ten.

Alternatively, a binary pulsar system with dissimilar objects, such as
a white dwarf or black hole companion, would provide potentially more
promising tests of dipole radiation.  Unfortunately, none has been
discovered to date; the dissimilar system B0655+64, with a white dwarf
companion is in a highly circular orbit, making measurement of the
periastron shift meaningless, and is not as relativistic as 1913+16.
From the upper limit on $\dot P_b$ (Table \ref{bpdata2}), one can
infer at best the weak bound $\omega_{BD} > 100$

\begin{figure}
\begin{center}
\leavevmode
\psfig{figure=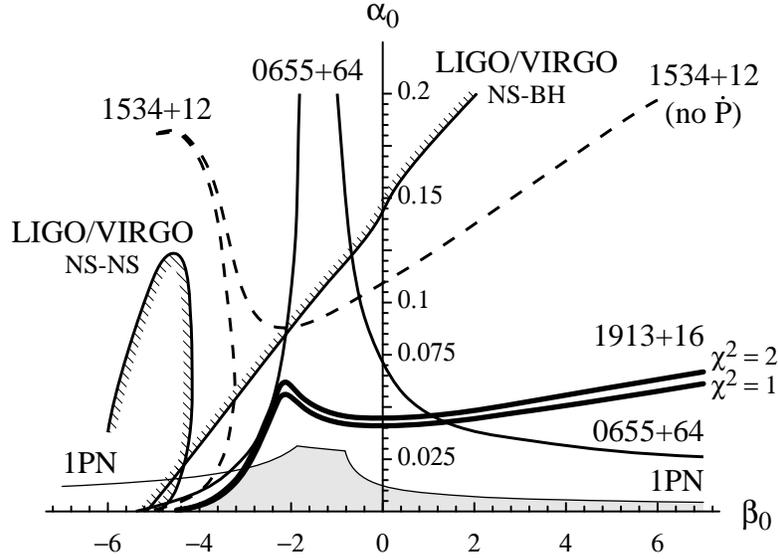,height=3.0in}
\end{center}
\caption{Region of the scalar-tensor theory $\alpha_0 - \beta_0$ plane
allowed by solar-system, binary-pulsar, and future gravitational-wave
observations.  A polytropic equation of state for the neutron stars
was assumed.  The shaded region is that allowed by all tests.  For
positive values of $\beta_0$, solar-system bounds (labelled 1PN) still
are the best. (From [Damour and Esposito-Far\`ese 1998], \copyright
1998 by the American Physical Society, reproduced by permission)}
\label{scalarbounds}
\end{figure}

Damour and Esposito-Far\`ese \cite{DamourEspo92} have generalized
these results to a broad class of scalar-tensor theories.  These
theories are characterized by a single function $\alpha(\varphi)$ of the
scalar field $\varphi$, which mediates the coupling strength of the
scalar field.  For application to the solar system or to binary systems, one
expands this function about a cosmological background field value $\varphi_0$:
\begin{equation}
\alpha(\varphi) = \alpha_0 (\varphi -\varphi_0) + {1 \over 2} \beta_0 
(\varphi -\varphi_0)^2 + \dots \,.
\end{equation}
A purely linear coupling function produces Brans-Dicke theory, with
$\alpha_0^2 = 1/(2 \omega_{BD} +3)$.  The function $\alpha(\varphi)$ acts
as a potential function for the scalar field $\varphi$, and, if $\beta_0
>0$, during cosmological evolution,
the scalar field naturally evolves toward the minimum of the
potential, \ie toward $\alpha_0 \approx 0$, $\omega_{BD} \to \infty$, or toward
a theory close to, though not precisely GR 
\cite{DamourNord93a,DamourNord93b}.  
Bounds on the parameters $\alpha_0$ and $\beta_0$ from solar-system,
binary-pulsar and gravitational-wave observations 
(see Sec. \ref{backreaction})
are shown in Figure \ref{scalarbounds} \cite{DamourEspo98}.  Negative
values of $\beta_0$ correspond to an unstable scalar potential; in
this case, objects such as neutron stars can experience a
``spontaneous scalarization'', whereby the interior values of $\varphi$
can take on values very different from the exterior values, through
non-linear interactions between strong gravity and the scalar field,
dramatically affecting the stars' internal structure and the consequent
violations of SEP.  On the other hand, $\beta_0 <0$ is of little
practical interest, because, with an unstable $\varphi$ potential,
cosmological evolution would presumably drive the system away from the
peak where $\alpha_0 \approx 0$, 
toward parameter values that could easily be excluded
by solar system experiments.  On the $\alpha_0 - \beta_0$ plane shown
in Figure \ref{scalarbounds}, the $\alpha_0$ axis corresponds to pure
Brans-Dicke theory, while the origin corresponds to pure GR.  As
discussed above, solar system bounds (labelled ``1PN'' in 
Figure \ref{scalarbounds})
still beat the binary pulsars.
The bounds labelled ``LIGO-VIRGO'' are discussed in Sec.
\ref{backreaction}.

%==================================================
% Section 6
%==================================================

\section{Gravitational-wave tests of gravitational theory}
\label{gwaves}

\subsection{Gravitational-wave observatories}

Some time in the next decade, a new opportunity for testing
relativistic gravity will be realized, with
the commissioning and operation of kilometer-scale, laser interferometric
gravitational-wave observatories in the U.S. (LIGO pro\-ject), Europe
(VIRGO and GEO600 projects) and Japan (TAMA300 project).  
Gravit\-at\-ion\-al-wave searches at these observatories are scheduled to
commence around 2002.  The LIGO 
broad-band antennae will have the capability of detecting and
measuring the gravitational waveforms from astronomical sources in a
frequency band between about 10 Hz (the seismic noise cutoff) and 
500 Hz (the photon counting noise cutoff), with a maximum
sensitivity to strain at around 100 Hz of $h \sim \Delta l/l \sim 10^{-22}$
(rms).  The most promising source for detection and study of
the gravitational-wave signal is the ``inspiralling compact binary''
-- a binary system of neutron stars or black holes (or one of each) in
the final minutes of a death dance leading to a violent merger.
Such is the fate, for example, of the Hulse-Taylor binary pulsar PSR
1913+16 in about 300 million years.  Given the expected sensitivity of the
``advanced LIGO'' (around 2007), which could see such sources out to
hundreds of megaparsecs, it has been estimated that from 3 to
100 annual inspiral events could be detectable.
Other sources, such as supernova core collapse events, instabilities
in rapidly rotating nascent neutron stars, signals from
non-axisymmetric pulsars, and a stochastic background of waves, may be
detectable (for reviews, see \cite{LIGO,snowmass}; for updates on
the status of various projects, see \cite{fritschel,brillet}).    

A similar
network of cryogenic resonant-mass gravitational antennae have  been  in
operation for many years, albeit at lower levels of sensitivity
($h \sim 10^{-19}$).  While modest improvements in sensitivity may be
expected in the future, these resonant detectors are not expected to
be competitive with the large interferometers, unless new designs
involving bars of spherical, or nearly spherical shape come to
fruition.  These systems are primarily sensitive to waves in
relatively narrow bands about frequencies in the hundreds to thousands
of Hz range \cite{rome,allegro,niobe,auriga}.

In addition, plans are being developed for an orbiting laser
interferometer space antenna (LISA for short).  Such a system,
consisting of three spacecraft separated by millions of kilometers,
would be sensitive primarily in the very low frequency band between
$10^{-4}$ and $10^{-1}$ Hz, with peak strain sensitivity of order $h \sim
10^{-23}$ \cite{danzmann}.   

In addition to opening a new astronomical window, the 
detailed observation of gravitational waves by such observatories may
provide the means to test general relativistic predictions for the
polarization and speed of the waves, and for gravitational radiation
damping. 

\subsection{Polarization of gravitational waves}
   
\begin{figure}
\begin{center}
\leavevmode
\psfig{figure=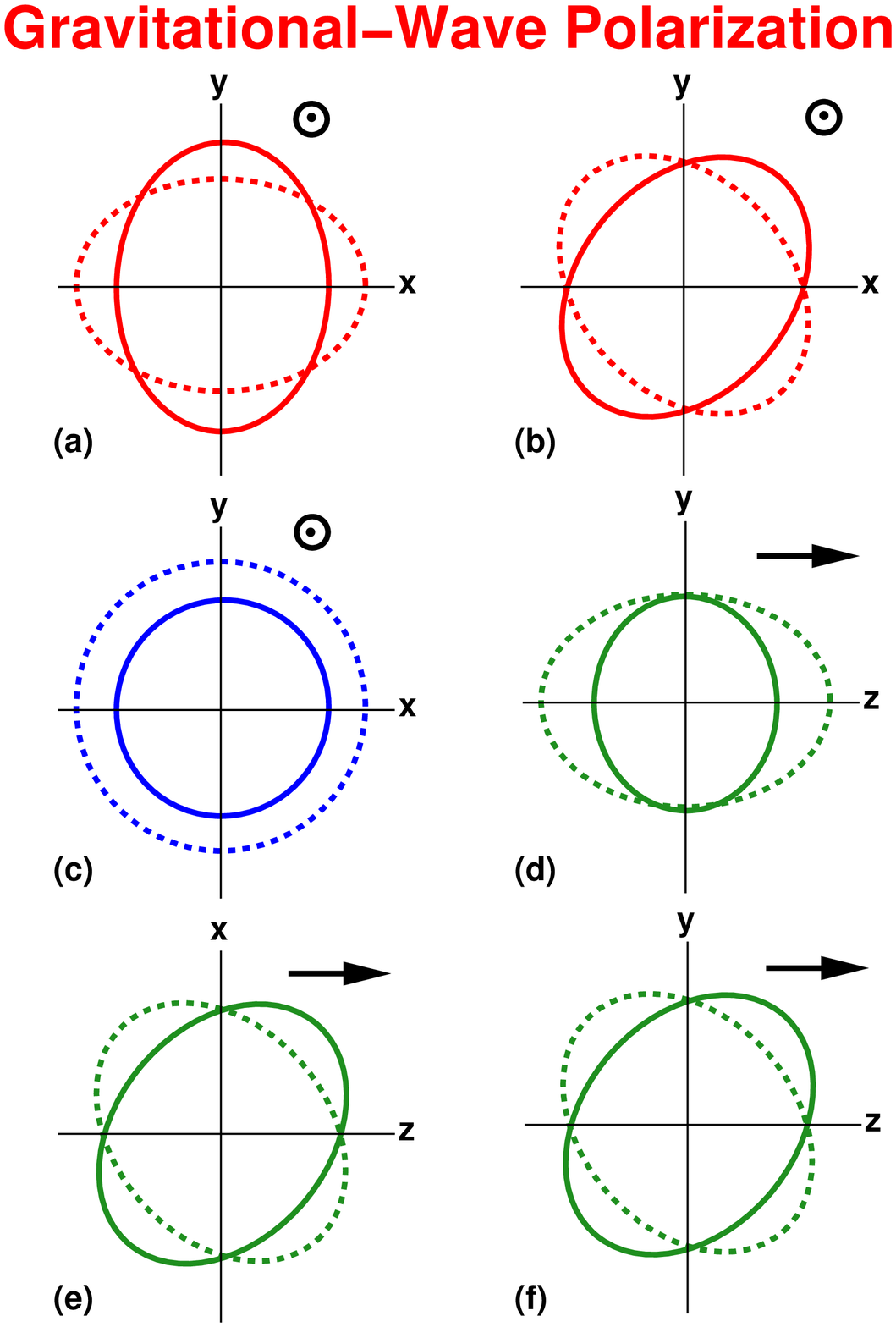,height=5.0in}
\end{center}
\caption{Six polarization modes for gravitational waves permitted in
any metric theory of gravity.  Shown is the displacement that each
mode induces on a ring of test particles.  The wave propagates in the
$+z$ direction.  There is no displacement out of the plane of the
picture.  In (a), (b) and (c), the wave propagates out of the plane;
in (d), (e), and (f), the wave propagates in the plane.  In general
relativity, only
(a) and (b) are present; in scalar-tensor gravity, (c) may also be
present.}
\label{wavemodes}
\end{figure}

A laser-interferometric or resonant bar gravitational-wave detector 
measures the local components of a symmetric $3\times3$ tensor which
is composed of the ``electric'' components of the Riemann curvature tensor,
$R_{0i0j}$, via the equation of geodesic deviation, given, for a pair
of freely falling particles by
$ {\ddot x}^i =  - R_{0i0j} x^j $,
where $x^i$ denotes the spatial separation.  
In general there are six independent components, which can be expressed in
terms of polarizations (modes with specific transformation properties
under rotations and boosts).  Three are transverse to the direction of
propagation, with two representing quadrupolar deformations and one
representing a monopole ``breathing'' deformation.  Three modes are
longitudinal, with one an axially symmetric 
stretching mode in the propagation direction,
and one quadrupolar mode in each of the two orthogonal planes containing the
propagation direction.  Figure \ref{wavemodes} shows the displacements
induced on a ring of freely falling test particles by each of these
modes.  
General relativity predicts only the first two
transverse quadrupolar modes (a) and (b) independently of the source; these
correspond to the waveforms $h_+$ 
and $h_\times$ 
discussed earlier (note the $\cos 2 \phi $ and $\sin 2 \phi $
dependences of the displacements) . 
Scalar-tensor gravitational waves
can in addition contain the transverse breathing mode (c).  More general
metric theories predict additional longitudinal modes,
up to the full complement of 
six (TEGP 10.2).

A suitable array of gravitational antennas could delineate or limit
the number of modes present in a given wave.  The strategy depends on
whether or not the source direction is known.
In general there are eight unknowns (six polarizations and two direction
cosines), but only six measurables ($R_{0i0j}$).  If the direction can
be established by either association of the waves with optical or
other observations, or by time-of-flight measurements between
separated detectors, then six suitably oriented detectors suffice to
determine all six components.  If the direction cannot be established,
then the system is underdetermined, and no unique solution can be
found.  However, if one assumes that only transverse waves are
present, then there are only three unknowns if the source direction is
known, or five unknowns otherwise.  Then the corresponding number
(three or five) of detectors can determine the polarization.  If
distinct evidence were found of any mode other than the two 
transverse quadrupolar modes of GR, the result would be disastrous for
GR.  On the other hand, the absence of a breathing mode would not
necessarily rule out scalar-tensor gravity, because the strength
of that mode depends on the nature of the source.  

Some
of the details of implementing such polarization observations have
been worked out for arrays of resonant cylindrical, disk-shaped, 
spherical and truncated icosahedral
detectors (TEGP 10.2, for recent reviews see \cite{lobo,wagoner});
initial work has been done to assess whether
the ground-based or space-based
laser-interferometers (or combinations of the two types) could perform
interesting polarization measurements 
\cite{wagoner2,brunetti,maggiore,gasperini}.  
Unfortunately for this purpose, the two LIGO observatories (in Washington
and Louisiana states, respectively) have been constructed to have their
respective arms as parallel as possible, apart from the curvature of the
Earth; while this maximizes the joint sensitivity of the two detectors to
gravitational waves, it minimizes their ability to detect two modes of
polarization.

\subsection{Gravitational radiation back-reaction}
\label{backreaction}

In the binary pulsar, a test of GR was made possible by measuring at
least three relativistic effects that depended upon only two unknown
masses.  The evolution of the orbital phase under the damping effect
of gravitational radiation played a crucial role.  Another situation
in which measurement of orbital phase can lead to tests of GR is that
of the inspiralling compact binary system.  The key differences are
that here gravitational radiation itself is the detected signal,
rather than radio pulses, and the phase evolution alone carries all
the information.  In the binary pulsar, the first derivative of the
binary frequency, $\dot f_b$, was measured; here the full nonlinear
variation of $f_b$ as a function of time is measured. 

Broad-band laser interferometers
are especially sensitive to the phase evolution of the gravitational
waves, which carry the information about the orbital phase evolution.
The analysis of gravitational-wave data from such sources will involve
some form of matched filtering of the noisy detector output against an
ensemble of theoretical ``template'' waveforms which depend on the
intrinsic parameters of the inspiralling binary, such as the component
masses, spins, and so on, and on its inspiral evolution.  
How accurate must a template be in order to
``match'' the waveform from a given source (where by a match we mean
maximizing the cross-correlation or the 
signal-to-noise ratio)?  In the total accumulated phase
of the wave detected in the sensitive bandwidth, the template must
match the signal to a fraction of a cycle.  For two inspiralling
neutron stars, around 16,000 cycles should be detected during the final few
minutes of inspiral; this implies a
phasing accuracy of $10^{-5}$ or better.  Since $v \sim 1/10$ during
the late inspiral, this means that correction terms in the phasing
at the level of
$v^5$ or higher are needed.  More formal analyses confirm this
intuition \cite{3min,finnchern,cutlerflan,poissonwill}.

Because it is a slow-motion system ($v \sim 10^{-3}$), the binary
pulsar is sensitive only to the lowest-order effects of gravitational
radiation as predicted by the quadrupole formula.  Nevertheless, the
first correction terms of order $v$ and $v^2$ to the quadrupole formula,
were
calculated as early as 1976 (\cite{wagwill}, see TEGP 10.3).  

But for laser-interferometric observations of gravitational waves,
the bottom line is that, in order to measure the astrophysical
parameters of the source and to test the properties of the
gravitational waves, it is necessary to derive the
gravitational waveform and the resulting radiation back-reaction on the orbit
phasing at least to 2PN
order beyond the quadrupole
approximation, and probably to 3PN
order.  

For the special case of 
non-spinning bodies moving on quasi-circular orbits ({\it i.e.}
circular apart from a slow inspiral), the evolution of the
gravitational wave frequency $f = 2f_b$ through 2PN
order has the 
form 
\begin{eqnarray}
\dot f &=&{96\pi \over 5} f^2 
(\pi {\cal M} f)^{5/3}
\biggl [ 1 
- \left ( {743 \over 336} + {11 \over 4} \eta \right ) (\pi mf)^{2/3}
+ 4\pi(\pi mf)  \nonumber \\
&&  + \left ( {34103 \over 18144} + {13661 \over 2016} \eta + {59 \over
18} \eta^2 \right )  (\pi mf)^{4/3} 
+ O[(\pi mf)^{5/3}] \biggr ] \,,
\label{fdot2PN}
\end{eqnarray}
where $\eta= m_1m_2/m^2$.  The first term is the quadrupole
contribution [compare Eq. (\ref{fdotGR})], 
the second term is the 1PN contribution, the
third term, with the coefficient $4\pi$, 
is the ``tail'' contribution,
and the fourth term is the 2PN
contribution, first reported jointly by Blanchet {\it et al.}
\cite{bdiww,bdi2pn,opus}.  
Calculation of the higher-order contributions is nearing completion.

Similar expressions can be derived for the loss of angular momentum
and linear momentum. (For explicit formulas for non-circular
orbits, see \cite{gopu}). These losses react
back on the orbit to circularize it and cause it to inspiral.  The
result is that the orbital phase (and consequently the 
gravitational-wave 
phase) evolves non-linearly with time.  It is the sensitivity of the
broad-band LIGO and VIRGO-type detectors to phase that makes the  
higher-order contributions to $df/dt$ so observationally relevant.
A ready-to-use set
of formulae for the 2PN gravitational waveform template, including the
non-linear evolution of the gravitational-wave frequency (not
including spin effects) have been published\cite{biww}  and
incorporated into the Gravitational Radiation Analysis and Simulation
Package (GRASP), a software toolkit used in LIGO.

If the coefficients of each of the powers of $f$ in Eq. (\ref{fdot2PN})
can be measured, then one again obtains more than two constraints on
the two unknowns $m_1$ and $m_2$, leading to the possibility to test
GR.  For example, 
Blanchet and Sathyaprakash \cite{lucsathya,lucsathya2} have shown that, by
observing a source with a sufficiently strong signal, an interesting
test of the $4\pi$ coefficient of the ``tail'' term could be
performed.  

Another possibility involves gravitational waves from a small mass
orbiting and inspiralling into a (possibly supermassive) spinning black hole.
A general non-circular, non-equatorial orbit will precess around the
hole, both in periastron and in orbital plane, 
leading to a complex gravitational waveform that carries
information about the non-spherical, strong-field spacetime around the hole.
According to GR, this spacetime must be the Kerr spacetime of a
rotating black hole, uniquely specified by its mass and angular
momentum, and consequently, observation of the waves could test this
fundamental hypothesis of GR \cite{ryan,poissonBH}.

Thirdly, the dipole gravitational radiation predicted by scalar-tensor
theories will result in a modification of the  gravitational-radiation
back-reaction, and thereby of the phase evolution.
Including only the leading quadrupole and dipole contributions, one
obtains, in Brans-Dicke theory,
\begin{eqnarray}
\dot f &=&{96\pi \over 5} f^2
(\pi {\cal M} f)^{5/3}
\biggl [ 1 + b (\pi mf)^{-2/3}  \biggr ] \,,
\label{fdotBD}
\end{eqnarray}
where ${\cal M} = (\chi^{3/5} {\cal G}^{-4/5})\eta^{3/5} m$, 
and $b$ is the coefficient of the dipole term, given by
$b= (5/48)(\chi^{-1} {\cal G}^{4/3} )\xi {\cal S}^2$, where
$\chi$, $\cal G$, $\cal S$ are given by Eqs. (\ref {BDcoefficients}),
and $\xi = 1/(2+\omega_{BD})$.
The effects are strongest for systems involving a neutron star and a
black hole.   Double neutron star systems are less
promising because the small range of masses available near
$1.4~M_\odot$ results in suppression of dipole radiation by symmetry
(the sensitivity $s$ turns out to be a relatively weak function of
mass near $1.4~M_\odot$, for typical equations of state).
For black holes, $s=0.5$ identically, consequently
double black-hole systems turn out to be observationally
identical in the two theories.  

\begin{figure}
\begin{center}
\leavevmode
\psfig{figure=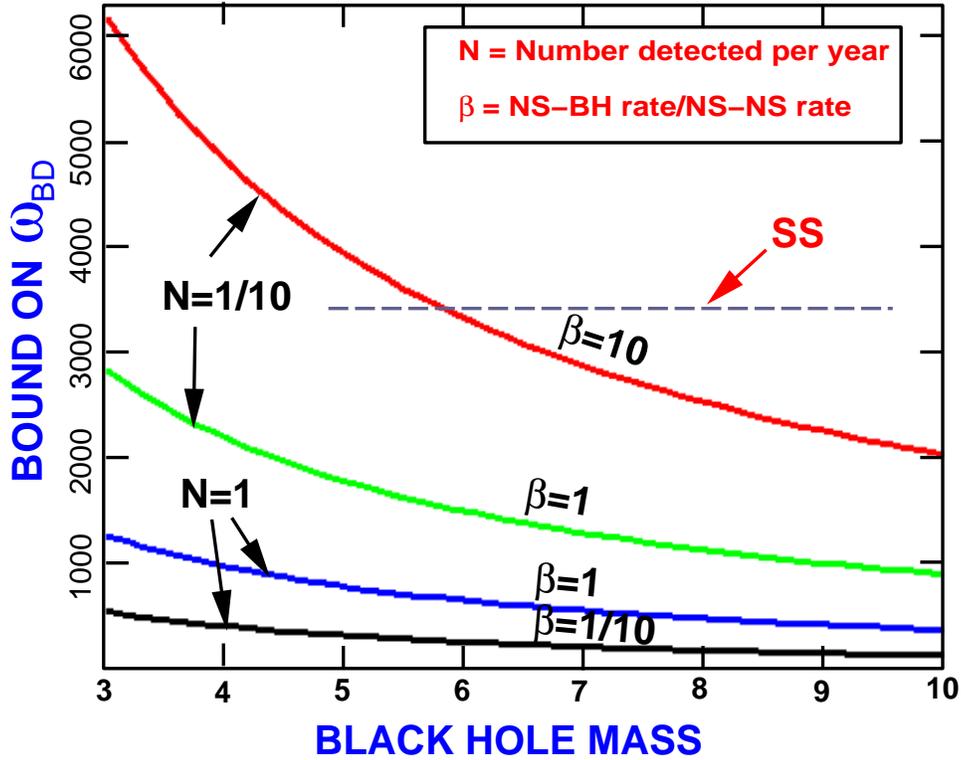,height=4.0in}
\end{center}
\caption{Bounds on the scalar-tensor coupling constant $\omega_{\rm
BD}$ from
gravitational-wave observations of inspiralling black-hole
neutron-star systems.  The solar system bound is around
$\omega_{\rm BD}=3500$.} 
\label{BDbounds}
\end{figure}

But for a $1.4 ~M_\odot$  
neutron star and a $10~ M_\odot$ ($3~ M_\odot$) black 
hole at 200 Mpc, the bound on
$\omega_{\rm BD}$ could be 600 (1800) (using advanced LIGO noise curves).  
The bound increases linearly with
signal-to-noise ratio \cite{willbd}.  If 
one demands that this test be performed
annually, thus requiring observation of frequent, and therefore
more distant, 
weaker sources, the bounds on $\omega_{\rm BD}$ will be too weak to compete with
existing solar-system bounds (this corresponds to the ``LIGO-VIRGO''
bound on the $\alpha_0$ axis in Figure \ref{scalarbounds}, 
which assumes a signal-to-noise ratio
of 10).
However, if one is prepared to wait 10 years for the
lucky observation of a nearby, strong source, the resulting bound
could exceed the current solar-system bound.  
The bounds are illustrated in Figure
\ref{BDbounds} by the curves marked $N=1$ and $N=1/10$.  
Figure \ref{BDbounds} assumes a
double-neutron-star inspiral rate of $10^{-6}$ per year per galaxy,
and a black-hole-neutron-star rate $\beta$ times that, where $\beta$
is highly uncertain.  
Values of $\beta = 1/10 \,,\, 1, \, {\rm and} \, 10$ are shown.
For the general class of scalar-tensor theories, the corresponding bounds
are plotted on the $\alpha_0 - \beta_0$ plane in Figure
\ref{scalarbounds}, under the restrictive assumption of a signal-to-noise
ratio (S/N) of 10 \cite{DamourEspo98}.   
All other factors being equal, the bound achievable on the 
$\alpha_0 - \beta_0$
parameters of scalar-tensor gravity is inversely proportional to S/N.

\subsection{Speed of gravitational waves}

According to GR, in the limit in which the wavelength of gravitational
waves is small compared to the radius of curvature of the background
spacetime, the waves propagate along null geodesics of the background
spacetime, {\it i.e.} they have the same speed, $c$, as light (in this
section, we do not set $c=1$).  In other
theories, the speed could differ from $c$ because of coupling of
gravitation to ``background'' gravitational fields.  For example, in
the Rosen bimetric theory with a flat background metric
$\mbox{\boldmath$\eta$}$,
gravitational waves follow null geodesics of $\mbox{\boldmath$\eta$}$,
while light follows null geodesics of ${\bf g}$ (TEGP 10.1).

Another way in which the speed of gravitational waves could differ
from $c$ is if gravitation were propagated by a massive field (a
massive graviton), in which case, $v_g$ would
be given by, in a local inertial frame,
\begin{equation}
{v_g^2 \over c^2} = 1- {m_g^2c^4 \over E^2} \,,
\label{eq1} 
\end{equation}
where $m_g$ and $E$ are the graviton rest mass and energy,
respectively.  For a recent review of the idea of a massive graviton
along with a model theory,
see \cite{visser}.

The most obvious way to test this is to
compare the arrival times of a gravitational wave and an
electromagnetic
wave from the same event, {\it e.g.} a supernova.
For a source at a distance $D$, the
resulting value of the difference $1-v_g/c$ is
\begin{equation}
1- {v_g \over c}= 5 \times 10^{-17} \left (
{{200 {\rm Mpc}} \over D} \right ) \left ( {{\Delta t}
\over {1 {\rm s}}} \right )
\,,
\label{eq2}
\end{equation}
where
$\Delta t \equiv \Delta t_a - (1+Z) \Delta t_e $ is the ``time difference'', 
where $\Delta t_a$ and $\Delta t_e$ are the differences
in arrival time and emission time, respectively, of the
two signals, and $Z$ is the redshift of the source.
In many cases, $\Delta t_e$ is unknown,
so that the best one can do is employ an upper bound on
$\Delta t_e$ based on observation or modelling.
The result will then be a bound on $1-v_g/c$.

For a massive graviton, if 
the frequency of the gravitational waves is such that $hf \gg
m_gc^2$,
where $h$ is Planck's constant, then $v_g/c \approx 1- {1 \over 2}
(c/\lambda_g
f)^{2}$, where $\lambda_g= h/m_gc$ is the graviton Compton wavelength,
and
the bound on $1-v_g/c$ can be converted to a bound on $\lambda_g$,
given
by
\begin{equation}
\lambda_g > 3 \times 10^{12}~{\rm km} \left ( {D \over {200 ~{\rm
Mpc}}}
{{100 ~{\rm Hz}} \over f} \right )^{1/2} \left ({1 \over
{f\Delta t}} \right )^{1/2} \,.
\label{eq4}
\end{equation}

The foregoing discussion assumes that the source emits {\it both}
gravitational and electromagnetic radiation in detectable amounts, and
that the relative time of emission can be established 
to sufficient accuracy, or can be shown to be sufficiently
small.

However, there is a situation in which a bound on the graviton mass
can be set using gravitational radiation alone \cite{graviton}.
That is the case of
the inspiralling compact binary.  Because the frequency of the
gravitational radiation sweeps from low frequency at the initial
moment of observation to higher frequency at the final moment, the
speed of the gravitons emitted will vary, from lower speeds initially
to higher speeds (closer to $c$) at the end.  This will cause a
distortion of the observed phasing of the waves and result in a
shorter than expected
overall time $\Delta t_a$ of passage of a given number of cycles.
Furthermore, through the technique of matched filtering, the
parameters of the compact binary can be measured accurately,
(assuming that GR is a good approximation to the orbital evolution, even in
the presence of a massive graviton), and
thereby the emission time $\Delta t_e$ can be determined accurately.
Roughly speaking, the ``phase interval'' $f\Delta t$ in Eq.
(\ref{eq4}) can be
measured to an accuracy $1/\rho$, where $\rho$ is the
signal-\-to-\-noise
ratio.

Thus one can estimate the bounds on $\lambda_g$ achievable for various
compact inspiral systems, and for various detectors.   For
stellar-mass inspiral (neutron stars or black holes) observed by the
LIGO/VIRGO class of ground-based interferometers, $D \approx
200 {\rm Mpc}$, $f \approx 100 {\rm Hz}$, and $f\Delta t \sim
\rho^{-1} \approx 1/10$.
The result is $\lambda_g > 10^{13}
{\rm km}$.  For supermassive binary black holes ($10^4$ to $10^7
M_\odot$) observed by the proposed laser-interferometer space antenna
(LISA), $D \approx 3 {\rm Gpc}$, $f \approx 10^{-3} {\rm Hz}$,
and $f\Delta t \sim
\rho^{-1} \approx 1/1000$. The result is $\lambda_g >
10^{17}~ {\rm km}$.

A full noise analysis using proposed noise curves for the advanced
LIGO and for LISA weakens these crude bounds by factors between two and 10.
These potential bounds can be compared
with the solid bound $\lambda_g > 2.8 \times 10^{12} \, {\rm km}$, 
\cite{talmadge} 
derived from solar system dynamics, which limit
the presence of a Yukawa modification of Newtonian gravity of the form
\begin{equation}
V(r) =(GM/r)\exp(-r/\lambda_g) \,,
\end{equation}
and with the model-dependent bound 
$\lambda_g > 6\times 10^{19} \,{\rm km}$ from consideration of
galactic and cluster dynamics \cite{visser}.

\subsection{Other strong-gravity tests}\label{other}

One of the central difficulties of testing general relativity in the
strong-field regime is the possibility of contamination by uncertain
or complex physics.  In the solar system, weak-field gravitational effects 
could
in most cases be measured cleanly and separately from
non-gravitational effects.  The remarkable cleanliness of the binary
pulsar permitted precise measurements of gravitational phenomena
in a strong-field context.  

Unfortunately, nature is rarely so kind.
Still, under suitable conditions, qualitative and even quantitive
strong-field tests of general relativity can be carried out.

One example is in cosmology.  From a few seconds after the big bang
until the present, the underlying physics of the universe is well
understood, although significant uncertainties remain (amount of dark
matter, value of the cosmological constant, the number of light
neutrino families, \etc.).  Some alternative theories of gravity that
are qualitatively different from GR fail to produce cosmologies that
meet even the minimum requirements of agreeing qualitatively with big-bang
nucleosynthesis (BBN) or the properties of the cosmic microwave background
(TEGP 13.2).
Others, such as Brans-Dicke theory, are sufficiently close to GR (for
large enough $\omega_{BD}$) that they conform to all cosmological
observations, given the underlying uncertainties.  The generalized
scalar-tensor theories, however, could have small $\omega_{BD}$ at
early times, while evolving through the attractor mechanism to large
$\omega_{BD}$ today.  One way to test such theories is through
big-bang nucleosynthesis, since the abundances of the light elements
produced when the temperature of the universe was about 1 MeV are
sensitive to the rate of expansion at that epoch, which in turn
depends on the strength of interaction between geometry and the
scalar field.  Because the universe is radiation-dominated at that
epoch, uncertainties in the amount of cold dark matter or of the
cosmological constant are unimportant.  The nuclear reaction rates are
reasonably well understood from laboratory experiments and theory, and
the number of light neutrino families (3) conforms to evidence from
particle accelerators.  Thus, within modest uncertainties, one can
assess the quantitive difference between the BBN predictions of GR and
scalar-tensor gravity under strong-field conditions and compare with
observations.
The most sophisticated recent analysis
\cite{damourpichon} places bounds on the parameters $\alpha_0$ and
$\beta_0$ of the generalized framework of Damour and Esposito-Far\`ese
(see Sec. \ref{binarypulsarsscalar} and Fig. \ref{scalarbounds}) that
are weaker than solar-system bounds for $\beta_0 < 0.3$, but
substantially stronger for  $\beta_0 > 0.3$.

Another example is the exploration of the spacetime near black holes
via accreting matter.  Observations of low-luminosity binary X-ray
sources suggest that a form of accretion known as advection-dominated
accretion flow (ADAF) may be important.  In this kind of flow, the
accreting gas is too thin to radiate its energy efficiently, but
instead transports (advects) it inward toward the central object.  If
the central object is a neutron star, the matter hits the surface and
radiates the energy away; if it is a black hole, the matter and its
advected energy disappear.   Systems in which the accreting object is
believed to be a black hole from estimates of its mass are indeed
observed to be underluminous, compared to systems where the object is
believe to be a neutron star.  This has been regarded as the first
astrophysical evidence for the existence of black hole event horizons
(for a review, see \cite{narayan}).  While supporting one of the
critical strong-field predictions of GR, the observations and models are not
likely any time soon to be able to distinguish one gravitational 
theory from another (except for theories that do not predict black
holes at all).

Another example involving accretion purports to explore the
strong-field region just outside massive black holes in active
galactic nuclei.  Here, iron in the
inner region of a thin
accretion disk is irradiated by X-ray-emitting material above or below the
disk, and fluoresces in the $K\alpha$ line.  The spectral shape of the
line depends on relativistic Doppler and curved-spacetime effects as
the iron orbits the black hole near the innermost stable circular
orbit, and could be used to determine whether the hole is a
non-rotating Schwarzschild black hole, or a rotating Kerr black hole.
Because of uncertainties in the detailed models, the results are
inconclusive to date, but the combination of higher-resolution
observations and better modelling could lead to striking tests of
strong-field predictions of GR.

\section{Conclusions} \label{S5}

We find
that general relativity has held up under extensive experimental scrutiny.
The question then arises, why bother to continue to test it?  One
reason is that gravity is a fundamental interaction of nature, and as
such requires the most solid empirical underpinning we can provide.
Another is that all attempts to quantize gravity and to unify it with
the other forces suggest that the standard general relativity of Einstein is
not likely to be the last word.
Furthermore, the predictions of general relativity are fixed; 
the theory contains 
no adjustable constants so nothing can be changed.  Thus every test 
of the theory is either a potentially deadly test or a possible probe for
new physics.  Although it is remarkable 
that this theory, born 80 years ago out of almost pure thought, 
has managed to survive every test, the possibility of finding 
a discrepancy will continue to drive experiments for years to come.

\section*{Acknowledgments}

This work was supported in part by the National Science Foundation,
Grant Number PHY 96-00049.

\end{document}